\pdfoutput=1

\documentclass[11pt]{article}
\usepackage{jheppub}

\usepackage{amsmath}
\usepackage{amssymb}
\usepackage{graphicx,xcolor,slashed}
\usepackage{enumerate}
\usepackage{ifpdf}
\usepackage[english]{babel}
\usepackage{url}
\usepackage{booktabs}
\usepackage{xspace}
\usepackage{xinttools}
\usepackage{mathrsfs}

\notoc

\renewcommand{\baselinestretch}{1.18}

\definecolor{darkgreen}{rgb}{0.0, 0.26, 0.15}
\definecolor{darkred}{rgb}{0.65,0.15,0}

\usepackage{float}

\hypersetup{bookmarksnumbered,bookmarksdepth=3}

\allowdisplaybreaks[1]

\DeclareFontFamily{U}{mathx}{\hyphenchar\font45}
\DeclareFontShape{U}{mathx}{m}{n}{
      <5> <6> <7> <8> <9> <10>
      <10.95> <12> <14.4> <17.28> <20.74> <24.88>
      mathx10
      }{}
\DeclareSymbolFont{mathx}{U}{mathx}{m}{n}
\DeclareFontSubstitution{U}{mathx}{m}{n}
\DeclareMathAccent{\widecheck}{0}{mathx}{"71}

\restylefloat{figure}
\definecolor{dgreen}{rgb}{0,0.70,0.30}
\definecolor{gold}{rgb}{0.85,.66,0}
\definecolor{purple}{rgb}{1.0,0.3,0.6}

\usepackage{tikz}
\usetikzlibrary{calc} \usetikzlibrary{patterns} \usetikzlibrary{decorations.pathreplacing} \usetikzlibrary{decorations.markings} \usetikzlibrary{decorations.pathmorphing} \usetikzlibrary{positioning} \usetikzlibrary{bending}

\newcommand{\bea}{\begin{eqnarray}}
\newcommand{\eea}{\end{eqnarray}}
\def\beq{\begin{equation}}
\def\eeq{\end{equation}}

\newcommand{\tree}{{\rm tree}}
\newcommand{\oneloop}{^{(1)}}
\newcommand{\maxsusy}{_{\text{max}}}

\newcommand{\hyper}{{(0,1,2,D=6)}}

\newcommand{\trv}{{\rm tr}_{\rm V}}
\newcommand{\trs}{{\rm tr}_{\rm S}}
\newcommand{\trodd}[1]{{\rm tr}_{\rm odd}(#1)}
\newcommand{\treven}[1]{\ensuremath{\tr(#1)\Big|_{\text{even}}}}
\newcommand{\PT}{{\rm PT}}
\newcommand{\FWL}{{\rm FWL}}

\newcommand{\al}{\alpha}
\newcommand{\be}{\beta}
\newcommand{\ga}{\gamma}
\newcommand{\la}{\lambda}
\newcommand{\ep}{\epsilon}

\newcommand{\dd}{\mathrm{d}}
\newcommand{\te}{\textrm}

\newcommand{\ZZ}{\mathbb Z}

\newcommand{\QQ}{\mathbb Q}

\newcommand{\varep}{\varepsilon}

\newcommand{\oslashed}[1]{\ensuremath{\slash \! \! \! #1}}

\DeclareMathOperator{\ord}{ord}
\DeclareMathOperator{\sgn}{sgn}
\DeclareMathOperator{\Pf}{Pf}
\DeclareMathOperator{\coef}{coef}
\DeclareMathOperator{\tr}{tr}
\DeclareMathOperator{\sk}{skip}

\DeclareRobustCommand{\eulern}{\genfrac{\langle}{\rangle}{0pt}{}}

\newcommand{\TE}[2][]{\ensuremath{%
    t_8^{#1}(%
    \xintFor ##1 in {#2} \do{%
      \xintifForLast{%
        ##1}{%
        ##1,%
      }
    }%
    )%
  }}

\newcommand{\TEf}[2][]{\ensuremath{%
    t_8^{#1}(%
    \xintFor ##1 in {#2} \do{%
      \xintifForLast{%
        f_{##1}}{%
        f_{##1},%
      }
    }%
    )%
  }}

\newcommand{\miss}[1]{\tilde{#1}}
\newcommand{\ferm}[1]{\hat{#1}}

\newcommand{\eg}{{e.g.~}}
\newcommand{\ie}{{i.e.~}}
\newcommand{\wrt}{w.r.t.~}

\usepackage[format=plain,
            labelfont=bf,
            textfont=it]{caption}

\title{One-loop Correlators and BCJ Numerators from Forward Limits}

\author[a]{Alex Edison,}
\author[b,c,d]{Song He,}
\author[a]{Oliver Schlotterer,}
\author[a]{Fei Teng}

\affiliation[a]{Department of Physics and Astronomy, Uppsala University, SE-75108 Uppsala, Sweden}
\affiliation[b]{CAS Key Laboratory of Theoretical Physics, Institute of Theoretical Physics, Chinese Academy of Sciences, Beijing 100190, China}
\affiliation[c]{School of Physical Sciences, University of Chinese Academy of Sciences, No.19A Yuquan Road, Beijing 100049, China}
\affiliation[d]{School of Fundamental Physics and Mathematical Sciences, Hangzhou Institute for Advanced Study, UCAS, Hangzhou 310024 and
International Centre for Theoretical Physics Asia-Pacific, Beijing/Hangzhou, China}

\emailAdd{alexander.edison@physics.uu.se}
\emailAdd{songhe@itp.ac.cn}
\emailAdd{oliver.schlotterer@physics.uu.se}
\emailAdd{fei.teng@physics.uu.se}

\date{\today}

\abstract{We present new formulas for one-loop ambitwistor-string
  correlators for gauge theories in any even dimension with arbitrary
  combinations of gauge bosons, fermions and scalars running in the
  loop. Our results are driven by new all-multiplicity expressions for tree-level
  two-fermion correlators in the RNS formalism that closely resemble the purely bosonic ones. After taking
  forward limits of tree-level correlators with an additional pair of
  fermions/bosons, one-loop correlators become combinations of Lorentz
  traces in vector and spinor representations. Identities between these
  two types of traces manifest all supersymmetry cancellations and the power 
  counting of loop momentum. We also obtain parity-odd contributions from forward
  limits with chiral fermions. One-loop numerators satisfying the
  Bern-Carrasco-Johansson (BCJ) duality for diagrams with linearized
  propagators can be extracted from such correlators using the
  well-established tree-level techniques in Yang-Mills theory coupled to biadjoint scalars. 
  Finally, we obtain streamlined expressions for BCJ
  numerators up to seven points using multiparticle fields.}

\preprint{UUITP--11/20}

\begin{document}

\maketitle{}

\newpage

\setcounter{page}{1}
\pagenumbering{roman}

\setcounter{tocdepth}{3}

%Tweak this value to change the line spacing in the TOC

\renewcommand{\baselinestretch}{.99}\normalsize

\tableofcontents
% Leave this value as 1.18 to reset the text to "correct"
% spacing
\renewcommand{\baselinestretch}{1.18}\normalsize

\numberwithin{equation}{section}

 \newpage

%%%%%%%%%%%%%%%%%%%%%%%%%%%%%%%%%%%%%%%%%%%%%%%% 
%%%%%%%%%%%%%%%%%%%%%%%%%%%%%%%%%%%%%%%%%%%%%%%% 
%%%%%%%%%%%%%%%%%%%%%%%%%%%%%%%%%%%%%%%%%%%%%%%% 
%%%%%%%%%%%%%%%%%%%%%%%%%%%%%%%%%%%%%%%%%%%%%%%% 
%%%%%%%%%%%%%%%%%%%%%%%%%%%%%%%%%%%%%%%%%%%%%%%% 

\section{Introduction}
\label{sec:intro}

\setcounter{page}{1}
\pagenumbering{arabic}

Recent years have witnessed enormous progress in understanding novel
structures and symmetries of scattering amplitudes in various
theories, as well as surprising connections between them. One
important example is the Bern-Carrasco-Johansson (BCJ) duality between
color and kinematics in gauge theories, and double-copy relations to
corresponding gravity theories~\cite{BCJ, loopBCJ, Bern:2017yxu}, see
\cite{Bern:2019prr} for a review.

The color-kinematic duality states that in a trivalent-diagram
expansion of gauge-theory amplitudes, kinematic factors can be
arranged to satisfy the same algebraic relations as color factors.
Kinematic factors with this property are known as BCJ numerators. 
Based on this, a remarkable conjecture is that loop integrands for 
gravity amplitudes can be obtained from gauge-theory ones by
simply substituting color factors for another copy of such BCJ
numerators.  At tree level, the double copy is equivalent to the
field-theory limit of the famous Kawai-Lewellen-Tye (KLT) relations
between open- and closed-string amplitudes~\cite{Kawai:1985xq}, and
the BCJ duality has also been proven directly~\cite{Bern:2010yg}. In the
quantum regime, this double-copy construction has led to great
advances in the study of the ultraviolet behavior of supergravity
amplitudes~\cite{Bern:2012uf, Bern:2012cd, Bern:2013uka, Bern:2014sna,
  Bern:2017ucb, Bern:2018jmv}. However, it remains a conjecture and
the principle behind it is poorly understood.

Apart from the original KLT relations, string theory has provided
constructions of BCJ numerators at tree and loop
level~\cite{Mafra:2011kj,Mafra:2011nv, Mafra:2014gja, He:2015wgf,
  Mafra:2015mja}.\footnote{Similarly, the gauge invariant reformulation of the color-kinematics
duality via BCJ relations can be elegantly derived from monodromy properties
of open-string worldsheets \cite{BjerrumBohr:2009rd, Stieberger:2009hq}. See \cite{Tourkine:2016bak, 
Hohenegger:2017kqy, Ochirov:2017jby, Tourkine:2019ukp, Casali:2019ihm} for loop-level extensions of 
monodromy relations among string amplitudes.} Relatedly, worldsheet methods originating from the
Cachazo-He-Yuan (CHY) formulation~\cite{Cachazo:2013hca,
  Cachazo:2013iea} have been a major driving force in understanding
and extending BCJ duality and the double copy. Based on scattering
equations~\cite{Cachazo:2013gna}, CHY formulas express tree amplitudes
in a large class of massless theories as worldsheet integrals which
can often be derived from ambitwistor string theories~\cite{Mason:2013sva,
  Berkovits:2013xba, Adamo:2013tsa, Adamo:2015hoa,
  Casali:2015vta}. These methods have not only led to new double-copy
realizations and connections for various
theories~\cite{Cachazo:2014xea,
  Chiodaroli:2014xia,Chiodaroli:2017ngp}, using loop-level
CHY/ambitwistor strings~\cite{Adamo:2013tsa,Geyer:2015bja,
  Geyer:2015jch, Cachazo:2015aol,Geyer:2016wjx,Geyer:2017ela,
  Geyer:2018xwu, Geyer:2019hnn}, they have also extended KLT and BCJ
double copy to one-loop level~\cite{He:2016mzd,He:2017spx}. Based on
nodal Riemann spheres, loop-level CHY/ambitwistor-string formulas yield
loop amplitudes in a new representation of their Feynman integrals
with propagators linear in loop momenta; alternatively they can be
understood as forward limits of tree amplitudes with a pair of momenta
in higher dimensions~\cite{He:2015yua, Cachazo:2015aol}.

In this paper, we continue the study of the loop-level BCJ duality and
double copy based on worldsheet methods. In particular, we present new results on all-multiplicity one-loop BCJ numerators for Feynman integrals 
with propagators linear in loop momentum, which extends and offers a new perspective on the algorithm of~\cite{He:2017spx}. 
Starting from the worldsheet correlator with external gluons, one can obtain $n$-gon master numerators by extracting the coefficient of Parke-Taylor factors with all possible orderings. As reviewed in appendix \ref{app:linprop}, 
BCJ representations of one-loop integrands with linearized propagators
arise naturally from one-loop CHY formulas \cite{Geyer:2015bja, Geyer:2015jch},\footnote{See \cite{Cardona:2016bpi, Cardona:2016wcr} for an alternative approach to one-loop CHY formulas based on the 
$\Lambda$ scattering equations~\cite{Gomez:2016bmv}.} 
also see \cite{Gomez:2016cqb, Gomez:2017lhy, Gomez:2017cpe, Ahmadiniaz:2018nvr, Agerskov:2019ryp}
for the relation between linearized and quadratic propagators. 
%%%

In the RNS formulation of the ambitwistor string \cite{Mason:2013sva,
  Adamo:2013tsa}, the correlator takes the form of a one-loop Pfaffian,
where the amount of spacetime supersymmetry is reflected by the
relative weights of different spin structures\footnote{Spin structures
  refer to the boundary conditions of the worldsheet fermions in the
  RNS formalism as they are taken around the two homology cycles of
  the worldsheet torus. The contributions of individual spin
  structures to the one-loop correlators are weighted by partition
  functions that reflect the amount of spacetime supersymmetry. The
  interplay between different spin structures in multiparticle
  correlators has been studied in the context of conventional strings
  \cite{Tsuchiya:1988va, Stieberger:2002wk, Bianchi:2006nf,
    Broedel:2014vla, Berg:2016wux} and ambitwistor strings
  \cite{He:2017spx}.} \cite{Geyer:2015jch}. The key of the algorithm
in \cite{He:2017spx} is to reduce the dependence on worldsheet punctures to Parke-Taylor factors
via repeated use of one-loop scattering equations, which can be rather
tedious in practice.
It is thus highly desirable to tame this technical difficulty by using
a representation of the one-loop correlator that is more suitable for
extracting BCJ numerators. This is one of the major achievements of the
current paper.

The simplest one-loop correlators arise when the states of $D=10$
super Yang--Mills (SYM) circulate in the loop. As we will review
shortly, these one-loop correlators receive contributions from forward
limits of tree-level correlators with an additional pair of bosons
(gluons) and those with fermions (gluinos)\footnote{We remark that
  tree-level correlators and BCJ numerators for any combination of
  external bosons and fermions can be extracted from their
  representation in pure-spinor superspace
  \cite{Mafra:2011kj,Mafra:2011nv, Gomez:2013wza}, see
  \cite{Berkovits:2013xba, Adamo:2015hoa} for a pure-spinor
  incarnation of the ambitwistor string. Even though the
  extraction of components can be obtained for any number of legs
  \cite{Lee:2015upy, Mafra:2015vca}, these are not the correlator
  representations that we will use in the forward-limit analysis of
  this work. One-loop correlators in pure-spinor superspace up to
  and including seven external legs can be found in \cite{Mafra:2016nwr,  
  Mafra:2018nla, Mafra:2018qqe}.}.
Since tree-level correlators of bosons can be simplified to the
well-known Pfaffian~\cite{Cachazo:2013hca}, it is highly desirable to
also bring the two-fermion correlators into Pfaffian form in order to
control the supersymmetry cancellations between their forward
limits. For this purpose we will derive a new representation of the
two-fermion correlator tailored to expose its interplay with the
bosonic correlator under forward limits. This representation realizes
the gluing-operator prescription of Roehrig and Skinner
\cite{Roehrig:2017gbt}.

Similarly we will derive one-loop correlators for general gauge
theories in even dimension $D<10$ via forward limits in an arbitrary
combination of scalars, fermions and gauge bosons in the
loop\footnote{The algorithm for one-loop BCJ numerators in
  \cite{He:2017spx} has been formulated for gauge theories with at
  least four supercharges, and non-supersymmetric four-point BCJ
  numerators have been derived from forward limits in
  \cite{Geyer:2017ela}. The method here certainly applies to
 the non-supersymmetric case in absence of fermion correlators.}. The
main advantage of our new representations of fermionic correlators is
that the extraction of BCJ numerators becomes a problem that has been
solved at tree level: the dependence on worldsheet punctures of
one-loop correlators turns out to be identical to that of tree
correlators for single-trace amplitudes in Yang-Mills coupled to
biadjoint scalars (YM${+}\phi^3$)~\cite{Cachazo:2014xea}. The
reduction of the latter to Parke-Taylor factors (or equivalently extracting
BCJ numerators for such amplitudes~\cite{Fu:2017uzt,Chiodaroli:2017ngp}) has been studied
extensively~\cite{Cardona:2016gon, Nandan:2016pya, Bjerrum-Bohr:2016axv, 
Schlotterer:2016cxa, Teng:2017tbo, AlexFei},
and we can directly apply these results to our one-loop problem.

As a result, we will present new expressions for BCJ numerators, not
only for ten-dimensional SYM but also for lower-dimensional gauge
theories with reduced or without spacetime supersymmetry. The
numerators of this work manifest the power counting of loop momenta by
representation-theoretic identities between Lorentz traces in vector
and spinor representations. Moreover, our construction preserves
locality, \ie the BCJ numerators do not involve any poles in momentum
invariants.

Furthermore, we will also present two new results on one-loop
correlators and BCJ numerators. First, we will compute parity-odd
contributions to the correlators by taking forward limits with chiral
fermions, both in $D=10$ SYM and in the $D=6$ case with a chiral
spectrum. In addition, we will simplify the BCJ numerators using the
so-called multiparticle fields~\cite{Mafra:2014oia, Mafra:2015vca},
which can be viewed as numerators of Berends-Giele
currents~\cite{Berends:1987me} that respect color-kinematics duality,
derived in the BCJ gauge~\cite{Lee:2015upy, Bridges:2019siz}.

\subsection{Conventions}

In the conventions of this paper, the CHY representation of tree-level amplitudes with a double-copy
structure is given by 
\begin{align}\label{eq:CHY}
{\cal M}^{\text{tree}}_{L \otimes R} = \mathscr{N} 
% -2\left(-\frac{g}{\sqrt{2}}\right)^{n-2} 
 \int \dd \mu_n^{\text{tree}}
\,\mathcal{I}_L^{\text{tree}}\,\mathcal{I}_R^{\text{tree}}\,,& &\dd\mu_n^{\text{tree}}\equiv\frac{\dd^n\sigma}{\text{Vol}[\text{SL}(2,\mathbb{C})]}\sideset{}{^{\,\prime}}\prod_{i=1}^{n}\delta(E_i)\,,
\end{align}
where the theory-dependent normalization factor $\mathscr{N} $ for instance specializes
to $-2 (-\frac{g}{\sqrt{2}})^{n-2} $ for gauge-theory amplitudes
with YM coupling $g$.\footnote{The combination $g/\sqrt{2}$ in the
normalization factor $\mathscr{N} $ of gauge-theory amplitudes
can be understood as rescaling the color factors.}
Inside the CHY measure $\dd\mu_n^{\text{tree}}$, the prime along with the product $\sideset{}{^{\,\prime}}\prod$ instructs to only impose the $n{-}3$ independent scattering equations for the punctures $\sigma_j \in \mathbb C$
on the Riemann sphere, 
\begin{align}
  E_i\equiv\sum_{\substack{j=1 \\ j\neq i}}^{n}\frac{k_i\cdot k_j}{\sigma_{ij}}=0\,,& &\sigma_{ij}\equiv\sigma_i-\sigma_j\,,
\end{align}
see~\cite{AlexFei} for additional details. Depending on the choice of the half-integrands 
$\mathcal{I}_{L,R}^{\text{tree}}$, (\ref{eq:CHY}) can be specialized to yield tree amplitudes in gauge 
theories, (super-)gravity and a variety of further theories \cite{Cachazo:2014xea, Cachazo:2016njl}. 
Color-ordered gauge-theory amplitudes are obtained from a Parke-Taylor factor 
$\mathcal{I}_L^{\text{tree}} \rightarrow ( \sigma_{12} \sigma_{23}\ldots \sigma_{n1})^{-1}$ and taking
$\mathcal{I}^{\text{tree}}_R$ to be the reduced Pfaffian given in \eqref{eq2.6}. The one-loop analogue
of the amplitude prescription (\ref{eq:CHY}) is reviewed in appendix~\ref{app:linprop}.

%%%%%%%%%%%%%%%%%%%%%%%%%%%%%%%%%%%%%%%%%%%%%%%% 
%%%%%%%%%%%%%%%%%%%%%%%%%%%%%%%%%%%%%%%%%%%%%%%% 
%%%%%%%%%%%%%%%%%%%%%%%%%%%%%%%%%%%%%%%%%%%%%%%% 
%%%%%%%%%%%%%%%%%%%%%%%%%%%%%%%%%%%%%%%%%%%%%%%% 
%%%%%%%%%%%%%%%%%%%%%%%%%%%%%%%%%%%%%%%%%%%%%%%% 

\subsection{Summary}
\label{sec:1.1}

The main results of the paper can be summarized as follows. 
\begin{itemize}
\item We present new expressions for tree-level correlators with two and four fermions and any number of bosons. By taking forward limits in a pair of bosons/fermions, we obtain a new formula (\ref{eq3.22}) for one-loop correlators in $D=10$ SYM.
\item By combining building blocks with vector bosons, fermions or
  scalars circulating the loop, we obtain a similar formula
  (\ref{eq4.6}) for one-loop correlators in general, possibly
  non-supersymmetric gauge theories in $D<10$.
\item Since the worldsheet dependence is identical to that of single-trace
  correlators for (YM${+}\phi^3$) tree amplitudes, we can recycle
  tree-level results to extract one-loop BCJ numerators in these
  theories.
\item We will derive parity-odd contributions (\ref{eq2.8new}) to 
one-loop correlators from forward limits with chiral fermions.
\item We present various BCJ numerators at $n\leq 7$ points in a
  compact form by using the multiparticle fields.
\end{itemize}

%%%%%%%%%%%%%%%%%%%%%%%%%%%%%%%%%%%%%%%%%%%%%%%% 
%%%%%%%%%%%%%%%%%%%%%%%%%%%%%%%%%%%%%%%%%%%%%%%% 
%%%%%%%%%%%%%%%%%%%%%%%%%%%%%%%%%%%%%%%%%%%%%%%% 
%%%%%%%%%%%%%%%%%%%%%%%%%%%%%%%%%%%%%%%%%%%%%%%% 
%%%%%%%%%%%%%%%%%%%%%%%%%%%%%%%%%%%%%%%%%%%%%%%% 

The paper is organized as follows. We start in sec.~\ref{sec:rev} by
collecting some results which will be used in the subsequent: First we
spell out tree-level correlators with $n$ bosons and those
with $n{-}2$ bosons and $2$ fermions in the RNS formalism for
ambitwistor string theory. Then we review how the tree-level input 
can be used to construct one-loop correlators by taking the forward limit 
in a pair of bosons or fermions with momenta in higher dimensions.

Next, we study one-loop correlators and BCJ numerators in $D=10$ SYM
in sec.~\ref{sec:loop} and express them as combinations of vector traces and spinor
traces of linearized field strengths with accompanying Pfaffians. We then
propose a key formula (\ref{eq3.11}) for
converting spinor traces to vector traces, which allows us to simplify
the one-loop correlators of $D=10$ SYM.
In particular, the power counting in loop momentum follows from
representation-theoretic identities between vector and spinor
traces. Once the correlator is written in this form, it is straightforward
produce BCJ numerators as the problem is equivalent to that for
tree-level amplitudes in YM${+}\phi^3$.

We move to general gauge theories in even dimensions $D<10$ in sec.~\ref{sec:gen}. 
By also including one-loop correlators from forward
limits in two scalars, we obtain a general formula for the case with
${\bf n_{\rm v}}$ vectors, ${\bf n_{\rm f}}$ Weyl fermions and ${\bf n_{\rm s}}$
scalars. In particular, we apply the general formula to obtain explicit
results for specific theories in $D=6$ and $D=4$.

In sec.~\ref{sec:odd}, we derive parity-odd contributions to one-loop
correlators from forward limits in chiral fermions, which are
parity-odd completions of correlators in $D=10$ SYM and those in lower
dimensions. Finally, in sec.~\ref{sec:MPF}, by using multiparticle
fields, we provide particularly compact expressions for the BCJ
numerators in various theories, which combine contributions from the
Pfaffians and the field-strength traces in the correlators.  

The discussion in the main text is complemented by three appendices:
Our representation of one-loop integrands will be reviewed in appendix \ref{app:linprop}; 
we review CFT basics and give the derivation for tree-level correlators with zero, two and 
four fermions in appendix \ref{bigapp:CFT}; we also prove the identity for reducing 
spinor traces to vector traces in appendix \ref{app:traces}. 

%%%%%%%%%%%%%%%%%%%%%%%%%%%%%%%%%%%%%%%%%%%%%%%% 
%%%%%%%%%%%%%%%%%%%%%%%%%%%%%%%%%%%%%%%%%%%%%%%% 
%%%%%%%%%%%%%%%%%%%%%%%%%%%%%%%%%%%%%%%%%%%%%%%% 
%%%%%%%%%%%%%%%%%%%%%%%%%%%%%%%%%%%%%%%%%%%%%%%% 
%%%%%%%%%%%%%%%%%%%%%%%%%%%%%%%%%%%%%%%%%%%%%%%% 

\section{Basics}
\label{sec:rev}

In this section, we use the RNS formulation of the ambitwistor string
in $D=10$ dimensions \cite{Mason:2013sva, Adamo:2013tsa} (see
\cite{Ramond:1971gb, Neveu:1971rx, DHoker:1988pdl} for the RNS superstring) to review
tree-level correlators with $n$ gluons (bosons). The latter evaluate
to the well-known Pfaffian in the CHY
formulation~\cite{Cachazo:2013hca}, and we will present new
representations for correlators with $2$ gluinos (fermions) and
$n{-}2$ gluons, also see appendix \ref{app:4ferm} for four-fermion correlators. 
On the support of scattering equations, the Pfaffian
can be expanded into smaller ones dressed by Lorentz contractions of
field strengths with two polarizations. As we will see, the correlator
with $2$ gluinos and $n{-}2$ gluons can be simplified to a similar
form, which features smaller Pfaffians dressed by gamma-matrix contracted
field strengths, with wave functions for the two fermions. We will see
that these representations of correlators are most suitable for
combining the forward limits in two gluons/gluinos and studying the
resulting supersymmetry cancellations.

%%%%%%%%%%%%%%%%%%%%%%%%%%%%%%%%%%%%%%%%%%%%%%%% 
%%%%%%%%%%%%%%%%%%%%%%%%%%%%%%%%%%%%%%%%%%%%%%%% 
%%%%%%%%%%%%%%%%%%%%%%%%%%%%%%%%%%%%%%%%%%%%%%%% 
%%%%%%%%%%%%%%%%%%%%%%%%%%%%%%%%%%%%%%%%%%%%%%%% 
%%%%%%%%%%%%%%%%%%%%%%%%%%%%%%%%%%%%%%%%%%%%%%%% 

\subsection{Vertex operators}
\label{sec:vop}

Let us first review the underlying vertex operators for the gluon with
momentum $k_\mu$ and polarization vector $\epsilon^\mu$ with
$\mu=0,1,\ldots,9$ which satisfy on-shell constraint
$k_\mu \epsilon^\mu=0$: 
\begin{align}
V^{(-1)}(\sigma) &\equiv \ep_\mu \psi^\mu(\sigma) e^{-\phi(\sigma)}e^{ik\cdot X(\sigma)}  \ , \ \ \ \ \ \ V^{(0)}(\sigma) \equiv \ep_\mu (P^\mu(\sigma) + (k\cdot \psi) \psi^\mu(\sigma) ) e^{ik\cdot X(\sigma)} \ .
\label{eq2.1}
\end{align}
The superscripts indicate the superghost charges ($-1$ and $0$), and
refer to the contributions from the superghost system by means of a
chiral boson $\phi$ \cite{Friedan:1985ey, Friedan:1985ge}. We work in
conventions where the factors of $\bar \delta(k \cdot P(\sigma))$
enforcing scattering equations \cite{Mason:2013sva} are attributed to
the integration measure in (\ref{eq:CHY}) when assembling amplitudes from the
correlators in this section.

We also introduce the vertex operators for the gluino in $D=10$
spacetime dimensions
\beq V^{(-1/2)}(\sigma) \equiv 2^{-\frac{1}{4}} \chi^\alpha
S_\alpha(\sigma) e^{-\frac{\phi(\sigma)}{2}} e^{ik\cdot X(\sigma)} \ ,
\ \ \ \ \ \ V^{(-3/2)}(\sigma) \equiv 2^{\frac{1}{4}} \xi_\al
S^\al(\sigma) e^{- \frac{3\phi(\sigma)}{2}} e^{ik\cdot X(\sigma)} \ ,
\label{eq2.2}
\eeq
where the superghost charges are $-\frac 1 2$ and $-\frac 3 2$,
respectively, and the normalization factors $2^{\pm \frac{1}{4}}$ are
chosen for later convenience. The fermion wave function $\chi^\alpha$
obeys the on-shell constraint
$k_\mu \gamma^\mu_{\alpha \beta} \chi^\beta = 0$, where Weyl-spinor
indices $\alpha,\beta=1,2,\ldots,16$ in an uppercase and lowercase
position are left-handed and right-handed, respectively.  The dual
wave function $\xi_\al$ in the expression (\ref{eq2.2}) for
$V^{(-3/2)}(\sigma)$ is defined to reproduce
\beq
\chi^\alpha = k^\mu \gamma_\mu^{\alpha \beta}  \xi_\beta\,.
\label{eq2.4}
\eeq
Note that $P^\mu$ and $\psi^\mu$ are the free worldsheet fields of the
RNS model, and $S_\alpha$ is the spin field \cite{Knizhnik:1985ke,
  Cohn:1986bn} (all depending on a puncture $\sigma$ on a Riemann
sphere).  Their operator-product expansions (OPEs) and the resulting
techniques to evaluate tree-level correlators of the vertex operators
(\ref{eq2.1}) and (\ref{eq2.2}) are collected in appendix
\ref{app:CFT}.

%%%%%%%%%%%%%%%%%%%%%%%%%%%%%%%%%%%%%%%%%%%%%%%% 
%%%%%%%%%%%%%%%%%%%%%%%%%%%%%%%%%%%%%%%%%%%%%%%% 
%%%%%%%%%%%%%%%%%%%%%%%%%%%%%%%%%%%%%%%%%%%%%%%% 
%%%%%%%%%%%%%%%%%%%%%%%%%%%%%%%%%%%%%%%%%%%%%%%% 
%%%%%%%%%%%%%%%%%%%%%%%%%%%%%%%%%%%%%%%%%%%%%%%% 

\subsection{Tree-level correlator for external bosons}
\label{sec:bos}

Given gluon vertex operators, one can compute the tree-level correlator for $n$ bosons
\beq
{\cal I}^\tree_{\rm bos}(1,2,\ldots, n 
) = \langle V_1^{(-1)}(\sigma_1) 
V_2^{(0)}(\sigma_2) V_3^{(0)}(\sigma_3)  \ldots V_{n-1}^{(0)}(\sigma_{n-1}) 
V_n^{(-1)}(\sigma_n)  \rangle\, ,
\label{eq2.5}
\eeq
where we have chosen two legs, $1$ and $n$, to have $-1$ superghost
charges. Correlators of this type serve as half-integrands in the CHY formula
(\ref{eq:CHY}) for tree amplitudes. A remarkable feature of the correlator (\ref{eq2.5}) 
is that on the support of scattering equations, it is equivalent to the well-known (reduced) Pfaffian
\beq
{\cal I}^\tree_{\rm bos}(1,2, \ldots, n 
) =\frac{1}{\sigma_{1,n}}
\Pf |\Psi_{ \{12\ldots n\} 
} (\{\sigma, k, \epsilon\}) |_{1,n}\,, \ \ \ \ \ \  \sigma_{i,j}\equiv \sigma_i - \sigma_j\,. 
\label{eq2.6}
\eeq
The $2n\times 2n$ antisymmetric matrix $\Psi$ was first introduced
in~\cite{Cachazo:2013hca}, with columns and rows labelled by the $n$
momenta $k_i$ and polarizations $\epsilon_i$ for $i=1,2, \cdots, n$,
and it also depends on the punctures $\sigma_i$. The entries of $\Psi$
are reviewed in appendix \ref{revpsi} to fix our conventions.

The reduced Pfaffian $\Pf |\ldots |_{1,n}$ in (\ref{eq2.6}) is
defined by deleting two rows and columns $1,n$ of the matrix $\Psi$
with a prefactor $1/\sigma_{1,n}$. More generally, one can
define it by deleting any two columns and rows $1\leq i<j\leq n$ and inserting
a prefactor $ (-1)^{i+j+n-1} /\sigma_{i,j}$: this amounts to having the
gluons $i,j$ the $-1$ picture, and while the correlator is manifestly
symmetric in the remaining $n{-}2$ particles, on the support of
scattering equations it becomes independent of $i,j$ thus completely
symmetric as required by Bose symmetry.

By the definition of the Pfaffian, one can derive a useful (recursive)
expansion, which was originally considered in \cite{Lam:2016tlk} and
used extensively in \eg~\cite{Fu:2017uzt,Teng:2017tbo}: %
\beq {\cal I}^\tree_{\rm bos}(1,2,\ldots,n) = \sum_{\{23\ldots n-1\}
  \atop{= A \cup B}} \Pf(\Psi_A) \sum_{\rho \in S_{|B|}}
\PT(1,\rho(B),n) (\ep_1{\cdot} f_{\rho(b_1)} {\cdot}
f_{\rho(b_2)}{\cdot} \ldots {\cdot} f_{\rho(b_{|B|})} {\cdot}
\ep_n)\,.
\label{eq2.7}
\eeq
Here the notation $\{2, 3, \ldots, n{-}1\} = A \cup B$ in
(\ref{eq2.7}) instructs to sum over all the $2^{n-2}$ splittings of
the set $\{2, 3, \ldots, n{-}1\} $ into disjoint sets $A$ and $B$ with
$|A|$ and $|B|$ elements. In each term, we have the Pfaffian of the
matrix with particle labels in $A$ (which is of size
$2 |A| \times 2 |A|$), times a sum over permutations $\rho \in S_{|B|}$ of labels in the
complement $B\equiv \{b_1, b_2, \cdots, b_{|B|}\}$. We define
$\Pf \Psi_{\emptyset}=1$ for the case of empty $A$. Moreover,
(\ref{eq2.7}) features Parke--Taylor factors
\beq
\PT(1,2, \cdots, n) =
\frac{1}{\sigma_{12}\sigma_{23}\ldots \sigma_{n-1,n} \sigma_{n,1}}  
\label{eq2.PT}
\eeq
in the cyclic ordering $(1, \rho(B), n) = (1,\rho(b_1),\rho(b_2),\ldots,\rho(b_{|B|}),n)$.
Finally, the kinematic coefficient of the Parke--Taylor factors in (\ref{eq2.7}) are
Lorentz contraction of $\epsilon_1$, $\epsilon_n$ and
linearized field strengths 
\beq
f_j^{\mu \nu} = k_j^\mu \ep_j^\nu - k_j^\nu \ep_j^\mu \, .
\label{deff}
\eeq
The dot products in
$(\ep_1{\cdot} f_{\rho(b_1)} {\cdot} \ldots {\cdot} f_{\rho(b_{|B|})}
{\cdot} \ep_n)$ are understood in the sense of matrix multiplication,
\eg
$(\ep_1\cdot f_{2} \cdot \ep_n)= \ep_1^\mu (f_2)_{\mu \nu} \ep_n^\nu$,
so we reproduce the well-known three-point example
\beq
{\cal I}^\tree_{\rm bos}(1,2,3)  = \frac{(k_3 \cdot \ep_2)(\ep_1\cdot \ep_3) +(\ep_1\cdot k_2)(\ep_2 \cdot \ep_3) -  (\ep_1\cdot \ep_2)(k_2 \cdot \ep_3)}{ \sigma_{1,2} \sigma_{2,3} \sigma_{3,1} } \, .
\eeq
At $n=4$, for instance, $\sum_{\{23\} = A \cup B}$ yields four
contributions with $(A,B) = (\{2,3\},\emptyset)$, $(\{2\},\{3\})$,
$ (\{3\},\{2\})$ and $(\emptyset, \{2,3\})$, which are given by
\beq \Pf \Psi_{\{2,3\}}\frac{\epsilon_1 \cdot
  \epsilon_4}{\sigma_{1,4}\sigma_{4,1}}, \quad \Pf \Psi_{\{2\}}
\frac{\epsilon_1 \cdot f_3 \cdot
  \epsilon_4}{\sigma_{1,3}\sigma_{3,4}\sigma_{4,1}}, \quad
(2\leftrightarrow 3), \quad \frac{\epsilon_1 \cdot f_2 \cdot f_3 \cdot
  \epsilon_4}{\sigma_{1,2} \sigma_{2,3} \sigma_{3,4} \sigma_{4,1}} +
(2\leftrightarrow 3)\,,
\label{bosterms}
\eeq
respectively.

It has been known since~\cite{Cachazo:2013iea} that using scattering
equations, one can expand the correlator as a linear combination of
Parke-Taylor factors, say in the partial-fraction independent set
$\{  \PT(1,\sigma(2,3, \cdots,n{-}1), n), \ \sigma \in S_{n-2} \}$, and the
coefficients are BCJ master numerators for the corresponding
$(n{-}2)!$ half-ladder diagrams. One way of doing so is to start from
\eqref{eq2.7}, and the challenge is identical to extracting BCJ
numerators for single-trace $({\rm YM} + \phi^3)$ amplitudes. See
\cite{Nandan:2016pya, Schlotterer:2016cxa, Teng:2017tbo, Chiodaroli:2017ngp, AlexFei} for more
details.

%%%%%%%%%%%%%%%%%%%%%%%%%%%%%%%%%%%%%%%%%%%%%%%% 
%%%%%%%%%%%%%%%%%%%%%%%%%%%%%%%%%%%%%%%%%%%%%%%% 
%%%%%%%%%%%%%%%%%%%%%%%%%%%%%%%%%%%%%%%%%%%%%%%% 
%%%%%%%%%%%%%%%%%%%%%%%%%%%%%%%%%%%%%%%%%%%%%%%% 
%%%%%%%%%%%%%%%%%%%%%%%%%%%%%%%%%%%%%%%%%%%%%%%% 

\subsection{Tree-level correlator for two external fermions}
\label{sec:ferm}

In the subsequent, we will cast two-fermion correlators involving two spin
fields $S_\alpha$ \cite{Knizhnik:1985ke,Cohn:1986bn} into simple forms
by virtue of the current algebra generated by $\psi^\mu \psi^\nu$
along the lines of \cite{Kostelecky:1986xg}. Note that such
simplifications are partly motivated by \eqref{eq2.7} since such a
correlator with external fermions can also be expanded in a similar
form.

In the first representation, we have the two fermions, say, leg $1$
and $n{-}1$, both in the $-\frac 1 2$ ghost picture, and one of the
gluons, say leg $n$, in the $-1$ picture. Throughout this work, we
will use the subscript ``f'' to denote fermions (gluinos) and
suppress any subscript for the vector bosons (gluons). On the support
of scattering equations, one can show that the tree-level
correlator can be simplified to (see appendix \ref{app:2ferm} for details)
\begin{align}
  &{\cal I}^\tree_{\rm 2f}(1_{\rm f}, 2,\ldots, (n{-}2) ,(n{-}1)_{\rm f}, \ferm{n})  \notag \\
& \quad = \langle V_1^{(-1/2)}(\sigma_1) 
V_2^{(0)}(\sigma_2) V_3^{(0)}(\sigma_3)  \ldots V_{n-2}^{(0)}(\sigma_{n-2}) V_{n-1}^{(-1/2)}(\sigma_{n-1}) 
V_n^{(-1)}(\sigma_n)  \rangle
\notag \\
 &\quad = \frac{1}{2}
\sum_{ \{23 \ldots n-2\} \atop {=A\cup B\cup C}} \textrm{Pf} (\Psi_{A})
 \sum_{\rho \in S_{|B|}} \sum_{\tau \in S_{|C|}} 
 \PT(1,\rho(B),n,\tau(C),n{-}1)  \label{eq2.8} \\
  &\qquad  \times (\chi_1  \oslashed{f_{\rho(b_{1})}} \oslashed{f_{\rho(b_{2})}}\ldots \oslashed{f_{\rho(b_{|B|})}}  \oslashed{\ep_n}
    \oslashed{f_{\tau(c_{1})}} \oslashed{f}_{\tau(c_{2})}\ldots \oslashed{f_{\tau(c_{|C|})}} \chi_{n-1})\, ,
  \notag
\end{align}
where we sum over all the splittings of the set
$\{2,3,\cdots, n{-}2\}$ into disjoint sets $A, B$ and $C$, with again
$\Pf (\Psi_A)$ times a sum over permutations $\rho$ and $\tau$ of the
labels in $B$ and $C$, respectively. Similar to \eqref{eq2.7}, we have
a Parke-Taylor factor $\PT(1,\rho(B), n, \tau (C),n{-}1)$ defined by
(\ref{eq2.PT}) for each term. The main difference is that instead of
the vector-index contraction, the linearized field strengths
(\ref{deff}) are now contracted into gamma matrices. More
specifically, with the conventions
\beq
\oslashed{\ep}_n = \ep_n^\mu \gamma_\mu\, , \ \ \ \ \ \ 
\oslashed{f}_j = \frac{1}{4} f_j^{\mu \nu} \gamma_{\mu \nu} = \frac{1}{2} k_j^\mu \ep_j^\nu \gamma_{\mu \nu} = \frac{1}{2} \oslashed{k_j} \oslashed{\ep_j}\, ,
\label{eq2.10}
\eeq
the last line of (\ref{eq2.8}) features gamma-matrix products with the
gluons in $\rho(B)$, $\tau(C)$ entering via $\oslashed{f}_j $, gluon
$n$ entering via $\oslashed{\ep}_n$, and the fermion wavefunctions
$\chi_1,\chi_{n{-}1}$ contracting the free spinor induces, \eg
$ (\chi_1 \oslashed{f_{2}} \oslashed{\ep_n} \chi_{n-1})=
\frac{1}{4}(\chi_1 \gamma_{\mu \nu} \gamma_\la \chi_{n-1}) f_2^{\mu
  \nu} \ep_n^\la$. In view of their contractions with Weyl spinors
$\chi_1,\chi_{n-1}$, the gamma matrices in (\ref{eq2.8}) are
$16\times 16$ Weyl-blocks within the Dirac matrices in 10
dimensions. Our conventions for their Clifford algebra and
antisymmetric products
are
\beq
\gamma^\mu \gamma^\nu + \gamma^\nu \gamma^\mu = 2 \eta^{\mu \nu} \, , \ \ \ \ \ \ 
\gamma^{\mu \nu} \equiv \frac{1}{2}(\gamma^\mu \gamma^\nu - \gamma^\nu \gamma^\mu)\,.
\label{eq2.9}
\eeq
A variant of (\ref{eq2.8}) with $\oslashed{\ep_n} $ moved adjacent to $\chi_{n-1}$ 
has been studied by Frost \cite{Frost:2017} along with its implication for the forward limit
in the fermions.

At $n=3$ points, the two-fermion correlator (\ref{eq2.8}) specializes to
\beq
{\cal I}^\tree_{\rm 2f}(1_{\rm f}, 2_{\rm f}, \ferm{3}) = \frac{(\chi_1 \oslashed{\ep_3} \chi_2) }{2 \sigma_{1,3} \sigma_{3,2} \sigma_{2,1} } \, ,
\label{eq2.ex}
\eeq
and the sum over $\{2\} = A \cup B \cup C$ in its $(n=4)$-point
instance gives rise to the following three terms instead of the four
terms in the bosonic correlator (\ref{bosterms}) (also see \cite{Adamo:2013tsa}):
\beq
\Pf \Psi_{ \{2\}} \frac{(\chi_1 \oslashed{\ep_4} \chi_3) }{2 \sigma_{1,4} \sigma_{4,3} \sigma_{3,1} }
, \quad \frac{(\chi_1 \oslashed{f_2} \oslashed{\ep_4} \chi_3) }{2 \sigma_{1,2} \sigma_{2,4} \sigma_{4,3} \sigma_{3,1} }
, \quad \frac{(\chi_1  \oslashed{\ep_4} \oslashed{f_2} \chi_3) }{2 \sigma_{1,4} \sigma_{4,2} \sigma_{2,3} \sigma_{3,1} }
\label{eq2.ex1}
\eeq
The formula (\ref{eq2.8}) for the two-fermion correlator is manifestly
symmetric in most of the gluons $2,3,\cdots, n{-}2$ except for the
last one $n$ which is earmarked through the hat notation in
${\cal I}^\tree_{\rm 2f}(\ldots , \ferm{n})$.  On the support of
scattering equations and the kinematic phase space of $n$ massless
particles, one can show that (\ref{eq2.8}) is also symmetric in all
of $\ferm{n}$ and $2,3,\ldots,n{-}2$. But this no longer the case in the
forward-limit situation of sec.~\ref{sec:odd}, where we extract
parity-odd contributions to one-loop correlators from (\ref{eq2.8}).

Note that the expression (\ref{eq2.8}) for the two-fermion correlator
can be straightforwardly generalized to any even spacetime dimension
since the structure of the underlying spin-field correlators is universal (see
appendix \ref{app:2ferm}). However, only $D=2 \ \textrm{mod} \ 4$
admit $\chi_1$ and $\chi_{n-1}$ of the same chirality since the
charge-conjugation matrix in these dimensions is off-diagonal in its
$2^{D/2-1}\times 2^{D/2-1}$ Weyl blocks. In order to extend
(\ref{eq2.8}) to $D=0 \ \textrm{mod} \ 4$ dimensions, 
$\chi_1$ and $\chi_{n-1}$ need to be promoted to Weyl spinors of opposite chirality.

%%%%%%%%%%%%%%%%%%%%%%%%%%%%%%%%%%%%%%%%%%%%%%%% 
%%%%%%%%%%%%%%%%%%%%%%%%%%%%%%%%%%%%%%%%%%%%%%%% 
%%%%%%%%%%%%%%%%%%%%%%%%%%%%%%%%%%%%%%%%%%%%%%%% 
%%%%%%%%%%%%%%%%%%%%%%%%%%%%%%%%%%%%%%%%%%%%%%%% 
%%%%%%%%%%%%%%%%%%%%%%%%%%%%%%%%%%%%%%%%%%%%%%%% 

\subsection{Alternative representation of the two-fermion correlator}
\label{sec:ferm2}

In this section, we present an alternative representation of the
two-fermion correlator which is manifestly symmetric in all its
$n{-}2$ gluons. To do that, we put the two fermions, say leg $1$ and
$n$, in the $-\frac 1 2$ and $-\frac 3 2$ picture, respectively, and on the
support of scattering equations we find (see appendix \ref{app:2ferm}
for details)
\begin{align}
{\cal I}^\tree_{\rm 2f}(1_{\rm f},2,\ldots, (n{-}1), n_{\rm f})&=\langle V_1^{(-1/2)}(\sigma_1) 
V_2^{(0)}(\sigma_2) V_3^{(0)}(\sigma_3)  \ldots V_{n-1}^{(0)}(\sigma_{n-1}) V_{n}^{(-3/2)}(\sigma_{n})  \rangle
  \! \!\label{eq2.11}\\
& = \sum_{\{23\ldots n-1\} \atop{= A \cup B}} \! \!
\Pf(\Psi_A) \sum_{\rho \in S_{|B|}} \!
\PT(1,\rho(B),n) (\chi_1  \oslashed{f_{\rho(b_1)}}
 \oslashed{f_{\rho(b_2)}}  \ldots  \oslashed{f_{\rho(b_{|B|})}}  \xi_n) 
\notag
\end{align}
which takes a form even closer to \eqref{eq2.7} since we also sum over
partitions $\{2,3,\cdots ,n{-}1\}=A \cup B$ with disjoint $A,B$. All the (gamma-matrix
contracted) field strengths (\ref{eq2.10}) in $\rho(B)$ are sandwiched
between $\chi_1$ and $\xi_n$.

At $n=3$, the sum over $\{2\} = A \cup B$ in (\ref{eq2.11}) involves two terms:
\begin{align}
{\cal I}^\tree_{\rm 2f}(1_{\rm f},2,3_{\rm f}) &=  \Pf \Psi_{ \{2\} } \frac{ \chi_1 \xi_3 }{\sigma_{1,3} \sigma_{3,1} } + \frac{ \chi_1 \oslashed{ f_2} \xi_3}{  \sigma_{1,2} \sigma_{2,3} \sigma_{3,1} } \label{eq2.ex2} \\
&= \frac{ -1}{  \sigma_{1,2} \sigma_{2,3} \sigma_{3,1} } \Big\{ (\ep_2\cdot k_1) (\chi_1 \xi_3) + \frac{1}{2} (\chi_1 \oslashed{k_3} \oslashed{\ep_2} \xi_3) \Big\}
\notag
\end{align}
In order to relate this to the earlier result (\ref{eq2.ex}) for the
fermionic three-point correlator, we have rewritten
$ \Pf \Psi_{ \{2\} } = \frac{ (\ep_2\cdot k_1) \sigma_{1,3}
}{\sigma_{2,1} \sigma_{2,3}}$ and
$\chi_1 \oslashed{ f_2} \xi_3 = - \frac{1}{2} \chi_1 \oslashed{k_3}
\oslashed{\ep_2} \xi_3$ in passing to the second line. These
identities are based on both momentum conservation and the
physical-state conditions
$\ep_2\cdot k_2 = \chi_1 \oslashed{k_1} = 0$. Finally, the Clifford
algebra (\ref{eq2.9}) gives rise to
$\chi_1 \oslashed{k_3} \oslashed{\ep_2} \xi_3 = 2 (\ep_2\cdot k_3)
\chi_1 \xi_3 - \chi_1 \oslashed{\ep_2} \oslashed{k_3} \xi_3$, and one
can identify the wavefunction $\chi_3 = \oslashed{k_3} \xi_3$ by
(\ref{eq2.4}).  In this way, we reproduce the permutation
\beq
{\cal I}^\tree_{\rm 2f}(1_{\rm f},2,3_{\rm f}) =  \frac{(\chi_1 \oslashed{\ep_2} \chi_3) }{2 \sigma_{1,2} \sigma_{2,3} \sigma_{3,1} }
\eeq
of the earlier three-point result (\ref{eq2.ex}). Even though this may
appear to be a detour in the computation of the three-point
correlator, the similarity of (\ref{eq2.11}) with the bosonic
correlator (\ref{eq2.7}) will be a crucial benefit for the computation
of forward limits.

At $n=4$ we have the four contributions similar to \eqref{bosterms}:
\beq
\Pf \Psi_{\{2,3\}} \frac{\chi_1 \xi_4}{\sigma_{1,4} \sigma_{4,1}}, \quad \Pf \Psi_{\{2\}} \frac{\chi_1 \oslashed{f_3} \xi_4}{\sigma_{1,3}\sigma_{3,4} \sigma_{4,1}}, \quad (2 \leftrightarrow 3), \quad \frac{\chi_1 \oslashed{f_2}\oslashed{f_3}\xi_4}{\sigma_{1,2}\sigma_{2,3}\sigma_{3,4}\sigma_{4,1}} + (2\leftrightarrow 3)\,.
\eeq
We remark that again we can further expand the $\Pf \Psi_A$ in both
cases, and on the support of scattering equations eventually one can
expand the correlator as a linear combination of (length-$n$)
Parke-Taylor factors.  Their coefficients can be identified with BCJ
numerators \cite{Cachazo:2013iea, Nandan:2016pya, Schlotterer:2016cxa,
  Teng:2017tbo}, now involving two external fermions on top of $n{-}2$
bosons. In the following, we will mostly work with the second
representation (\ref{eq2.11}) of the two-fermion correlator when we
take the forward limit in the two fermions and combine it with the
bosonic forward limit of (\ref{eq2.7}). The parity-odd part of
one-loop numerators in chiral theories in turn will be derived from
the first representation (\ref{eq2.8}) of the two-fermion correlator,
see sec.~\ref{sec:odd}.

Similar to the results of the previous section, the two-fermion
correlator (\ref{eq2.11}) generalizes to any even spacetime
dimension. The chiralities of $\chi_1$ and $\xi_n$ remain opposite in
any $D=2 \ \textrm{mod} \ 4$, whereas dimensions
$D=0 \ \textrm{mod} \ 4$ require a chirality flip in one of $\chi_1$
or $\xi_n$.

As detailed in appendix \ref{app:4ferm}, four-fermion correlators with
any number of bosons can be brought into a very similar form. Six or
more fermions, however, necessitate vertex operators in the $+1/2$
superghost picture that feature excited spin fields and give rise to
more complicated $n$-point correlators \cite{Atick:1986rs,
  Kostelecky:1986ab, Lee:2017ujn}. Still, the results are available
from the manifestly supersymmetric pure-spinor formalism
\cite{Berkovits:2000fe}, where $n$-point correlators in Parke--Taylor form
are available in superspace \cite{Mafra:2011kj,Mafra:2011nv}. Their components
for arbitrary combinations of bosons and fermions can be conveniently extracted
through the techniques of \cite{Lee:2015upy, Mafra:2015vca}.

%%%%%%%%%%%%%%%%%%%%%%%%%%%%%%%%%%%%%%%%%%%%%%%% 
%%%%%%%%%%%%%%%%%%%%%%%%%%%%%%%%%%%%%%%%%%%%%%%% 
%%%%%%%%%%%%%%%%%%%%%%%%%%%%%%%%%%%%%%%%%%%%%%%% 
%%%%%%%%%%%%%%%%%%%%%%%%%%%%%%%%%%%%%%%%%%%%%%%% 

\subsection{Forward limits and gluing operators}
\label{sec:sub23}

Finally, we review the prescription for taking forward limits in a pair
of legs, which can be both bosons or both fermions. The momenta of the
two legs are $+\ell$ and $-\ell$, respectively, which should be taken
off shell, \ie $\ell^2 \neq 0$.\footnote{This can be realized by
  allowing only these two momenta to have non-vanishing components in
  certain extra dimension. For example, in $D{+}1$ dimensions the
  momenta for the two additional legs are taken to be
  $\pm (\ell; |\ell |)$, while those for others are $(k_i; 0)$ for
  $i=1,2,\cdots, n$. } Moreover, we need to sum over the polarization
states and other quantum numbers of the two legs. For example, since we
consider all particles (both gluons and gluinos) to be in the adjoint
representation of \eg $U(N)$ color group, we have to sum over the
$U(N)$ degrees of freedom of the pair of legs. In this way, the
one-loop color-stripped amplitude can be obtained by summing over
tree-level ones with the two adjacent legs inserted in all possible
positions. This is the origin of the one-loop Parke-Taylor factors (\ref{fixgaugeint}),
also see~\cite{He:2015yua, Cachazo:2015aol} for more details.

We shall now define the kinematic prescription for forward limits in two
bosonic or fermionic legs. For that in bosonic legs $i$ and $j$, we define
\beq
\ep^\mu_i \ep^\nu_j \rightarrow   \FWL_{i,j}(\ep^\mu_i \ep^\nu_j )= \eta^{\mu \nu} - \ell^\mu \bar \ell^\nu - \ell^\nu \bar\ell^\mu \, , \ \ \ \ \ \ 
\FWL_{i,j}(k_i,k_j)= (+\ell,-\ell) 
\label{eq2.12}
\eeq
with an auxiliary vector $\bar \ell^\mu$ subject to
$\ell \cdot \bar \ell=1$. Note that we have used the completeness
relation of polarization vectors.

For the forward limit in fermionic legs $i$ and $j$, we define
\beq
(\chi_i)^\al (\xi_j)_\be \rightarrow
 \FWL_{i,j}\big(
(\chi_i)^\al (\xi_j)_\be \big)
=
\delta^\al{}_\be
\, , \ \ \ \ \ \ 
\FWL_{i,j}(k_i,k_j)= (+\ell,-\ell)  \, ,
\label{eq2.13}
\eeq
where we have used the completeness relation for fermion wave
functions. When applied to a pair of vertex operators with total
superghost charge $-2$, the prescriptions (\ref{eq2.12}) and
(\ref{eq2.13}) implement the gluing operators of Roehrig and Skinner
\cite{Roehrig:2017gbt}.

Before proceeding, we remark that after taking the forward limit in a pair
of gluons/gluinos in the tree-level correlator, \eqref{eq2.7} and
\eqref{eq2.11}, the only explicit dependence on loop momentum $\ell$
is in $\Pf\Psi_A$ through diagonal entries of the submatrix $\mathsf{C}_A$;
there is no loop momentum in other parts of $\Pf\Psi_A$ or factors
involving particles in $B$. We will see in the subsequent that this
observation immediately yields the power counting of loop momentum for
BCJ numerators in various gauge theories.

%%%%%%%%%%%%%%%%%%%%%%%%%%%%%%%%%%%%%%%%%%%%%%%%
%%%%%%%%%%%%%%%%%%%%%%%%%%%%%%%%%%%%%%%%%%%%%%%%
%%%%%%%%%%%%%%%%%%%%%%%%%%%%%%%%%%%%%%%%%%%%%%%%
%%%%%%%%%%%%%%%%%%%%%%%%%%%%%%%%%%%%%%%%%%%%%%%%
 
\section{One loop correlators and numerators of ten-dimensional SYM}
\label{sec:loop}

In this section, we study one-loop correlators with external bosons
for ten-dimensional SYM, which in turn give explicit BCJ numerators at
one-loop level. We begin by taking the forward limit of tree-level
correlators with two additional bosons and fermions, \eqref{eq2.7} and
\eqref{eq2.11}, respectively; in order to combine them, we present a
key result of the section, namely a formula to express a spinor trace
with any number of particles in terms of vector traces. Moreover, the
relative coefficient is fixed by maximal supersymmetry, thus we can
write a formula for the one-loop correlator with all the supersymmetry
cancellations manifest at any multiplicity.

Even though this section is dedicated to ten-dimensional SYM, we will
retain a variable number $D$ of spacetime dimensions in various
intermediate steps. This is done in preparation for the analogous
discussion of lower-dimensional gauge theories in section
\ref{sec:gen} and justified by the universality of the form
(\ref{eq2.11}) of two-fermion correlators.

%%%%%%%%%%%%%%%%%%%%%%%%%%%%%%%%%%%%%%%%%%%%%%%%%%% 
%%%%%%%%%%%%%%%%%%%%%%%%%%%%%%%%%%%%%%%%%%%%%%%%%%% 
%%%%%%%%%%%%%%%%%%%%%%%%%%%%%%%%%%%%%%%%%%%%%%%%%%% 
%%%%%%%%%%%%%%%%%%%%%%%%%%%%%%%%%%%%%%%%%%%%%%%%%%% 

\subsection{The forward limit of two bosons/fermions}
\label{sec:loop.1}

Implementing the forward limits (\ref{eq2.12}) and (\ref{eq2.13}) via
gluing operators \cite{Roehrig:2017gbt} sends the presentation
(\ref{eq2.7}) and (\ref{eq2.11}) of the tree-level correlators to
\begin{align}
\FWL_{1,n}\big[{\cal I}^\tree_{\rm bos}(1,2,\ldots,n) \big] &=
 \! \sum_{\{23\ldots n-1\} \atop{= A \cup B}} 
\Pf(\Psi_A) \sum_{\rho \in S_{|B|}}
\PT(1,\rho(B),n) \label{eq3.1} \\
&\qquad \times \left\{ \begin{array}{cl}  (D-2) &: \ B = \emptyset
  \\ (f_{\rho(b_1)} {\cdot} f_{\rho(b_2)}{\cdot} \ldots {\cdot} f_{\rho(b_{|B|})} )^{\mu \nu} \eta_{\mu \nu} &: \ B \neq \emptyset
\end{array} \right.
\notag
\\
\FWL_{1,n}\big[{\cal I}^\tree_{\rm 2f}(1_{\rm f}, 2,\ldots,n{-}1,n_{\rm f}) \big]&= \! \sum_{\{23\ldots n-1\} \atop{= A \cup B}} 
\Pf(\Psi_A) \sum_{\rho \in S_{|B|}}
\PT(1,\rho(B),n) \notag \\
&\qquad \times \left\{ \begin{array}{cl} 2^{D/2-1}&: \ B = \emptyset
  \\ (\oslashed{f_{\rho(b_1)}}  \oslashed{f_{\rho(b_2)}}  \ldots  \oslashed{f_{\rho(b_{|B|})}})_\alpha{}^\beta  \delta^\alpha_\beta   &: \ B \neq \emptyset
\end{array} \right.   \, .
 \label{eq3.2}
\end{align}
The contribution of $B=\emptyset$ stems from contractions
$\eta_{\mu \nu}( \eta^{\mu \nu} - \ell^\mu \bar \ell^\nu - \ell^\nu
\bar\ell^\mu ) =D-2$ and
$\delta^\alpha_\beta \delta_\alpha^\beta = 2^{D/2-1}$ in
(\ref{eq2.12}) and (\ref{eq2.13}), the latter being the dimension of a
chiral spinor representation in even spacetime dimensions $D$. In
spelling out the contributions of $B\neq \emptyset$ to the bosonic
forward limit, we have exploited that the terms
$\sim \ell^\mu \bar \ell^\nu + \ell^\nu \bar\ell^\mu$ in
(\ref{eq2.12}) do not contribute upon contraction of with vectors
different from $\ep_{i},\ep_{j}$ \cite{Roehrig:2017gbt}.

We shall introduce some notation for the frequently reoccurring traces
over vector and spinor indices, relegating the discussion of parity-odd
pieces to sec.~\ref{sec:odd}:
\begin{align}
\trv(1, 2, \ldots, p) &= (f_{1}
\cdot f_{2}\cdot \ldots \cdot f_{p} )^{\mu \nu} \eta_{\mu \nu}
= (f_1)^{\mu_1}{}_{\mu_2} (f_2)^{\mu_2}{}_{\mu_3}\ldots (f_p)^{\mu_p}{}_{\mu_1}
\label{eq3.3}
\\
%%%%%
\trs(1, 2, \ldots, p)  &=
(\slash \! \! \!  f_{1}
 \slash \! \! \!  f_{2}  \ldots  \slash \! \! \!  f_{p})_\alpha{}^\beta  \delta^\alpha_\beta \, \big|_{\te{even}}
\label{eq3.4}
\\
&= \frac{1}{4^p} f_1^{\mu_1 \nu_1}f_2^{\mu_2 \nu_2}
\ldots f_p^{\mu_p \nu_p} 
(\gamma_{\mu_1 \nu_1})_{\al_1}{}^{\al_2} (\gamma_{\mu_2 \nu_2} )_{\al_2}{}^{\al_3}\ldots 
(\gamma_{\mu_p \nu_p})_{\al_p}{}^{\al_1} \, \big|_{\te{even}} \, .
\notag
\end{align}
We remark that the spinor trace in (\ref{eq3.2}) would in principle
contain parity-odd terms, but here we define $\trs (1, 2, \ldots, p)$
to be the parity-even part by manually discarding parity-odd
terms.\footnote{For $n=5$, the parity-odd term
  $\varepsilon_{10}(f_1,f_2,f_3,f_4,f_5)$ in a chiral spinor trace vanishes by momentum
  conservation, in contrast to the one in (\ref{odd.6}) due to a
  different prescription.}  Note that the $B = \emptyset$ contribution to
(\ref{eq3.2}) formally arises from $\trs(\emptyset) = 2^{D/2-1}$, and
non-empty traces are cyclic and exhibit the parity properties
\beq
\trv(1, 2, \ldots, p) = (-1)^p \trv(p, \ldots, 2, 1) \, , \ \ \ \ \ \ 
\trs(1, 2, \ldots, p) = (-1)^p \trs(p, \ldots, 2, 1) \, .
\label{eq3.p}
\eeq
In order to study the supersymmetry cancellations in one-loop
correlators, we will be interested in linear combinations of bosonic
and fermionic forward limits with theory-dependent relative weights.
The main results of this work are driven by the observation that most
of the structure in (\ref{eq3.1}) and (\ref{eq3.2}) is preserved in
combining bosons and fermions such that the linear combinations are
taken at the level of the field-strength traces: with an a priori undetermined weight
factor $\alpha \in \QQ$, we have
\begin{align}
{\cal I}\oneloop_{{\rm bos},\alpha}&(1, 2, \ldots, n) = \FWL_{+,-}\big[ {\cal I}^\tree_{\rm bos}(+,1, 2, \ldots, n,-) +  \alpha \cdot {\cal I}^\tree_{\rm 2f}(+_f,1, 2, \ldots,n,-_f) \big] \, \big|_{\te{even}} \notag
\\
&=  \!  \sum_{\{12\ldots n\} \atop{= A \cup B}} 
\Pf(\Psi_A) \sum_{\rho \in S_{|B|}}
\PT(+,\rho(B),-)  \label{eq3.5} \\
&\ \ \ \ \ \ \ \times \left\{ \begin{array}{cl}  \big[D{-}2 + \alpha \cdot  2^{D/2-1} \big]  &: \ B = \emptyset \\
%%%
\big[
\trv( {\rho(b_1)}, {\rho(b_2)},\ldots, {\rho(b_{|B|})})
+ \alpha \cdot \trs({\rho(b_1)}, {\rho(b_2)}, \ldots, {\rho(b_{|B|})})
\big] &: \ B \neq \emptyset \end{array} \right. \, .
  \notag
\end{align}
The $B = \emptyset$ contribution $\sim \Pf(\Psi_{\{12\ldots n\} })$
will be proportional to at least one power of loop momentum since a
plain Pfaffian in a tree-level context is known to vanish on the
support of the scattering equations. The diagonal entries $\mathsf{C}_{jj}$ in
the expansion of $ \Pf(\Psi_A) $ within (\ref{eq3.5}) still involve terms
$\ep_j^\mu(\frac{ \ell_\mu }{\sigma_{j,+}} - \frac{ \ell_\mu
}{\sigma_{j,-}} )$ which would be absent in the naive tree-level
incarnation of $ \Pf(\Psi_{\{12\ldots n\} })$ without any reference to
extra legs $+,-$.

%%%%%%%%%%%%%%%%%%%%%%%%%%%%%%%%%%%%%%%%%%%%%%%%%%% 
%%%%%%%%%%%%%%%%%%%%%%%%%%%%%%%%%%%%%%%%%%%%%%%%%%% 
%%%%%%%%%%%%%%%%%%%%%%%%%%%%%%%%%%%%%%%%%%%%%%%%%%% 
%%%%%%%%%%%%%%%%%%%%%%%%%%%%%%%%%%%%%%%%%%%%%%%%%%% 

\subsection{From spinor traces to vector ones}
\label{sec:loop.2}

In this subsection we propose the identities which allow us to convert
any spinor trace to vector ones. Our result will be useful in the
subsequent sections when we study the one-loop correlator and BCJ
numerators for various gauge theories.

Our starting point is the well-known formula for traces of chiral gamma matrices
\beq
\trs(\gamma^{\mu \nu}) = 0 \, , \ \ \ \ \ \ \trs(\gamma^{\mu \nu} \gamma^{\la \rho}) = 2^{D/2-1}
 (\eta^{\nu \la} \eta^{\mu \rho} -\eta^{\mu \la} \eta^{\nu \rho} ) \, , \ \ \ \ \ \ \te{etc.}
  \label{eq3.6}
\eeq
We will review a recursion for such traces in appendix
\ref{app:traces}, and based on that it is easy to show that
$\trs(\emptyset) = 2^{D/2-1}$ generalizes to %
\begin{align}
\trs(1) = 0 \, , \ \ \ \ \ \ \trs(1, 2) = 2^{D/2-4} \trv(1, 2)
\, , \ \ \ \ \ \ \trs(1, 2, 3) = 2^{D/2-4} \trv(1, 2, 3)\, ,
  \label{eq3.7}
\end{align}
where the numbers enclosed in $(\ldots)$ label the external particles
according to our conventions (\ref{eq3.3}) and (\ref{eq3.4}). Starting
from four points, more permutations appear: for $n=4$ the result is
given by a sum of single traces and double traces \wrt vector
indices,
\begin{align}
\trs(1, 2, 3, 4) = \quad & 2^{D/2-5} \big\{ \trv(1, 2, 3, 4)
 - \trv(1, 3, 2, 4) - \trv(1, 2, 4, 3) \big\} \label{eq3.8} \\
 \quad + & 2^{D/2-7} \big\{ \trv(1, 2 ) \trv(3, 4)
 +\trv(1, 3 ) \trv(2, 4)
 +\trv(1, 4 ) \trv(2, 3) \big\} \, , \notag
 %%%%%
 \end{align}
 where we have used the parity properties (\ref{eq3.p}): for
 single-trace terms we have $6$ (cyclically inequivalent) permutations
 but only $3$ of them are independent under parity.

 Moving to the $n=5$ case, we find that $\trs(1,2,3,4,5)$ is again
 given by combinations of single and double traces,
 \begin{align} 
2^{D/2-6} \big\{ &\trv(1,2,3,4,5)
 -  \trv(1,2,3,5,4)
 - \trv(1,2,4,3,5)  - \trv(1,2,4,5,3)\notag \\
  -&\trv(1,2,5,3,4)
 + \trv(1,2,5,4,3) 
 - \trv(1,3,2,4,5)
 + \trv(1,3,2,5,4)\notag \\
 - &\trv(1,4,2,3,5)  - \trv(1,3,5,2,4)
 + \trv(1,4,3,2,5)
 - \trv(1,3,4,2,5) \big\} \label{eq3.9} \\
 +2^{D/2-7} \big\{ &\trv(1,2) \trv(3,4,5)
 +\trv(1,3 ) \trv(2,4,5) +\trv(1,4) \trv(2,3,5)
 +\trv(1,5) \trv(2,3,4) \notag\\
   +&\trv(2,3) \trv(1,4,5) +\trv(2,4) \trv(1,3,5)+\trv(2,5) \trv(1,3,4)
 +\trv(3,4) \trv(1,2,5) \notag\\
  +& \trv(3,5) \trv(1,2,4) +\trv(4,5) \trv(1,2,3) \big\}\,, \notag
\end{align}
where only $4!/2=12$ single-trace terms, and ${5 \choose 2}=10$
double-trace terms are independent under parity.

As we will show recursively in Appendix \ref{app:traces}, in general
the $n$-point spinor trace can be written as a sum of terms with
$j=1, 2, \cdots, \lfloor n/2\rfloor$ vector traces with suitable
prefactors,
\begin{align}
\trs(1,2,\ldots,n)&= 2^{D/2-1-n} \sum_{j=1}^{\lfloor n/2 \rfloor} \frac{1}{2^j j!}  \! \! \! \! \! \sum_{\{12 \ldots n\} \atop{=A_1\cup A_2\cup \ldots \cup A_j}}  \! \! \prod_{i=1}^j \bigg(
\sum_{\sigma \in  S_{|A_i|}/\ZZ_{|A_i|} }
\trv( \sigma(A_i) )  
\ord_\sigma^{id}
\bigg)\, ,
\label{eq3.11}
\end{align}
where for each $j$, we sum over partitions of $\{1,2,\ldots,n\}$ into
$j$ disjoint subsets $A_1,A_2,\ldots,A_j$, and the factor
$\frac{1}{j!}$ compensates for the overcounting of partitions due to
permutations of $A_1,A_2,\ldots,A_j$; for each subset $A_i$ we sum
over all cyclically inequivalent permutations
$\sigma \in S_{|A_i|}/\ZZ_{|A_i|}$, \eg by fixing the first element in
$\trv(\sigma(A_i))$ to be the smallest one in $A_i$; finally the sign
$\ord_\sigma^{id}$ counts the number of descents in $\sigma$ (compared
to the identity permutation). For example, $\ord_{132}^{id}=-1$,
$\ord_{1243}^{id}=\ord_{1324}^{id}=-1$, and
$\ord_{1432}^{id}=1$. An alternative representation of the parity-even spinor
trace (\ref{eq3.4}) in terms of a Pfaffian can be found in (4.35a) of \cite{Roehrig:2017gbt}.

More generally, if the spinor trace has an ordering $\rho$, one can
choose the first element $\sigma_1$ to be the smallest in $\rho$, and
the sign $\ord_\sigma^\rho$ can be factorized as
\beq
\ord_\sigma^{\rho} = \sgn^{\rho}_{\sigma_p,\sigma_{p-1}} \sgn^{\rho}_{\sigma_{p-1},\sigma_{p-2}}
\ldots \sgn^{\rho}_{\sigma_4,\sigma_{3}} \sgn^{\rho}_{\sigma_3,\sigma_{2}} \, ,
\label{eq3.12}
\eeq
where the $\sgn_{ij}^\rho$ factors are defined to be $\pm1$ according
to the conventions of \cite{He:2017spx}
\beq
\sgn_{ij}^\rho = \left\{ \begin{array}{rl}
+1 &: \ i \ \te{is on the right of $j$ in} \ \rho(1,2,\ldots,p) \\
-1 &: \ i \ \te{is on the left of $j$ in} \ \rho(1,2,\ldots,p) 
\end{array} \right. \, .
\label{eq3.13}
\eeq
For example, $\ord_{132}^{132}=1$ instead of $\ord_{132}^{id}=-1$ and
$\ord_{1432}^{1243}=-1$. Let's end the discussion with an example for
triple-trace contribution ($j=3$) of $\trs(1,2,3,4,5,6)$, which reads 
\begin{align}
\label{eq3.10} 
2^{D/2-10}& \big\{
\trv(1,2 ) [\trv(3,4) \trv(5,6)  +   \trv(3,5 ) \trv(4,6)
 +  \trv(3,6) \trv(4,5) ]\\
 &
 +  \trv(1,3 ) [ \trv(2,5) \trv(4,6)
 + \trv(2,6) \trv(4,5) +  \trv(2,5) \trv(4,6) ]\notag \\
  &
 +  \trv(1,4 ) [\trv(2,3) \trv(5,6)  +    \trv(2,5) \trv(3,6)
 +   \trv(2,6) \trv(3,5)] \notag \\
  &
 +  \trv(1,5) [\trv(2,3) \trv(4,6)
 +    \trv(2,4) \trv(3,6) +    \trv(2,6) \trv(3,4)]\notag \\
  &
 +  \trv(1,6) [\trv(2,3) \trv(4,5)  +  \trv(2,4) \trv(3,5)
 +   \trv(2,5) \trv(3,4)] \big\} \, .\notag
\end{align}

%%%%%%%%%%%%%%%%%%%%%%%%%%%%%%%%%%%%%%%%%%%%%%%%%%% 
%%%%%%%%%%%%%%%%%%%%%%%%%%%%%%%%%%%%%%%%%%%%%%%%%%% 
%%%%%%%%%%%%%%%%%%%%%%%%%%%%%%%%%%%%%%%%%%%%%%%%%%% 
%%%%%%%%%%%%%%%%%%%%%%%%%%%%%%%%%%%%%%%%%%%%%%%%%%% 

\subsection{Ten-dimensional SYM}
\label{sec:loop.3}

Since we have not been careful about the normalization of the
fermionic tree-level correlator (\ref{eq2.8}), the normalization
constant $\alpha$ in (\ref{eq3.5}) for a single Weyl fermion will be
fixed by the example of ten-dimensional SYM. The supersymmetry
cancellations are well-known to yield vanishing $(n\leq 3)$-point
one-loop integrands in $D=10$ SYM \cite{Green:1982sw}. Accordingly, there exists a choice
$\alpha= - \frac{1}{2}$ in (\ref{eq3.5}) such that both the
$B= \emptyset$ contributions
$D{-}2 + \alpha \cdot 2^{D/2-1} \, \big|_{D=10}$ and those with
$|B|=2,3$
vanish: 
\begin{align}
{\cal I}\oneloop_{D=10 \ {\rm SYM}}&(1,2,\ldots,n) =  \!  \sum_{\{12\ldots n\} \atop{= A \cup B}} 
\Pf(\Psi_A) \sum_{\rho \in S_{|B|}}
\PT(+,\rho(B),-)  \label{eq3.22} \\
&\ \ \ \ \ \ \ \times \left\{ \begin{array}{cl}  0  &: \ |B| \leq 3 \\
%%%
\big[
\trv({\rho(b_1)}, {\rho(b_2)}, \ldots, {\rho(b_{|B|})})
-\tfrac{1}{2} \trs({\rho(b_1)}, {\rho(b_2)}, \ldots, {\rho(b_{|B|})})
\big] &: \ |B| \geq 4 \end{array} \right. \, .
  \notag
\end{align}
Recall that the spinor trace $\trs$ is defined to contain the parity-even part
only.  We have used the relation (\ref{eq3.7}) between vector and
spinor two- and three-traces in $D=10$ dimensions,
$\trs(1, 2) = 2 \trv(1, 2)$ and $ \trs(1, 2, 3) = 2 \trv(1, 2,
3)$. Throughout this work, the external
  states of the one-loop correlators are gauge bosons.  Thus we will
  no longer specify $_{\rm bos}$ in the subscripts of
  ${\cal I}\oneloop$.
  
The first non-vanishing contribution to (\ref{eq3.22}) from the field strengths at
$|B|=4$ turns out to not depend on the permutation $\rho$ and
reproduces the famous $t_8$-tensor, {\it cf.}~(\ref{eq3.8}),
\begin{align}
\trv(1, 2, 3, 4)  &- \tfrac{1}{2} \trs(1, 2, 3, 4)  \, \big|_{D=10} = \trv(1, 2, 3, 4)  \label{eq3.23} \\
& \ \ \ \ \ \  - \tfrac{1}{2} \big\{ \trv(1, 2, 3, 4)
 - \trv(1, 3, 2, 4) - \trv(1, 2, 4, 3) \big\} \notag \\
 & \ \ \ \ \ \ - \tfrac{1}{8} \big\{ \trv(1, 2 ) \trv(3, 4)
 +\trv(1, 3 ) \trv(2, 4)
 +\trv(1, 4 ) \trv(2, 3) \big\}  \notag \\
 &\ \ \ = \tfrac{1}{2} \trv(1, 2, 3, 4) -  \tfrac{1}{8}   \trv(1, 2 ) \trv(3, 4) + {\rm cyc}(2,3,4) 
 =  \tfrac{1}{2} \TEf{1,2,3,4} \, ,
\notag
\end{align}
which is known from one-loop four-point amplitudes of the superstring
\cite{Green:1982sw} and defined by
\begin{align}
\TEf{1,2,3,4}&= \trv(1,2,3,4)+\trv(1,3,2,4)+\trv(1,2,4,3) \label{eq2.t8} \\
&
-\frac{1}{4} \big\{ \trv(1, 2 ) \trv(3, 4)
 +\trv(1, 3 ) \trv(2, 4)
 +\trv(1, 4 ) \trv(2, 3) \big\} \, . \notag
\end{align}
Hence, the four-point instance of (\ref{eq3.22}) is the well-known
permutation symmetric combination of Parke--Taylor factors,
\begin{align}
{\cal I}\oneloop_{D=10 \ {\rm SYM}}&(1,2,3,4) = \frac{1}{2} \TEf{1,2,3,4} \sum_{\rho \in S_{4}}
\PT(+,\rho(1,2,3,4),-) \, .
\label{eq3.24}
\end{align}
Starting at five points, we need the case with $|B|=5$, and a similar
expression can be given
\begin{align}
&\trv(1,2,3,4,5)  - \tfrac{1}{2} \trs(1,2,3,4,5)  \, \big|_{D=10}  \label{eq3.25}\\
&=
\tfrac{1}{4}
 \big\{3 \trv(1,2,3,4,5)
 +  \trv(1,2,3,5,4)
 + \trv(1,2,4,3,5)  + \trv(1,2,4,5,3)\notag \\
 & \ \ \ \ \ \
 +\trv(1,2,5,3,4)
 - \trv(1,2,5,4,3)  + \trv(1,3,2,4,5)
 - \trv(1,3,2,5,4)\notag \\
 & \ \ \ \ \ \
 + \trv(1,4,2,3,5)  + \trv(1,3,5,2,4)
 - \trv(1,4,3,2,4)
 + \trv(1,3,4,2,5) \big\} 
\notag \\
&\ \ \
-\tfrac{1}{8} \big\{ \trv(1,2) \trv(3,4,5)
 +\trv(1, 3) \trv(2,4,5) +\trv(1,4 ) \trv(2,3,5)
 +\trv(1,5 ) \trv(2,3,4)\notag\\
 & \ \ \ \ \ \
 +\trv(2,3) \trv(1,4,5) +\trv(2,4 ) \trv(1,3,5) +\trv(2,5) \trv(1,3,4)
 +\trv(3,4 ) \trv(1,2,5) \notag\\
 & \ \ \ \ \ \
+\trv(3,5 ) \trv(1,2,4)
 +\trv(4,5 ) \trv(1,2,3) \big\} \notag \\
 &= \tfrac{1}{4} \big\{ t_8([f_1,f_2],f_3,f_4,f_5)+
  t_8([f_1,f_3],f_2,f_4,f_5)  +t_8([f_1,f_4],f_2,f_3,f_5)+
  t_8([f_1,f_5],f_2,f_3,f_4) \notag \\
  & \ \ \ \ \ \ +t_8([f_2,f_3],f_1,f_4,f_5)+
  t_8([f_2,f_4],f_1,f_3,f_5)  +t_8([f_2,f_5],f_1,f_3,f_4)+
  t_8([f_3,f_4],f_1,f_2,f_5) \notag \\
  & \ \ \ \ \ \ +t_8([f_3,f_5],f_1,f_2,f_4)+
  t_8([f_4,f_5],f_1,f_2,f_3)   \big\} \, ,\notag
\end{align}
where we have used \eg
$[f_1,f_2]^{\mu \nu}\equiv f_1^{\mu}{}_\la f_2^{\la \nu} - f_2^{\mu}{}_\la
f_1^{\la \nu} $ inside the $t_8$ tensor. Let us already emphasize here
that (\ref{eq3.22}) {\it after} rewriting  $\trs(\ldots)$ in terms of
$\trv(\ldots)$ applies to any dimensional reduction of ten-dimensional
SYM, for instance ${\cal N}=4$ SYM in $D=4$ ({\it cf.}~section~\ref{sec:gen}).

By analogy with (\ref{eq3.23}), one may define higher-rank tensors
beyond $t_8$ in (\ref{eq2.t8}). We can use the difference of vector
and spinor traces to define higher-point extensions of (\ref{eq2.t8})
that will capture the kinematic factors besides the $\Pf(\Psi_A) $ in
the correlators (\ref{eq3.22}) $D=10$ SYM.  As exemplified by the
five-point case (\ref{eq3.25}), higher-point
$\trv(\ldots) -\tfrac{1}{2} \trs(\ldots)$ will involve $t_8$ tensors
with nested commutators of $f_j$ w.r.t.\ Lorentz indices in its
entries. The only new tensor structures that are not expressible in
terms of $t_8$ with commutators arise from the permutation symmetric
combination\footnote{At six points, for instance,
  $\trv(1,2,\ldots,6)- \frac{1}{2} \trs(1,2,\ldots,6)$ can be
  rewritten as its permutation symmetric part plus permutations of the
  two topologies $t_8([f_1,f_2], [f_3,f_4],f_5,f_6)$ and
  $t_8( [ [f_1,f_2], f_3],f_4, f_5,f_6)$.}
 \begin{equation}
t_{2n}(f_1,f_2,\ldots,f_n) = \frac{1}{(n{-}1)!} \! \sum_{\rho \in S_{n-1}} \! \! \!
\big[ 2\trv(1{,} {\rho(2)}{,} {\rho(3)}{,} \ldots{,} {\rho(n)})  
-  \trs(1{,} {\rho(2)}{,} {\rho(3)}{,} \ldots{,} {\rho(n)} )\big]  \, \big|_{D=10} 
\label{3.t2n} 
\end{equation}
involving an even number $n$ of field strengths. The permutation sum
vanishes for odd $n$ by the parity properties (\ref{eq3.p}). Rewriting
correlators of $D=10$ SYM in terms of (\ref{3.t2n}) is the kinematic
analogue of decomposing color traces in gauge-theory amplitudes into
contracted structure constants and symmetrized traces, where only the
latter can furnish independent color tensors
\cite{vanRitbergen:1998pn}.

The simplest instance of (\ref{3.t2n}) beyond $t_8$ is a rank-twelve
tensor $t_{12}$ occurring at $n=6$. As detailed in appendix
\ref{not12}, the case of $t_{12}$ admits an exceptional simplification
that is not possible for $t_{16}$ and any higher-rank tensor
(\ref{3.t2n}): One can reduce $t_{12}$ to products,
\begin{equation}
  t_{12}(f_1, f_2 ,\dots, f_6) = \frac{1}{24}\big[ \trv(1,2 ) \TEf{3,4,5,6} + (1,2|1,2,3,4,5,6)\big]\, ,
  \label{simpt12}
\end{equation}
where the four-traces and products
$ \trv(i_1,i_2 ) \trv(i_3,i_4 ) \trv(i_5,i_6 )$ conspire to
$t_8$. Here and throughout the rest of this work, the notation
$+ (1,2|1,2,\ldots,k)$ instructs to add all permutations of the
preceding expression where the ordered pair of labels $1,2$ is
exchanged by any other pair $i,j \in \{1,2,\ldots,k\}$ with $i<j$. A
similar notation $+ (1,2,\ldots,j|1,2,\ldots,k)$ with $j<k$ will be
used to sum over all possibilities to pick $j$ elements from a
sequence of $k$, for a total of ${k\choose j}$ terms.

The exceptional simplification of $t_{12}$ in (\ref{simpt12}) can be
anticipated from the fact that six-traces $\trv(1{,}2,\dots {,}6)$
cancel from the combination (\ref{3.t2n}) after rewriting the spinor
traces via (\ref{eq3.11}). For any higher-rank $t_{2n}$ at $n\geq 8$
in turn, the coefficient of $\trv(1,2,\dots ,n)$ is non-zero when
expressing the spinor traces of (\ref{3.t2n}) in terms of
$\trv(\ldots)$. These coefficients are worked out in terms of Eulerian
numbers in appendix \ref{not12}.

In summary, the tensor structure of the $n$-point correlators
(\ref{eq3.22}) in $D=10$ SYM is captured by $\Pf(\Psi_A)$ and
even-rank tensors $t_{2n}$ in (\ref{3.t2n}) including $t_8$ in
(\ref{eq2.t8}) contracting nested commutators of field strengths.

\subsection{BCJ numerators versus single-trace YM$+\phi^3$ at tree level}
\label{sec:BCJ}

Given the general formula \eqref{eq3.22} for the one-loop correlator
in ten-dimensional SYM, one can read off the BCJ master numerators $N\oneloop$ 
of an $n$-gon diagram as soon as all the $\sigma_j$-dependences of the
Parke--Taylor factors and the $\te{Pf}(\Psi_A)$ are lined up with
\beq
{\cal I}\oneloop(1,2,\ldots,n) = \frac{1}{2}\sum_{\omega \in S_n} \PT(+,\omega(1,2,\ldots,n),-)
N\oneloop(+,\omega(1,2,\ldots,n),-) \, ,
\label{eq3.21}  
\eeq
where we need to use scattering equations at $(n{+}2)$
points.\footnote{We have included the explicit factor of $\frac{1}{2}$
  such that the BCJ numerators $N^{(1)}$ discussed in section~\ref{sec:MPF} directly match
  those in the literature.  It is canceled by parts of the overall
  normalization factor $\mathscr{N}$ in \eqref{eq:CHY} between the amplitude ${\cal M}^{\rm tree}_{L \otimes R}$ 
  and the CHY integral.} More specifically, the numerator $N\oneloop(+,\omega(1,2,\ldots,n),-)$ refers to 
 one of the $(n{+}2)$-point half-ladder diagrams in the right panel of figure \ref{figcutting}
 that arises from the partial-fraction decomposition of the $n$-gon propagators reviewed
 in appendix \ref{app:linprop}.

For a given partition $\{1,2,\ldots ,n\}= A \cup B$ in \eqref{eq3.22}, the leftover task
 is to absorb the $\sigma_j$-dependence of
the Pfaffian into the $(|B|{+}2)$-point Parke--Taylor factors,
\beq
\Pf(\Psi_A) \PT(+,\rho(B),-)  = \sum_{\omega \in S_n} \PT(+,\omega(1,2,\ldots,n),-) K_A(\omega,\rho(B))
\label{eq3.22alt}  
\eeq
such as to form $(n{+}2)$-point Parke--Taylor factors. The kinematic
factors $K_A(\omega,\rho(B))$ are multilinear in the polarization
vectors of the set $A$ that enter via $\Pf(\Psi_A)$.

The identical challenge arises at tree level when computing the BCJ
master numerators of single-trace (YM$+\phi^3$) amplitudes. Recall
that both gluons and scalars in (YM$+\phi^3$) amplitudes are in the
adjoint representation of a color group, and the scalars are
additionally in the adjoint representation of a flavor group. A
color-stripped amplitude has all the $n$ particles in an ordering,
thus the CHY half-integrand is given by a (length-$n$) Parke-Taylor
factor. In addition, by ``single-trace" we mean the scalars are also
in an ordering after stripping off the flavors, and the other CHY
half-integrand is given by a Parke-Taylor factor for scalars in legs
$1,\ldots, k$ and a Pfaffian for gluons in legs
$k{+}1, \ldots, n$~\cite{Cachazo:2014nsa}. BCJ master numerators are
obtained by reducing to Parke-Taylor factors using scattering
equations~\cite{Cachazo:2013iea}:
\begin{align}
\Pf(\Psi_{\{k+1\ldots n\} }) &\PT(1,\tau(2,\ldots,k{-}1),k) = 
\sum_{\pi \in S_{n-2}} \PT(1,\pi(2,\ldots,k{-}1,k{+}1,\ldots,n),k) 
 \label{eq4.23}   \\
&\ \ \ \ \ \ \ \ \ \ \  \times N^{\rm tree}_{{\rm YM}+\phi^3}(1,\pi(2,\ldots,k{-}1,k{+}1,\ldots,n),k| 1, \tau(2, \ldots, k{-}1), k) \, . \notag
\end{align}
Here without loss of generality, we have chosen the ordering for
scalars to be $1, \tau(2, \ldots, k{-}1), k$ with $\tau\in S_{k{-}2}$,
and the second line of (\ref{eq4.23}) can be attained by the
techniques of \cite{Nandan:2016pya, Chiodaroli:2017ngp, Teng:2017tbo}:
The Parke--Taylor coefficients $N^{\rm tree}_{{\rm YM}+\phi^3}(\ldots)$
are BCJ master numerators associated with a half-ladder diagram, with
$1$ and $k$ on opposite ends and the permutations
$\pi\in S_{n-2}$ acting on the remaining particles 
(the second ordering $1, \tau(2, \ldots, k{-}1), k$
is for the scalars \wrt the flavor group).

By matching \eqref{eq3.22alt} with \eqref{eq4.23}, one can identify
the kinematic factors $K_A(\omega,\rho(B))$ in a one-loop context with
a (YM$+\phi^3$)-master numerator at tree level. One needs to pick the
scalars to be in $+, -, B$ and the gluons to be in $A$, and choose the
two orderings to match the permutations $\omega, \rho$:
\beq\label{3.26} K_A(\omega,\rho(B)) = N^{\rm tree}_{{\rm
    YM}+\phi^3}(+,\omega(1,2,\ldots,n),-| +, \rho(B),-) 
\, .
\eeq
Two of the current authors present an improved method of computing the
necessary $N^{\text{tree}}$ in Ref.~\cite{AlexFei}.

As an illustration, let us consider the simplest case with one gluon,
\ie $k=n{-}1$, then $\Pf\Psi_{\{n\}}=\mathsf{C}_{n,n}$ and
partial-fraction manipulations are sufficient to show that \cite{Nandan:2016pya},
\beq
\mathsf{C}_{n,n}\, \PT(1,2,\ldots,n{-}1)=\sum_{a=1}^{n{-}2} \PT(1, 2, \ldots,
a, n, a{+}1, \ldots, n{-}1) \sum_{i=1}^a \epsilon_{n}\cdot k_i \, ,
\eeq 
where the ordering for the scalars has been chosen 
as $1, 2,\ldots, n{-}1$ for simplicity.
By \eqref{3.26}, we have $A=\{n\}$ and $B=\{1,2,\ldots, n{-}1\}$
for the one-loop case; if we choose $\omega=id$, then
the non-vanishing one-loop master numerators only occur for the ordering
$\rho_a(B)\equiv(1, \ldots, a, n, a{+}1, \ldots, n{-}1)$ (for
$a=1, \ldots, n{-}2$), which read
$K(\omega, \rho_a(B))=\sum_{i=1}^a \epsilon_n \cdot k_i$.

One can proceed similarly in case of more gluons: for $k=n{-}2$,
by expanding ${\rm Pf}\Psi_{\{n-1,n\}}$ and using scattering equation
of leg $n$, after some algebra we obtain \cite{Nandan:2016pya}
\begin{align}
&{\rm Pf}\Psi_{\{n-1,n\}}{\rm PT}(1,2,\ldots,n-2)
\\
&=
\sum_{a=1}^{n{-}3}
\sum_{b=0}^{n{-}3{-}a}{\rm PT}(1,\ldots,a, n, \ldots, a{+}b, n{-}1,\ldots,n{-}2) 
\left(
\sum_{i=1}^a \epsilon_{n-1}{\cdot} f_n{\cdot} k_i
+
\sum_{i=1}^a \epsilon_n {\cdot} k_i \sum_{j=1}^{a+b} \epsilon_{n-1} {\cdot} k_j 
\right)
\nonumber
\\
&\ \ \ \ \ \ +
\sum_{a=1}^{n{-}3}
\sum_{b=0}^{n{-}3{-}a}{\rm PT}(1,\ldots,a,n{-}1,%a{+}1,
\ldots,a{+}b,n,%a{+}b+1,
\ldots,n{-}2) 
\sum_{i=1}^a \epsilon_{n-1}{\cdot} k_i
\left(\epsilon_{n} {\cdot} k_{n-1} +
\sum_{j=1}^{a+b} \epsilon_n {\cdot} k_j  
\right)\, ,
\nonumber
\end{align}
where we have Parke-Taylor factors with label $n$ and $n{-}1$ inserted
at various positions. In this way, one can continue with more and more
gluons and obtain BCJ master numerators for single-trace amplitudes in
YM$+\phi^3$ \cite{Chiodaroli:2017ngp}. Similar techniques have been 
used in \eg\cite{Nandan:2016pya, Schlotterer:2016cxa, Teng:2017tbo}, and more
recently in~\cite{He:2018pol, He:2019drm,AlexFei}.

%%%%%%%%%%%%%%%%%%%%%%%%%%%%%%%%%%%%%%%%%%%%%%%%
%%%%%%%%%%%%%%%%%%%%%%%%%%%%%%%%%%%%%%%%%%%%%%%%
%%%%%%%%%%%%%%%%%%%%%%%%%%%%%%%%%%%%%%%%%%%%%%%%
%%%%%%%%%%%%%%%%%%%%%%%%%%%%%%%%%%%%%%%%%%%%%%%%

%%%%%%%%%%%%%%%%%%%%%%%%%%%%%%%%%%%%%%%%%%%%%%%%
%%%%%%%%%%%%%%%%%%%%%%%%%%%%%%%%%%%%%%%%%%%%%%%%
%%%%%%%%%%%%%%%%%%%%%%%%%%%%%%%%%%%%%%%%%%%%%%%%
%%%%%%%%%%%%%%%%%%%%%%%%%%%%%%%%%%%%%%%%%%%%%%%%
 
\section{General gauge theories}
\label{sec:gen}

In this section, we move to more general gauge theories in even
dimensions whose spectrum may involve an arbitrary combination of
adjoint scalars, fermions and gauge bosons.  Accordingly, their
one-loop correlators are built from forward limits not only in vectors
and Weyl fermions but also in scalars. As we will review, the tree-level
correlator with $2$ scalars and $n{-}2$ gluons can be obtained from
dimension reduction of the $n$-gluon one~\cite{Cachazo:2015aol}. By
combining all the building blocks from forward limits, we then have a
formula for one-loop correlators with ${\bf n_{\rm v}}$ vectors,
${\bf n_{\rm f}}$ Weyl fermions and ${\bf n_{\rm s}}$ scalars in $D$
dimensions\footnote{We denote these numbers of different species by
  boldface ${\bf n}$, to avoid confusion with the $n^{\rm th}$
  external leg.}. We will present examples of such correlators in
various theories in $D=6$ and $D=4$.

%%%%%%%%%%%%%%%%%%%%%%%%%%%%%%%%%%%%%%%%%%%%%%%%%%% 
%%%%%%%%%%%%%%%%%%%%%%%%%%%%%%%%%%%%%%%%%%%%%%%%%%% 
%%%%%%%%%%%%%%%%%%%%%%%%%%%%%%%%%%%%%%%%%%%%%%%%%%% 
%%%%%%%%%%%%%%%%%%%%%%%%%%%%%%%%%%%%%%%%%%%%%%%%%%% 

\subsection{Forward limits in general gauge theories}
\label{sec:gen.1}

Before we present a formula for general one-loop correlators in even
dimension, let us first review the tree-level correlator involving two
scalars. In fact, the bosonic tree-level correlators (\ref{eq2.7}) can
be straightforwardly adapted to two external scalars in legs $1,n$ by
taking their polarizations $\ep_1,\ep_n$ to satisfy
\begin{align}
\ep_1 \cdot \ep_{n} &= 1 \notag \\
\ep_1 \cdot \ep_{j} &= \ep_n \cdot \ep_j = 0 \ \ \forall \ \ j=2,3,\ldots,n{-}1  \label{eq4.1}\\
\ep_1 \cdot k_{p} &= \ep_n \cdot k_p = 0  \ \ \forall \ \ p=1,2,\ldots,n
 \,,
\notag
\end{align}
which can also be realized from dimensional reduction. The resulting
scalar correlator solely features the $B=\emptyset$ term of
(\ref{eq2.7}),
\beq
{\cal I}^\tree_{\rm 2{\rm s}}(1_{\rm s}, 2,\ldots, n{-}1, n_{\rm s})  =
\Pf(\Psi_{\{23\ldots n{-}1\}})\, ,
\label{eq4.3}
\eeq
where we have used the subscript ``$\rm s$" to denote scalars (recall that
gluons have no subscript).  The scalar forward limit analogous to
(\ref{eq2.12}) simply amounts to
$ \FWL_{i,j}(k_i,k_j)= (+\ell,-\ell) $,
\beq
\FWL_{1,n}\big[{\cal I}^\tree_{\rm 2s}(1_{\rm s}, 2,\ldots, n{-}1, n_{\rm s}) \big]= 
\Pf(\Psi_{\{2 3\ldots n{-}1\}})\, .
\label{eq4.5}
\eeq
We shall now combine the building blocks (\ref{eq3.1}), (\ref{eq3.2})
and (\ref{eq4.5}) for the forward limits in $D$-dimensional vectors,
$D$-dimensional Weyl fermions and scalars. In the presence of
${\bf n_{\rm v}}$, ${\bf n_{\rm f}}$ and ${\bf n_{\rm s}}$ species of
vectors, Weyl fermions and scalars, respectively, one arrives at the
following parity-even parts of one-loop correlators in even
dimensions,
\begin{align}
{\cal I}\oneloop_{({\bf n_{\rm v}}, {\bf n}_{\rm f}, {\bf n}_{\rm s},D)}&(1,2,\ldots,n) = \FWL_{+,-}\big[ {\bf n_{\rm v}} {\cal I}^\tree_{\rm bos}(+,1, 2, \ldots, n, -) \notag \\
&\qquad -\tfrac{\bf n_{\rm f} }{2} {\cal I}^\tree_{\rm 2f}(+_{\rm f}, 1, 2,\ldots, n,-_{\rm f})
+ {\bf n}_{\rm s} {\cal I}^\tree_{\rm 2s}(+_{\rm s},1, 2, \ldots, n,-_{\rm s}) \big]  \, \big|_{\te{even}} \notag
\\
&\! \! \! \! \! \! \! \!  \! \! \! \! \! \! \! \!  =  \!  \sum_{\{12\ldots n\} \atop{= A \cup B}} 
\Pf(\Psi_A) \sum_{\rho \in S_{|B|}}
\PT(+,\rho(B),-)  \label{eq4.6} \\
&\qquad \! \! \! \! \! \! \! \!  \! \! \! \! \! \! \! \!  \times \left\{ \begin{array}{cl}  \big[ {\bf n_{\rm v}} \,(D{-}2) + {\bf n}_{\rm s} -  {\bf n_{\rm f} }\,  2^{D/2-2}  \big] &: \ B = \emptyset \\
%%%
\big[
{\bf n_{\rm v}} \trv(\rho(b_1), \rho(b_2), \ldots, \rho(b_{|B|}))
-\tfrac{ \bf n_{\rm f}}{2} \trs(\rho(b_1), \rho(b_2), \ldots, \rho(b_{|B|}))
\big] &: \ B \neq \emptyset \end{array} \right. \, .
  \notag
\end{align}
Note that ${\bf n_{\rm s}}$ only appears in the term with
$B=\emptyset$ and $A=\{1,2,\ldots, n\}$, and we again have the loop-momentum
dependence $\mathsf{C}_{jj}=\ep_j^\mu(\frac{ \ell_\mu }{\sigma_{j,+}} - \frac{ \ell_\mu
}{\sigma_{j,-}} )+\ldots$ in $\Pf(\Psi_A)$ for any choice of $A$. Moreover, the 
coefficient ${\bf n_{\rm v}} \,(D{-}2) + {\bf n}_{\rm s} - {\bf n_{\rm f} }\,
2^{D/2-2} $ of $\Pf(\Psi_{\{12\ldots n \} })$ can be recognized as the
difference of bosonic and fermionic on-shell degrees of freedom:
$D$-dimensional vector bosons and Weyl fermions have $D{-}2$ and
$2^{D/2-2}$ physical degrees of freedom, respectively.  Hence, the
$B=\emptyset$ contribution to (\ref{eq4.6}) is absent in
supersymmetric theories.

Given that $\trv(\ldots)$ and $\trs(\ldots)$ vanish at $|B|=1$,
supersymmetric theories admit at most $|A|=n{-}2$ particles in
$\Pf(\Psi_A)$. As a consequence, the maximum power of loop momenta in
the parity-even part of supersymmetric correlators is $\ell^{n-2}$,
reproducing the power counting of \cite{He:2017spx} (such
power-counting has been studied since the early days of unitarity
methods \cite{Bern:1994zx,Bern:1993tz,Bern:1992ad}). As will be
detailed below, the parity-odd contributions to $D=4$ correlators with
four supercharges may exceed this bound and involve up to $n{-}1$
powers of $\ell$.

The $|B|=2$ and $|B|=3$ contributions to (\ref{eq4.6}) are
proportional to ${\bf n_{\rm v}} - 2^{D/2-5} {\bf n_{\rm f} }$ by the
relative factor between vector and spinor traces in (\ref{eq3.7}).
These terms are absent whenever the ratio of ${\bf n_{\rm v}}$ and
${\bf n_{\rm f} }$ fits to the maximally supersymmetric gauge
multiplet in the respective dimension, \ie
$({\bf n_{\rm f} },{\bf n_{\rm s} })=({\bf n_{\rm v} },0)$ in $D=10$,
$({\bf n_{\rm f} },{\bf n_{\rm s} })=(4{\bf n_{\rm v} },0)$ in $D=6$ and
$({\bf n_{\rm f} },{\bf n_{\rm s} })=(8{\bf n_{\rm v} },6{\bf n_{\rm v} })$ in $D=4$, respectively.  Like
this, (\ref{eq4.6}) manifests that 16 supercharges are a necessary and
sufficient condition for the $|B|\leq 3$ contributions to vanish and
for the maximum power of loop momentum to be $\ell^{n-4}$.

Finally, the $|B|=4$ contribution to (\ref{eq4.6}) is proportional to
$({\bf n_{\rm v}} - 2^{D/2-5} {\bf n_{\rm f} })\trv(1,2,3,4) +
2^{D/2-6} {\bf n_{\rm f} }\TEf{1,2,3,4}$. In the maximally
supersymmetric situation where
${\bf n_{\rm v}} = 2^{D/2-5} {\bf n_{\rm f} }$, the correlator
contributions at $|B|=4$ are permutations of
$\TEf{1,2,3,4}\Pf \Psi_{ \{56\ldots n\} }$. Hence, the maximally supersymmetric
tensor numerators with the highest power of loop momentum $\ell^{n-4}$ are
built from a permutation sum over
$\TEf{1,2,3,4} \prod_{j=5}^n (\ep_j \cdot \ell)$.

%%%%%%%%%%%%%%%%%%%%%%%%%%%%%%%%%%%%%%%%%%%%%%%%%%% 
%%%%%%%%%%%%%%%%%%%%%%%%%%%%%%%%%%%%%%%%%%%%%%%%%%% 
%%%%%%%%%%%%%%%%%%%%%%%%%%%%%%%%%%%%%%%%%%%%%%%%%%% 
%%%%%%%%%%%%%%%%%%%%%%%%%%%%%%%%%%%%%%%%%%%%%%%%%%% 

\subsection{Examples in $D=6$ and $D=4$}
\label{sec:gen.3}

We shall now spell out several examples of the general correlator
(\ref{eq4.6}) with reduced supersymmetry.

\bigskip
\noindent
{\bf (i)}
A six-dimensional chiral gauge multiplet with half-maximal
supersymmetry (8 supercharges instead of 16) contains a single vector
${\bf n_{\rm v}}=1$ and two Weyl fermions ${\bf n_{\rm f}}=2$ with a total of
$4+4$ on-shell degrees of freedom 
\begin{align}
{\cal I}\oneloop_{(1,2,0,D=6)}&(1,2,\ldots,n) = \FWL_{+,-}\big[  {\cal I}^\tree_{\rm bos}(+,1, 2, \ldots, n, -) 
 -  {\cal I}^\tree_{\rm 2f}(+_{\rm f}, 1,2,\ldots, n,-_{\rm f}) \big]  \, \big|_{\te{even}}^{D=6}  \notag
\\
&\! \! \! \! \! \! \! \! \! \! \! \! \! \! \! \! =    \sum_{\{12\ldots n\} \atop{= A \cup B}} 
\Pf(\Psi_A) \sum_{\rho \in S_{|B|}}
\PT(+,\rho(B),-)  \label{eq4.6dim6} \\
&\qquad \! \! \! \! \! \! \! \! \! \! \! \! \! \! \! \!  \times \left\{ \begin{array}{cl}  0 &: \ |B| \leq 1 \\
%%%
\big[
 \trv(\rho(b_1), \rho(b_2), \ldots, \rho(b_{|B|}))
-  \trs(\rho(b_1), \rho(b_2), \ldots, \rho(b_{|B|}))
\big]\, \big|_{D=6} &: \ |B| \geq 2  \end{array} \right.   \, .
  \notag
\end{align}
The simplest $D=6$ spinor traces resulting from (\ref{eq3.7}) and
(\ref{eq3.8}) include $\trs(1,2) = \frac{1}{2} \trv(1,2)$ as
well as $ \trs(1,2,3) = \frac{1}{2} \trv(1,2,3)$ and
introduce the following contributions to (\ref{eq4.6dim6}):
\begin{align}
\trv(1,2)  - \trs(1, 2)  \, \big|_{D=6} &=   \frac{1}{2} \trv(1, 2) \notag \\
\trv(1, 2, 3)  - \trs(1, 2, 3)  \, \big|_{D=6} &=   \frac{1}{2} \trv(1, 2, 3) \notag \\
\trv(1, 2, 3, 4)  - \trs(1, 2, 3, 4)  \, \big|_{D=6} &= 
 \frac{1}{4} \big\{ 3 \trv(1, 2, 3, 4)
 + \trv(1, 3, 2, 4) + \trv(1, 2, 4, 3) \big\} \notag \\
 & \qquad - \frac{1}{16} \big\{ \trv(1, 2 ) \trv(3, 4)
 + {\rm cyc}(2,3,4) \big\}  \notag \\
 &=    \frac{1}{4} \TEf{1,2,3,4} + \frac{1}{2} \trv(1, 2, 3, 4)      \label{eq4.6a}  
\end{align}
As a result of the reduced supersymmetry, already the splittings with
$|B|=2,3$ contribute to (\ref{eq4.6dim6}) which were absent for the
ten-dimensional counterpart (\ref{eq3.24}) with maximal supersymmetry.
Similarly, the six-dimensional combination of four-traces in
(\ref{eq4.6a}) is no longer permutation invariant, \ie cannot be
expressed solely in terms of the $t_8$-tensor (\ref{eq2.t8}).

\bigskip
\noindent
{\bf (ii)}
A six-dimensional hypermultiplet \wrt 8 supercharges contains a
single Weyl fermion ${\bf n_{\rm f}}=1$ and two scalars ${\bf n_{\rm s}}=2$ with a
total of $2+2$ on-shell degrees of freedom,
\begin{align}
&{\cal I}\oneloop_\hyper(1,2,\ldots,n) = \FWL_{+,-}\big[ 
                2{\cal I}^\tree_{\rm 2s}(+_{\rm s},1{,} 2{,} \ldots{,} n, -_{\rm s})  -\tfrac{1 }{2} {\cal I}^\tree_{\rm 2f}(+_{\rm f}, 1{,} 2{,}\ldots{,} n,-_{\rm f}) \big]   \,  \big|_{\te{even}}^{D=6}  \notag \\
&\  = -\frac{ 1}{2}  \!  \sum_{\{12\ldots n\} \atop{= A \cup B}}  \!
\Pf(\Psi_A) \! \! \sum_{\rho \in S_{|B|}} \! \!
\PT(+,\rho(B),-) \left\{ \begin{array}{cl}  0&: \ |B| \leq 1 \\
%%%
 \trs( \rho(b_1),  \rho(b_2),\ldots, \rho(b_{|B|}))  \, \big|_{D=6}&: \ |B|  \geq 2 \end{array} \right. \, .
 \label{eq4.7}
\end{align}
The simplest contributions at $|B|=2,3,4$ are
\begin{align}
 - \frac{1}{2} \trs(1, 2)  \, \big|_{D=6} &=  - \frac{1}{4} \trv(1, 2) \notag \\
 - \frac{1}{2} \trs(1, 2, 3)  \, \big|_{D=6} &=  - \frac{1}{4} \trv(1, 2, 3) \notag \\
 -  \frac{1}{2} \trs(1, 2, 3, 4)  \, \big|_{D=6} &=    \frac{1}{8} \TEf{1,2,3,4} - \frac{1}{4} \trv(1, 2, 3, 4) \, .\label{eq4.6b}  
\end{align}
The expressions in (\ref{eq4.6a}) and (\ref{eq4.6b}) confirm the
decomposition of a ten-dimensional gauge multiplet into one vector
multiplet and two hypermultiplets in $D=6$: By adding two copies of
(\ref{eq4.6b}) to (\ref{eq4.6a}), the two- and three- traces drop out,
and one recovers the four-trace of $D=10$ SYM in (\ref{eq3.23}).
In sec.~\ref{sec:MPF.3}, we will spell out simplified expressions
for $(n\leq 5)$-point BCJ numerators resulting from (\ref{eq4.7}) in 
terms of multiparticle fields.

\bigskip
\noindent
{\bf (iii)}
Reducing all the way to $D=4$, we can examine a gauge multiplet
of $\mathcal{N}=1$ SYM, which has two fermionic degrees of freedom,
so with ${\bf n_{\rm v}}=1$ and ${\bf n_{\rm f}}=2$
\begin{align}
{\cal I}\oneloop_{(1, 2,0,D=4)}&(1,2,\ldots,n) = \FWL_{+,-}\big[  {\cal I}^\tree_{\rm bos}(+,1,2,\ldots,n,-) 
 -  {\cal I}^\tree_{\rm 2f}(+_{\rm f},1,2,\ldots,n,-_{\rm f}) \big]   \,  \big|_{\te{even}}^{D=4}  \notag
\\
& \! \! \! \! \! \! \! \! \! \! \! \! \! \! \! \!  =  \!  \sum_{\{12\ldots n\} \atop{= A \cup B}} 
\Pf(\Psi_A) \sum_{\rho \in S_{|B|}}
\PT(+,\rho(B),-)  \label{eq4.fourd} \\
&\qquad \! \! \! \! \! \! \! \! \! \! \! \! \! \! \! \!  \times \left\{ \begin{array}{cl}  0 &: \ |B| \le 1\\
    %%%
\big[
 \trv(\rho(b_1), \rho(b_2), \ldots, \rho(b_{|B|}))
- \trs(\rho(b_1), \rho(b_2), \ldots, \rho(b_{|B|}))
\big]\, \big|_{D=4} &: \ |B| \geq 2  \end{array} \right.   \, .
  \notag
\end{align}
The first three contributions in the $|B|\ge 2$ sector can be easily
read off from (\ref{eq3.7}) and (\ref{eq3.8}), as in the previous
examples,
\begin{align}
\trv(1, 2)  -  \trs(1, 2)  \, \big|_{D=4} &=   \frac{3}{4}  \trv(1, 2) \notag \\
\trv(1, 2, 3)  - \trs(1, 2, 3)  \, \big|_{D=4} &=   \frac{3}{4} \trv(1, 2, 3) \notag \\
\trv(1, 2, 3, 4)  - \trs(1, 2, 3, 4)  \, \big|_{D=4} &= 
 2^{-3} \big\{ 7 \trv(1, 2, 3, 4)
 + \trv(1, 3, 2, 4) + \trv(1, 2, 4, 3) \big\} \notag \\
 &\quad- 2^{-5} \big\{ \trv(1, 2) \trv(3, 4)
 + {\rm cyc}(2,3,4) \big\}  \notag \\
 &=    \frac{1}{8} \TEf{1,2,3,4} + \frac{3}{4} \trv(1, 2, 3, 4)  \, .    \label{eq4.fourda}  
\end{align}
Examinations of extended $\mathcal{N}=4,2$ supersymmetry in $D{=}4$
are redundant since the respective correlators are equivalent to the
$D{=}10$ example in (\ref{eq3.22}) and the $D{=}6$ example in
(\ref{eq4.6}). In absence of supersymmetry, the four-point instance
of (\ref{eq4.6}) has been used in \cite{Geyer:2017ela} to reproduce
the BCJ numerators of \cite{Bern:2013yya} with up to four powers of loop momentum for the 
box diagram.

Finally, we remark that the BCJ numerators in these general gauge
theories can be extracted from the same worldsheet techniques as for
$D=10$ SYM: In all cases, their $\sigma$-dependence exclusively enters
in the form of $\PT(+,1,2,\ldots,j,-) \Pf \Psi_{ \{j+1\ldots n\} }$
whose rewriting in terms of $n{+}2$-point Parke--Taylor factors can be
reduced to a solved tree-level problem as discussed in section
\ref{sec:BCJ}. We will present some examples for such BCJ numerators
in sec.~\ref{sec:MPF} and simplify them using multiparticle fields.

%%%%%%%%%%%%%%%%%%%%%%%%%%%%%%%%%%%%%%%%%%%%%%%%%%% 
%%%%%%%%%%%%%%%%%%%%%%%%%%%%%%%%%%%%%%%%%%%%%%%%%%% 
%%%%%%%%%%%%%%%%%%%%%%%%%%%%%%%%%%%%%%%%%%%%%%%%%%% 
%%%%%%%%%%%%%%%%%%%%%%%%%%%%%%%%%%%%%%%%%%%%%%%%%%% 
 
\section{Parity-odd contributions}
\label{sec:odd}

In this section, we derive parity-odd contributions to one-loop
correlators from forward limits in chiral fermions. More specifically,
this amounts to a parity-odd completion of the correlators
(\ref{eq3.22}) for $D=10$ SYM and those instances of (\ref{eq4.6})
with a chiral spectrum.

%%%%%%%%%%%%%%%%%%%%%%%%%%%%%%%%%%%%%%%%%%%%%%%%%%% 
%%%%%%%%%%%%%%%%%%%%%%%%%%%%%%%%%%%%%%%%%%%%%%%%%%% 
%%%%%%%%%%%%%%%%%%%%%%%%%%%%%%%%%%%%%%%%%%%%%%%%%%% 
%%%%%%%%%%%%%%%%%%%%%%%%%%%%%%%%%%%%%%%%%%%%%%%%%%% 

\subsection{General prescription and low-multiplicity validation}
\label{sec:odd.1}
The worldsheet prescription for the parity-odd sector of one-loop
amplitudes has been discussed in \cite{Clavelli:1986fj, Gross:1987pd,
  DHoker:1988pdl} for conventional strings and in \cite{Adamo:2013tsa}
for ambitwistor strings. Both approaches have in common that one of
the bosonic vertex operators needs to be inserted in the ghost picture
$-1$. This insertion of $V^{(-1)}$ in (\ref{eq2.1}) is essential for
zero-mode saturation in the ghost sector and gauge anomalies such as
the hexagon anomaly of $D=10$ SYM \cite{Frampton:1983ah,
  Frampton:1983nr, Zumino:1983rz}.

Accordingly, the forward-limit implementation of the parity-odd sector
should start from a tree-level correlator that also has an insertion
of $V^{(-1)}$. That is why the forward limit is performed in the
representation (\ref{eq2.8}) of the two-fermion correlator at tree
level, where both two fermions are in the $-1/2$ ghost picture. The
forward-limit prescription
\beq
(\chi_i)^\al (\chi_j)^\be \rightarrow
 \FWL_{i,j}\big(
(\chi_i)^\al (\chi_j)^\be \big)
= - \ell^\mu \gamma_\mu^{\al \be}
\label{eq2.13new}
\eeq
follows from (\ref{eq2.4}) \& (\ref{eq2.13}).  To ensure the correct
relative normalization between the parity-odd and parity-even sectors,
we repeat the exercise from sections~\ref{sec:loop.1} \&
\ref{sec:loop.3} of fixing the relative factor $\beta \in \mathbb Q$ between the bosonic and
fermionic forward limits using known properties,
\begin{align}
  &{\cal I}\oneloop_{\rm even,\beta}(\ferm{1}, 2,3,\ldots ,  n)  = \notag\\
  & \qquad {\rm FWL}_{+,-} 
                \big[{\cal I}^\tree_{\rm bos}(+,1, 2,3,\ldots, n,{-}) + \beta {\cal I}^\tree_{\rm 2f}(+_{\rm f},\ferm{1}, 2,3,\ldots, n,{-}_{\rm f}) \big] \, \Big|_{\te{even}} \,.
                \label{fivepointtwo}
\end{align}
With judicious application of scattering equations, the choice
$\beta=-1$ reproduces the $n=4$ result calculated in (\ref{eq3.22}) \&
(\ref{eq3.23}).\footnote{Also, one could in principle compute the parity-even
sector of higher-point correlators using this forward limit.
%Higher points for the parity-even sector
%  could in principle be done using this forward limit.  
  However, the presence of $\oslashed{\ell}$ and $\oslashed{\ep}_1$ (without an
  accompanying $\oslashed{k}_1$) obscure the supersymmetry
  cancelations, requiring increasingly complicated application of
  scattering equations.  As we will see shortly, the choice for
  $\beta$ is also reinforced by matching the expected relative factor
  between parity-odd and parity-even results.} 
The forward limit (\ref{fivepointtwo}) has also been studied by Frost \cite{Frost:2017},
where the singularities in $\sigma_{+,-}$ were demonstrated to cancel between
the bosonic and fermionic contribution. Also, Frost related the fermionic forward limit 
to the $\tau \rightarrow i \infty$ limit of the Ramond-sector contribution to bosonic one-loop 
correlators which generalizes the analysis of \cite{Roehrig:2017gbt} to ghost pictures 
 $(-\frac{1}{2}, -\frac{1}{2},-1)$.
  
The parity-odd forward limit
inherits this choice of $\beta$, converting the $D$-dimensional version 
of the tree-level correlator (\ref{eq2.8}) into
\begin{align}
&{\cal I}\oneloop_{\rm odd}(\ferm{1}, 2,3,\ldots ,  n)  = -{\rm FWL}_{+,-} 
                \big[ {\cal I}^\tree_{\rm 2f}(+_{\rm f},\ferm{1}, 2,3,\ldots, n,{-}_{\rm f}) \big] \, \Big|_{\te{odd}}  \notag \\
& \quad  = \frac{1}{2}
\sum_{ \{23 \ldots n\} \atop {=A\cup B\cup C}} \textrm{Pf} (\Psi_{A})
 \sum_{\rho \in S_{|B|}} \sum_{\tau \in S_{|C|}} 
 \PT(+,\rho(B),n,\tau(C),-)  \label{eq2.8new} \\
  &\qquad  \times  \trodd{ \oslashed{\ell}  \, \oslashed{f_{\rho(b_{1})}} \oslashed{f_{\rho(b_{2})}}\ldots \oslashed{f_{\rho(b_{|B|})}}  \oslashed{\ep_1}
    \oslashed{f_{\tau(c_{1})}} \oslashed{f}_{\tau(c_{2})}\ldots \oslashed{f_{\tau(c_{|C|})}} } \, .
  \notag
\end{align}
The notation $\trodd{\ldots}$ instructs to only keep
the parity-odd part of the chiral trace\footnote{When contracting $\ell^\mu, f_i^{\mu \nu}$ and
  $\ep_1^\mu$ with $2^{D/2} \times 2^{D/2}$ Dirac gamma matrices
  $\Gamma^\mu$ instead of the $2^{D/2-1} \times 2^{D/2-1}$ Weyl blocks
  $\gamma^\mu$, one can obtain $\trodd{\ldots}$ by inserting the
  $D$-dimensional chirality matrix $\Gamma_{D+1}$ into the trace.}
proportional to the Levi--Civita symbol
$\varep^{\mu_1 \mu_2\ldots \mu_D}$,
\beq
\trodd{ \oslashed{\ell}  \, \oslashed{f_{i}} \ldots \oslashed{f_{j}}\,
\oslashed{\ep_{1}} \, \oslashed{f_{p}} \ldots  \oslashed{f_{q}} }
\equiv
\ell^\mu \gamma_\mu^{\alpha \beta} (\oslashed{f_{i}} \ldots \oslashed{f_{j}} \,
\oslashed{\ep_{1}}  \, \oslashed{f_{p}} \ldots \oslashed{f_{q}} )_{\alpha  \beta} \, \big|_{\te{odd}} \, .
\label{deftrodd}
\eeq
Accordingly,
$\trodd{\ldots}$ with less than $D$ gamma matrices in the ellipsis
automatically vanish,
\beq
\trodd{ \gamma^{\mu_1} \gamma^{\mu_2} \ldots \gamma^{\mu_p} } = 0 \ \ \ \forall \ p<D \, .
\label{odd.1}
\eeq
Hence, the partitions of $\{2,3,\ldots ,n\}$ into $A,B,C$ must have $|B|+|C| \geq \frac{D}{2}-1$
in order to allow for a non-vanishing trace, starting with
\beq
\trodd{\gamma^{\mu_1} \gamma^{\mu_2} \ldots \gamma^{\mu_D} } 
= i 2^{D/2-1} \varep^{\mu_1 \mu_2\ldots \mu_D} \, .
\label{odd.2}
\eeq
This implies a minimum multiplicity $n = \frac{ D}{2}$ to obtain
non-zero parity-odd correlators (\ref{eq2.8new})
\begin{align}
{\cal I}\oneloop_{\rm odd}(\ferm{1}, 2,\ldots ,n) \, \Big|_{n<D/2} &= 0
\label{odd.3}  \\
{\cal I}\oneloop_{\rm odd}(\ferm{1}, 2,\ldots ,n) \, \Big|_{n=D/2} &= 
 \frac{1}{2}\sum_{ \{23 \ldots D/2\} \atop {= B\cup C}} 
 \sum_{\rho \in S_{|B|}} \sum_{\tau \in S_{|C|}} 
 \PT(+,\rho(B),n,\tau(C),-) \notag \\
  &\qquad  \times  \trodd{ \oslashed{\ell}  \, \oslashed{f_{\rho(b_{1})}}  \ldots \oslashed{f_{\rho(b_{|B|})}}  \oslashed{\ep_{1}}
    \oslashed{f_{\tau(c_{1})}}  \ldots \oslashed{f_{\tau(c_{|C|})}} }
\notag \\
&= 2^{-D/2} i \varep_{\mu_1 \mu_2\ldots \mu_D}  \ell^{\mu_{1}} \ep_{1}^{\mu_2} f_2^{\mu_3 \mu_4}f_3^{\mu_5 \mu_6}
\ldots f_{D/2}^{\mu_{D-1} \mu_{D}} 
\notag \\
&\qquad  \times \sum_{\rho \in S_{D/2}}  \PT(+,\rho(1,2,\ldots,\tfrac{D}{2}),-)  \, ,
\label{odd.4}
\end{align}
in lines with the analysis of fermionic zero mode in one-loop
worldsheet prescriptions \cite{Clavelli:1986fj, Gross:1987pd,
  DHoker:1988pdl, Adamo:2013tsa}. Moreover, the tensor structure of
the $(\frac{D}{2})$-point correlator (\ref{odd.4}) is entirely
determined by the fermionic zero modes. Like this, the
permutation-symmetric sum over Parke--Taylor factors in (\ref{odd.4})
is consistent with the worldsheet derivation.  In order to avoid
proliferation of indices, we employ shorthands
\beq
\varep_D(v_1,v_2,\ldots,v_D) =  \varep_{\mu_1 \mu_2\ldots \mu_D}  v_1^{\mu_1}  v_2^{\mu_2} \ldots  v_D^{\mu_D}
\, , \ \ \ \ \ \
\varep^\mu_D(v_2,\ldots,v_D) =  \varep^{\mu}{}_{\lambda_2\ldots \lambda_D}   v_2^{\lambda_2} \ldots  v_D^{\lambda_D}
\label{odd.5}
\eeq for Levi--Civita contractions of $D$-dimensional vectors
$v_j$. In this notation, the permutation-symmetric BCJ-numerators
following from (\ref{odd.4}) are given by 
\begin{equation}
  N\oneloop_{\text{odd}}(+,\rho(1,2,\ldots,\tfrac{D}{2}),-) =
  i \varep_D(\ell,\ep_{1},k_2,\ep_2,k_3,\ep_3,\ldots,k_{D/2},\ep_{D/2}) 
  \label{odd.6}
\end{equation}
after absorbing the leading factor of $\frac{1}{2}$ following the
definition of $N\oneloop$ from (\ref{eq3.21}).

In $D=10$ dimensions, this becomes a five-point numerator that
reproduces the parity-odd part of the pentagon numerator
$i \varep_{10}(\ell,\ep_{1},k_2,\ep_2,k_3,\ep_3,\ldots,k_{5},\ep_{5})$
in ten-dimensional SYM \cite{Mafra:2014gja, He:2017spx}. 
With the normalization of (\ref{odd.6}) and (\ref{eq3.22}),
we arrive at the relative factor of parity-even and -odd terms known from \cite{Green:2013bza} that
plays an important role for S-duality of the five-point one-loop amplitude of type-IIB
superstrings.
Similarly,
(\ref{odd.6}) in $D=6$ yields the parity-odd term
$i \varep_6(\ell,\ep_{1},k_2,\ep_2,k_3,\ep_3)$ in the triangle
numerator of chiral six-dimensional SYM with eight supercharges
\cite{Berg:2016fui, He:2017spx}.

%%%%%%%%%%%%%%%%%%%%%%%%%%%%%%%%%%%%%%%%%%%%%%%%%%% 
%%%%%%%%%%%%%%%%%%%%%%%%%%%%%%%%%%%%%%%%%%%%%%%%%%% 
%%%%%%%%%%%%%%%%%%%%%%%%%%%%%%%%%%%%%%%%%%%%%%%%%%% 
%%%%%%%%%%%%%%%%%%%%%%%%%%%%%%%%%%%%%%%%%%%%%%%%%%% 

\subsection{Anomalies and their singled-out leg}
\label{sec:odd.2}

In order to reproduce the expected gauge anomalies from our parity-odd
correlators, we need to evaluate the forward-limit prescription
(\ref{eq2.8new}) at multiplicities $\geq \frac{D}{2}+1$. This requires
chiral gamma traces beyond (\ref{odd.1}) and (\ref{odd.2}) such
as\footnote{We use $\miss{\mu}_j$ to denote that $\mu_j$ is missing
  from the index list.  The standard convention is to use
  $\hat{\mu}_j$, but we wish to avoid confusion with the special
  fermion leg which we have also labeled with $\ferm{i}$.
  Additionally, it is worth noting that it is possible to use
  overantisymmetrization identity from (\ref{odd.over}) to 
  rewrite (\ref{odd.7}) in a more symmetric form and to extend the summation range
  to all of $1\leq i<j\leq D{+}2$,
  %restore the sum to the full range
  \begin{equation}
  \trodd{ \gamma^{\mu_1}\gamma^{\mu_2} \ldots \gamma^{\mu_{D+2}} } =  -i 2^{D/2-1}
\sum_{1 \le i < j \le D{+}2} (-1)^{(j-i)}\eta^{\mu_i \mu_j}
\varep^{\mu_1 \ldots \miss{\mu}_i \ldots \miss{\mu}_j \ldots \mu_{D{+}2}} \, .
\end{equation}
More generally, the overantisymmetrization identity can be used to remove any specific 
index label from the summation range, freezing it to only appear in the $\varep$ tensor.}
\begin{equation}
\trodd{ \gamma^{\mu_1}\gamma^{\mu_2} \ldots \gamma^{\mu_{D+2}} } =  -i 2^{D/2-1}
\sum_{2 \le i < j \le D{+}2} (-1)^{(j-i)}\eta^{\mu_i \mu_j}
\varep^{\mu_1 \ldots \miss{\mu}_i \ldots \miss{\mu}_j \ldots \mu_{D{+}2}}  
\label{odd.7}
\end{equation}
and its generalizations, details of which are provided
in appendix \ref{sec:po-trace} (also see (4.35b) of \cite{Roehrig:2017gbt} for an alternative form 
of the all-multiplicity result). We have checked for the six-point
correlator of $D=10$ SYM and for the four-point correlator of chiral
SYM in $D=6$ that the forward-limit prescription (\ref{eq2.8new})
reproduces the expressions of \cite{He:2017spx},
\begin{align}
{\cal I}\oneloop_{\rm odd}&(\ferm{1}, 2,\ldots , n) \, \Big|_{n=D/2+1} = 
i \ell_\mu \sum_{\rho \in S_{D/2+1}}  \PT(+,\rho(1,2,\ldots,\tfrac{D}{2}{+}1),-)  \notag \\
 &\times \Big\{ \big[ (\ell \cdot \ep_2) \varep_D^\mu(\ep_1,k_3,\ep_3,\ldots,k_{D/2+1},\ep_{D/2+1}) + (2\leftrightarrow 3,4,\ldots, \tfrac{D}{2}{+}1) \big] \notag \\
 &\ \ \ \ + \frac{1}{2}  \big[ \sgn_{23}^\rho E^\mu_{1|23,4,\ldots,D/2+1}  + (2,3|2,3,4,\ldots, \tfrac{D}{2} +1) \big]
 \notag \\
 &\ \ \ \ + \frac{1}{2} \big[ \sgn_{12}^\rho E^\mu_{12|3,4,\ldots,D/2+1}  + (2\leftrightarrow 3,4,\ldots, \tfrac{D}{2}+1) \big] \Big\} \, ,
\label{odd.8}
\end{align}
where the notation $(2,3|2,3,4,\ldots, \tfrac{D}{2} {+}1)$ is explained below
(\ref{simpt12}). The $\rho$-dependent signs $\sgn_{ij}^\rho$ are defined in
(\ref{eq3.13}), and we have introduced the following shorthands for
the tensor structures in the last two lines:
\begin{align}
E^\mu_{12|3,4,\ldots,p} &= (\ep_1 \cdot k_2) \varep_{D}^\mu(\ep_2,k_3,\ep_3,\ldots,k_p,\ep_p)
-(\ep_2 \cdot k_1)\varep_{D}^\mu(\ep_1,k_3,\ep_3,\ldots,k_p,\ep_p) \cr
&- (\ep_1 \cdot \ep_2)\varep_{D}^\mu(k_2,k_3,\ep_3,\ldots,k_p,\ep_p)
\label{odd.9}\\
E^\mu_{1|23,4,\ldots,p}  &= (\ep_2 \cdot k_3)\varep_{D}^\mu(\ep_1,k_{2}{+}k_{3},\ep_3,\ldots,k_p,\ep_p)
-(\ep_3 \cdot k_2)\varep_{D}^\mu(\ep_1,k_{23},\ep_2,\ldots,k_p,\ep_p) \cr
&-(\ep_2 \cdot \ep_3)\varep_{D}^\mu(\ep_1,k_{2},k_3,\ldots,k_p,\ep_p)
-(k_2 \cdot k_3)\varep_{D}^\mu(\ep_1,\ep_{2},\ep_3,\ldots,k_p,\ep_p) \, .
\label{odd.10}
\end{align}
Note that we have used the overantisymmetrization identity
\beq
\varepsilon^{[\mu_1 \mu_2 \ldots \mu_D} \eta^{\mu_{D+1}] \lambda} = 0
\label{odd.over}
\eeq
in deriving (\ref{odd.8}) from (\ref{eq2.8new}). As a major advantage
of the forward-limit prescription (\ref{eq2.8new}), it bypasses the
reference to the spurious position of the picture-changing operator in
the one-loop worldsheet prescription \cite{Adamo:2013tsa}. Like this,
the Parke--Taylor decomposition of $(n\geq \frac{D}{2}+1)$-point
correlators is greatly facilitated by the approach in this section.

On the support of the scattering equations, (\ref{odd.8}) vanishes
under linearized gauge variations $\ep_j \rightarrow k_j$ in all the
legs $j=2,3,\ldots,n$ except for the first one. The variation
$\ep_1 \rightarrow k_1$ in the leg which is singled out by the hat
notation in (\ref{odd.8}) is proportional to $\ell^2$
\beq
{\cal I}\oneloop_{\rm odd}(\ferm{1}, 2,\ldots ,n) \, \Big|^{\ep_1 \rightarrow k_1}_{n=D/2+1} = i \ell^2
\varep_{D}(k_2,\ep_{2},k_3,\ep_3,\ldots,k_{D/2+1},\ep_{D/2+1})
\label{odd.11}
\eeq
and therefore yields rational loop integrals, see section 5.5 of
\cite{He:2017spx} for details in a CHY context\footnote{See
  \cite{Chen:2014eva, Mafra:2014gja} for earlier discussions in a
  field-theory context and \cite{Clavelli:1986fj, Mafra:2016nwr,
    Berg:2016wux} in a string-theory context.}.

Given the asymmetric gauge variations, the $(\frac{D}{2}{+}1)$-point
correlator (\ref{odd.8}) cannot be permutation invariant, not even on
the support of scattering equations. Indeed, the difference between
singling out legs 1 and 2 through the ghost picture $(-1)$ in (\ref{eq2.8})
is given by \cite{He:2017spx}
\begin{align}
&{\cal I}\oneloop_{\rm odd}(\ferm{1}, 2,3,\ldots , n) - {\cal I}\oneloop_{\rm odd}( 1, \ferm{2},3,\ldots , n) \, \Big|_{n=D/2+1}  \\
&\ \ = - \ell^2 \varep_D(\ep_1,\ep_2,k_3,\ep_3,k_4,\ep_4,\ldots,k_{D/2+1},\ep_{D/2+1})
 \sum_{\rho \in S_{D/2+1}}  \PT(+,\rho(1,2,\ldots,\tfrac{D}{2}{+}1),-) \, , \notag
\end{align}
see \cite{Mafra:2016nwr} for the analogous asymmetry of the one-loop
six-point amplitude of the pure-spinor superstring.

%%%%%%%%%%%%%%%%%%%%%%%%%%%%%%%%%%%%%%%%%%%%%%%%
%%%%%%%%%%%%%%%%%%%%%%%%%%%%%%%%%%%%%%%%%%%%%%%%
%%%%%%%%%%%%%%%%%%%%%%%%%%%%%%%%%%%%%%%%%%%%%%%%
%%%%%%%%%%%%%%%%%%%%%%%%%%%%%%%%%%%%%%%%%%%%%%%%
 
\section{BCJ numerators in terms of multiparticle fields}
\label{sec:MPF}

In this section, we provide alternative representations of the BCJ
numerators, where the contributions from the Pfaffian and the
field-strength traces in the correlators (\ref{eq3.22}), (\ref{eq4.6})
and (\ref{eq2.8new}) are combined. The driving force for particularly
compact expressions are so-called {\it multiparticle fields} --
essentially the numerators of Berends--Giele currents
\cite{Berends:1987me} in BCJ gauge, where the color-kinematics duality
is manifest \cite{Lee:2015upy, Bridges:2019siz}. Multiparticle fields
were initially constructed in pure-spinor superspace \cite{Mafra:2014oia} 
(see \cite{Mafra:2011kj, Mafra:2011nv} for precursors in the context of superstring tree
amplitudes) and later on formulated 
in components for arbitrary combinations of bosons and fermions \cite{Mafra:2015vca}.  
They became central ingredients of BCJ numerators \cite{Mafra:2011kj, Mafra:2014gja, 
Mafra:2015mja,  He:2017spx} and correlators for multiparticle string amplitudes
\cite{Gomez:2013sla, Mafra:2016nwr, Berg:2016wux, Mafra:2018nla,
  Mafra:2018qqe}.

%%%%%%%%%%%%%%%%%%%%%%%%%%%%%%%%%%%%%%%%%%%%%%%%%%% 
%%%%%%%%%%%%%%%%%%%%%%%%%%%%%%%%%%%%%%%%%%%%%%%%%%% 
%%%%%%%%%%%%%%%%%%%%%%%%%%%%%%%%%%%%%%%%%%%%%%%%%%% 
%%%%%%%%%%%%%%%%%%%%%%%%%%%%%%%%%%%%%%%%%%%%%%%%%%% 

\subsection{Brief review}
\label{sec:MPF.1}

Multiparticle polarizations $\ep_{P}^\mu$ and field strengths
$f_P^{\mu \nu}$ will be indexed by words $P=12\ldots p$ or
multiparticle labels. This subsection simply collects the definitions
relevant to later equations, and the reader is referred to \cite{Mafra:2015vca, 
Berg:2016fui, Garozzo:2018uzj} for further background.

Two-particle versions of polarization vectors and field strengths are defined by 
\begin{align}
\ep_{12}^\mu &= (k_2\cdot \ep_1) \ep_2^\mu - (k_1\cdot \ep_2) \ep_1^\mu + \frac{1}{2}(\ep_1\cdot \ep_2)(k_1^\mu - k_2^\mu) \notag \\
&= \frac{1}{2} \big\{ (k_2\cdot \ep_1) \ep_2^\mu
+ (\ep_1)_\nu f_2^{\nu \mu}  -(1\leftrightarrow 2) \big\} 
\label{eq5.1} \\
f_{12}^{\mu \nu} &=  (k_2\cdot \ep_1) f_2^{\mu \nu} - (k_1\cdot \ep_2) f_1^{\mu \nu} 
+ f_1^{\mu}{}_\la f_2^{\la \nu} -  f_2^{\mu}{}_\la f_1^{\la \nu} \notag \\
&= k_{12}^\mu \ep_{12}^\nu - k_{12}^\nu \ep_{12}^\mu - (k_1 \cdot k_2) (\ep_1^\mu \ep_2^\nu - \ep_1^\nu \ep_2^\mu )
\label{eq5.2}
\end{align}
and obey $\ep_{12}^\mu=-\ep_{21}^\mu$ as well as $f_{12}^{\mu \nu}=-f_{21}^{\mu \nu}$.
Here and below, the notation for multiparticle momenta is
\beq
 k_{12\ldots p} = k_1 + k_2 +\ldots + k_p\,.
 \label{eq5.0}
\eeq
Three-particle polarizations are defined in two steps: Promoting
(\ref{eq5.1}) to a recursion with labels $(1,2) \rightarrow (12,3)$  
yields the intermediate expression
\beq
\widehat{\ep}_{123}^\mu =  \frac{1}{2} \big\{ (k_3\cdot \ep_{12}) \ep_3^\mu -  (k_{12}\cdot \ep_3) \ep_{12}^\mu
+( \ep_{12})_\nu f_3^{\nu \mu}-(\ep_{3})_\nu f_{12}^{\nu \mu}   \big\}  \label{eq5.3}
\eeq
subject to $\widehat{\ep}_{123}^\mu=- \widehat{\ep}_{213}^\mu$. Given that
$\widehat{\ep}_{123}^\mu+\widehat{\ep}_{231}^\mu+\widehat{\ep}_{312}^\mu=3 k_{123}^\mu h_{123}$
for some scalar $h_{123}$ given below,
an improved version that obeys the kinematic off-shell Jacobi 
identity $\ep_{123}^\mu+\ep_{231}^\mu+\ep_{312}^\mu=0$
on top of $\ep_{123}^\mu=-\ep_{213}^\mu$ follows from the redefinition
\beq
\ep_{123}^\mu = \widehat{\ep}_{123}^\mu - k_{123}^\mu h_{123} \, , \ \ \ \ \ \ 
h_{123} = \frac{1}{12} \ep_1^\mu (f_2)_{\mu \nu} \ep_3^\nu + {\rm cyc}(1,2,3) = - h_{213} = - h_{132} \, ,
\label{eq5.3a}
\eeq
which is part of a non-linear gauge transformation \cite{Lee:2015upy}.
The associated field strength subject to $f_{123}^{\mu \nu}=-f_{213}^{\mu \nu}$
and $f_{123}^{\mu \nu}+f_{231}^{\mu \nu} + f_{312}^{\mu \nu}=0$ is
\begin{align}
f_{123}^{\mu \nu} &= k_{123}^\mu \ep_{123}^\nu - (k_{12}\cdot k_3) \ep_{12}^\mu \ep_3^\nu
 - (k_1\cdot k_2) (\ep_1^\mu \ep_{23}^\nu - \ep_2^\mu \ep_{13}^\nu)
- (\mu \leftrightarrow \nu) \, , \label{eq5.4}
\end{align}
where $h_{123}$ drops out from $k_{123}^\mu \ep_{123}^\nu- (\mu \leftrightarrow \nu)$.
Some of the later numerators involve the four-particle field strength 
that can be assembled from
\begin{align}
\widehat{\ep}_{1234}^\mu &=  \frac{1}{2} \big\{ (k_4\cdot \ep_{123}) \ep_4^\mu -  (k_{123}\cdot \ep_4) \ep_{123}^\mu
+( \ep_{123})_\nu f_4^{\nu \mu}- (\ep_{4})_\nu f_{123}^{\nu \mu}   \big\} 
\notag \\
{\ep'}_{1234}^{\mu} &= \widehat{\ep}_{1234}^\mu - (k_{12}\cdot k_3) \ep_3^\mu h_{124} - (k_1\cdot k_2) (\ep_2^\mu h_{134} - \ep_1^\mu h_{234}) \label{eq5.4y} \\
f^{\mu \nu}_{1234} &= k_{1234}^\mu {\ep'}_{1234}^{\nu} 
- (k_{123}\cdot k_4) \ep_{123}^\mu \ep_4^\nu
- (k_{12}\cdot k_3) (\ep_{12}^\mu \ep_{34}^\nu + \ep_{124}^\mu \ep_{3}^\nu) \notag \\
& - (k_1\cdot k_2) (\ep_{13}^\mu \ep_{24}^\nu + \ep_{14}^\mu \ep_{23}^\nu + \ep_{134}^\mu \ep_{2}^\nu
-  \ep_{234}^\mu \ep_{1}^\nu)
- (\mu \leftrightarrow \nu)   \notag
\end{align}
and obeys $f_{1234}^{\mu \nu}=-f_{2134}^{\mu \nu}$
and $f_{1234}^{\mu \nu}+f_{2314}^{\mu \nu} + f_{3124}^{\mu \nu}=0$
as well as $f_{1234}^{\mu \nu}-f_{1243}^{\mu \nu} + f_{3412}^{\mu \nu} -  f_{3421}^{\mu \nu}=0$.

%%%%%%%%%%%%%%%%%%%%%%%%%%%%%%%%%%%%%%%%%%%%%%%%%%% 
%%%%%%%%%%%%%%%%%%%%%%%%%%%%%%%%%%%%%%%%%%%%%%%%%%% 
%%%%%%%%%%%%%%%%%%%%%%%%%%%%%%%%%%%%%%%%%%%%%%%%%%% 
%%%%%%%%%%%%%%%%%%%%%%%%%%%%%%%%%%%%%%%%%%%%%%%%%%% 

\subsection{$D=10$ examples}
\label{sec:MPF.2}

We shall now express $(n\geq 5)$-point examples of the $D=10$ SYM
correlators (\ref{eq3.22}) in terms of multiparticle fields and
provide a new seven-point result.  At six points, we spell out local
versions of the manifestly gauge-invariant BCJ numerators in
\cite{He:2017spx}, also see \cite{Mafra:2014gja} for their supersymmetrization. 
In the same way as the $t_8$-tensor
(\ref{eq2.t8}) furnishes the four-point BCJ numerators in
(\ref{eq3.24}), higher-point numerators will boil down to its
contraction with multiparticle field strengths such as 
(\ref{eq5.2}), (\ref{eq5.4}) and (\ref{eq5.4y}),
\beq
\TE{A,B,C,D} = \TEf{A,B,C,D} \, .
\label{eq5.4w}
\eeq
The symmetries of $f_A^{\mu \nu}$ in its multiparticle label
$A=12\ldots$ propagate to (\ref{eq5.4w}) in the obvious manner, \eg
$f_{12}^{\mu \nu}=-f_{21}^{\mu \nu}$ and
$f_{123}^{\mu \nu} + {\rm cyc}(1,2,3) = 0$ imply that
$\TE{12,3,4,5}= - \TE{21,3,4,5}$ and
$\TE{123,4,5,6} + {\rm cyc}(1,2,3) = 0$, respectively. We also
introduce vectorial generalizations
\begin{align}
\TE[\mu]{A,B,C,D,E} &=  \ep_A^\mu \TE{B,C,D,E} + 
\ep_B^\mu \TE{A,C,D,E}+
\ep_C^\mu \TE{A,B,D,E} \notag \\
&\ \ \ \
+\ep_D^\mu \TE{A,B,C,E}
+\ep_E^\mu \TE{A,B,C,D}
\label{eq5.6} 
\end{align}
and tensorial ones
\begin{align}
\TE[\mu \nu]{A,B,\ldots,F} &= ( \ep_A^\mu \ep_B^\nu  +  \ep_A^\nu \ep_B^\mu )  \TE{C,D,E,F}
+ (A,B|A,B,\ldots,F)
\label{eq5.6new}  \\
%%%
\TE[\mu \nu \lambda]{A,B,\ldots,G} &=  (\ep_A^\mu \ep_B^\nu \ep_C^\lambda + {\rm symm}(\mu,\nu,\lambda))  \TE{D,E,F,G} + (A,B,C|A,B,\ldots,G) \, .
\notag
\end{align}
These building blocks are symmetric under exchange of multiparticle labels,
say $\TE[\ldots]{A,B,\ldots}=\TE[\ldots]{B,A,\ldots}$, obey the symmetries of 
$\ep_A^\mu,f_B^{\mu \nu}$ within $A,B,\ldots$ and follow the
combinatorics of their counterparts in pure-spinor superspace
\cite{Mafra:2014gsa, Mafra:2018nla}.

From the decomposition (cf.\ (\ref{eq3.21}))
\beq
{\cal I}\oneloop_{D=10 \ {\rm SYM}}(1,2,\ldots ,n) = \frac{1}{2}\sum_{\rho \in S_{n}} N\oneloop\maxsusy(+,\rho(1,2,\ldots,n),-)
\PT(+,\rho(1,2,\ldots,n),-)\, ,
\eeq
we find the four- and five-point numerators
\begin{align}
N\oneloop\maxsusy(+,1,2,3,4,-) &= \TE{1,2,3,4}
\notag \\
N\oneloop\maxsusy(+,1,2,3,4,5,-) &= \ell_\mu \TE[\mu]{1,2,3,4,5}  - \frac{1}{2} \Big\{
\TE{{12},3,4,5}+\TE{{13},2,4,5}  \label{eq5.5} \\
&\qquad
+\TE{{14},2,3,5}+\TE{{15},2,3,4}
+\TE{{23},1,4,5}+\TE{{24},1,3,5} \notag \\
&\qquad
+\TE{{25},1,3,4}+\TE{{34},1,2,5}
+\TE{{35},1,2,4}+\TE{{45},1,2,3}
\Big\} \, .  \notag
\end{align}
Note that the contribution
$f_1^{\mu}{}_\la f_2^{\la \nu} - f_2^{\mu}{}_\la f_1^{\la \nu}$ to the
two-particle field strength in (\ref{eq5.2}) stems from the
commutators $[f_1,f_2]^{\mu \nu}$ in the $t_8$-representation of the
five-traces (\ref{eq3.25}). The remaining contributions to
$f^{\mu \nu}_{12}$ such as
$(k_2\cdot \ep_1) f_2^{\mu \nu} - (k_1\cdot \ep_2) f_1^{\mu \nu} $ are
due to
$\Pf(\Psi_{ \{1\} }) [\trv(f_2f_3f_4f_5) - \frac{1}{2}
\trs(f_2f_3f_4f_5)]$ and its permutations in (\ref{eq3.22}).  
%The worldsheet origin of the five-point numerators (\ref{eq5.5}) is also
%explained in appendix D of \cite{He:2017spx}, using the one-loop
%ambitwistor-string prescription in the RNS formalism, and its
%supersymmetrization can be found in \cite{Mafra:2014gja}.
Supersymmetric BCJ numerator on quadratic propagators with the structure of
(\ref{eq5.5}) have been constructed in \cite{Mafra:2014gja}. Moreover, the $t_8$-tensors 
in (\ref{eq5.5}) have been later on derived from the one-loop ambitwistor-string prescription in the RNS 
formalism, see appendix D of \cite{He:2017spx}.
Additionally, antisymmetrizing (\ref{eq5.5}) in 1,2, we find the
numerator of a massive box diagram (with legs 1,2 in a dangling tree)
to be $\TE{12,3,4,5}$.

On the support of scattering equations, the six-point analogues of
(\ref{eq5.5}) following from the correlators (\ref{eq3.22}) are
\begin{align}
&N\oneloop\maxsusy(+,1,2,3,4,5,6,-)=  \frac{1}{2} \ell_\mu \ell_\nu \TE[\mu\nu]{1,2,3,4,5,6}  - \frac{1}{2} \ell_\mu \Big\{
\TE[\mu]{12,3,4,5,6}  + (1,2|1,2,\ldots,6)   
\Big\} \notag \\
& \quad + \frac{1}{4}  \Big\{ \TE{{12},{34},5,6} +\TE{{13},{24},5,6}+\TE{{14},{23},5,6}+(5,6|1,2,\ldots,6) \Big\} \label{eq5.7} \\
& \quad + \frac{1}{6}  \Big\{ \TE{{123},{4},5,6} +\TE{{321},{4},5,6}+(1,2,3|1,2,\ldots,6) \Big\}  
+ \widehat t_{12}(1,2,3,4,5,6)  \, . \notag
\end{align}
The linear order in loop momentum follows the combinatorics of the
$\ell$-independent five-point numerator in (\ref{eq5.5}), i.e.\ with all 
$\TE[\mu]{ij,\ldots}$ subject to $1\leq i<j\leq 6$. 
The $\ell$-independent part features a total of 45
arrangements $t_8(ij,pq,\ldots) $ with
$1{\leq} i{<}j{\leq} 6,\ 1{\leq} p{<}q{\leq} 6$ and $i{<} p$ as well
as $20$ pairs of terms $ t_8({ijk},\ldots) +t_8({kji},\ldots)$ with
$1{\leq} i{<}j{<}k{\leq} 6$.  Finally, the last term of (\ref{eq5.7})
adds a permutation symmetric piece to the zeroth order in $\ell$:
\begin{align}
\widehat t_{12}(1,2,3,4,5,6) &= \frac{1}{12} \Big\{ (k_2^\mu {-} k_1^\mu) \TE[\mu]{12,3,4,5,6} +(1,2|1,2,3,4,5,6)
\Big\}  \notag \\
&-\frac{1}{24} \TE[\mu\nu]{1,2,3,4,5,6} \sum_{j=1}^6 k_j^\mu k_j^\nu  \, .
\label{eq5.81}
\end{align}
Note that the combinatorics of (\ref{eq5.7})
and (\ref{eq5.81}) also governs the supersymmetric and local hexagon
numerator in (4.35) of \cite{Mafra:2014gja}.  As indicated by the
widehat of $\widehat t_{12}$, (\ref{eq5.81}) is \emph{not} the
gauge-invariant $t_{12}$-quantity in (\ref{3.t2n}).

In the seven-point generalization of (\ref{eq5.7}), all
$\ell$-dependent terms can be anticipated by adjoining a vector index
to the building blocks of the above $N\oneloop\maxsusy(+,1,\ldots,6,-)$, 
see the first three lines of 
\begin{align}
&N\oneloop\maxsusy(+,1,2,\ldots,7,-)=  \frac{1}{6} \ell_\mu \ell_\nu \ell_\lambda \TE[\mu\nu\lambda]{1,2,\ldots,7}  - \frac{1}{4} \ell_\mu \ell_\nu \Big\{
\TE[\mu \nu]{12,3,\ldots,7}  +(1,2|1,2,\ldots,7) \Big\} \notag \\
& \quad + \frac{1}{4} \ell_\mu \Big\{ \TE[\mu]{{12},{34},5,6,7} +\TE[\mu]{{13},{24},5,6,7}+\TE[\mu]{{14},{23},5,6,7}+(5,6,7|1,2,\ldots,7) \Big\} \notag \\
& \quad + \frac{1}{6} \ell_\mu \Big\{ \TE[\mu]{{123},{4},\ldots,7} +\TE[\mu]{{321},{4},\ldots,7}+(1,2,3|1,2,\ldots,7) \Big\}  
+ \ell_\mu \widehat t^\mu_{12}(1,2,3,\ldots,7)  \notag \\
& \quad - \frac{1}{2} \Big\{ \widehat t_{12}(12,3,\ldots,7)  +(1,2|1,2,\ldots,7)  \Big\} 
 -  \frac{1}{8}  \Big\{ \TE{{12},{34},56,7} +104 \ \te{others} \Big\}  \notag \\
&\quad -  \frac{1}{12}  \Big\{  \TE{{123},45,6,7}  + \TE{{321},45,6,7}   +209 \ \te{other pairs} \Big\}  \notag \\
&\quad + \frac{1}{12} \Big\{ {-} \TE{1234,5,6,7} + 
\TE{4321,5,6,7} + \TE{1423,5,6,7} \notag \\
&\qquad\qquad + \TE{2314,5,6,7}  +(5,6,7|1,2,\ldots,7)\Big\} \notag \\
  &\quad - \frac{1}{96} \Big\{ \Delta_8(12|3,4,5,6,7) + (1,2|1,2,\ldots,7) \Big\}
    \, .\label{eq5.82}
\end{align}
The $\ell$-independent terms in the last five lines
contain the new seven-point information\footnote{The coefficients
  $ - \frac{1}{8} = (-\frac{1}{2})^3$ and
  $- \frac{1}{12} = \frac{1}{6} (-\frac{1}{2}) $ of
  $ \TE{{12},{34},56,7}$ and $ \TE{{123},45,6,7}$ can be anticipated
  by multiplying the combinatorial factors $-\frac{1}{2}$ and
  $\frac{1}{6}$ of the two- and three-particle slots in (\ref{eq5.5}) and
  (\ref{eq5.7}). The permutation sums include combinations
  $ \TE{{ijk},pq,\ldots} + \TE{{kji},pq,\ldots}$ with
  $1{\leq}i{<}j{<}k{<}7$ and $1{\leq } p{<}q{\leq}7$ as well as
  $ \TE{{ij},pq,rs,\ldots}$ with
  $1{\leq}i{<}j{<}7, \ 1{\leq}p{<}q{<}7, \ 1{\leq } r{<}s{\leq}7$ and
  $i{<}p{<}r$ as well as
  $ {-} \TE{ijkl,\ldots} + \TE{lkji,\ldots} + \TE{i ljk ,\ldots} +
  \TE{jkil ,\ldots} $ with $1{\leq}i{<}j{<}k{<}l{<}7$.}. We have
introduced a vectorial and a two-particle version of the permutation
symmetric hexagon building block (\ref{eq5.81}),
\begin{align}
\widehat t^\mu_{12}(1,2,3,\ldots,7) &=  \frac{1}{12} \Big\{ (k_2^\lambda {-} k_1^\lambda) t_8^{\mu \lambda}(12,3,\ldots,7) +(1,2|1,2,\ldots,7)
\Big\}  \notag \\
&-\frac{1}{24} t_8^{\mu\nu \lambda}(1,2,\ldots,7) \sum_{j=1}^7 k_j^\nu k_j^\lambda \notag
\\
\widehat t_{12}(12,3,\ldots,7) &= 
 \frac{1}{12} \Big\{ (k_3^\mu {-} k_{12}^\mu) t_8^\mu(123,4,5,6,7) +(3\leftrightarrow 4,5,6,7)\Big\} \label{eq5.83} \\
&+ \frac{1}{12} \Big\{ (k_4^\mu {-} k_3^\mu) t_8^\mu(12,34,5,6,7) + (3,4|3,4,5,6,7)\Big\}  \notag \\
&-\frac{1}{24} t_8^{\mu\nu}(12,3,\ldots,7) \Big\{ k_{12}^\mu k_{12}^\nu + \sum_{j=3}^7 k_j^\mu k_j^\nu  \Big\} \, . \notag
\end{align}
Additionally, we gather those terms which could not be lined up with multiparticle polarizations
in the new $\Delta$ building block in the last line of (\ref{eq5.82}),
\begin{align}
  \Delta_8(12|3,4,5,6,7) &= 2 (k_1 \cdot k_2) \Big( (\ep_1 \cdot \ep_3)(\ep_2 \cdot(k_3 - k_1)) - (\ep_2 \cdot \ep_3)(\ep_1 \cdot(k_3 - k_2)) \Big)\TE{4,5,6,7} \notag\\
  &\qquad + (3 \leftrightarrow 4,5,6,7)\,.  
  \label{newdel}          
\end{align}
This object is antisymmetric in the two labels to the left of the
vertical bar, and as such contributes to the seven-point hexagon
numerator where those two legs have been pulled out as the dangling
tree. It would be interesting to relate (\ref{newdel}) to a component version
of the so-called refined building blocks $J$ in pure-spinor 
superspace \cite{Mafra:2014gsa, Mafra:2018nla}.

%%%%%%%%%%%%%%%%%%%%%%%%%%%%%%%%%%%%%%%%%%%%%%%%%%% 
%%%%%%%%%%%%%%%%%%%%%%%%%%%%%%%%%%%%%%%%%%%%%%%%%%% 
%%%%%%%%%%%%%%%%%%%%%%%%%%%%%%%%%%%%%%%%%%%%%%%%%%% 
%%%%%%%%%%%%%%%%%%%%%%%%%%%%%%%%%%%%%%%%%%%%%%%%%%% 

\subsection{$D=6$ examples}
\label{sec:MPF.3}

Similar one-loop numerators can be given for $D=6$ SYM with
half-maximal supersymmetry.  We will focus on a hypermultiplet running
in the loop, whose correlators have been expressed in terms of traces in (\ref{eq4.7}).
As a reminder, the particle content for the hypermultiplet is given by
the parameter choices
${\bf n_{\rm v}} = 0,\, {\bf n_{\rm f}} = 1, \, {\bf n_{\rm s}} = 2$,
and the numerators will be defined via
\begin{equation}
{\cal I}\oneloop_\hyper = \frac{1}{2}\sum_{\rho \in S_{n}}
N\oneloop_{\text{hyp}}(+,\rho(1,2,\ldots,n),-)
\PT(+,\rho(1,2,\ldots,n),-)\, .
\end{equation}

As an analogue of the multiparticle $t_8$-tensor (\ref{eq5.4w}) that
governs maximally supersymmetric numerators, the basic scalar building
block for $D=6$ SYM is
\begin{equation}
t_4(A,B) = -\frac{1}{2}(f_A)^{\mu \nu} (f_B)_{\mu \nu} = \frac{1}{2} \trv(A,B)\, .
 \label{eq5.84}
\end{equation}
Its simplest instance
$t_4(1,2) =  (k_1\cdot
\ep_2)(k_2\cdot \ep_1)- (k_1\cdot k_2)(\ep_1\cdot \ep_2) $ vanishes in the momentum phase space of two
massless particles, but we will find non-vanishing multiparticle
examples. In particular, one can attain linearized gauge invariance at
the level of loop integrands by relaxing momentum conservation: The
numerators of this section are understood to rely on no Mandelstam
identity other than $s_{12\ldots n}=0$ at $n$ points. This proposal
goes back to work of Minahan in 1987 \cite{Minahan:1987ha} and will be
referred to as Minahaning (also see \cite{Berg:2016wux, Berg:2016fui,Bern:2012uf}
for four-point implementations).

At three points for instance, Minahaning amounts to keeping nonzero
$s_{ij}$ while imposing $s_{12}{+}s_{13}{+}s_{23}=0$, and it
introduces non-vanishing $s_{ijk}$ at four points. For
dot products with polarization vectors, transversality and momentum
conservation will be used as usual, \ie
$(\ep_1\cdot k_{12\ldots n})= (\ep_1\cdot k_{2\ldots n})=0$. These
choices lead to
$t_4(12,3)= (k_1\cdot k_2)(\ep_1\cdot \ep_2) (k_1\cdot \ep_3)$,
where the factor of $(k_1\cdot k_2)$ cancels the formally divergent
propagator $(k_1{+}k_2)^{-2}$ of a three-point diagram with an
external bubble. More generally, any potentially divergent propagator
introduced by Parke--Taylor integrals (\ie forward limits of
doubly-partial amplitudes) will be cancelled by the corresponding
Mandelstam invariant from the numerators of this section. 
However, this mechanism does not cure forward-limit divergences in the tree-level propagators
that arise when integrating non-supersymmetric correlators (\ref{eq4.6}) in
terms of doubly-partial amplitudes.

Similar to (\ref{eq5.6}) and (\ref{eq5.6new}), the subsequent
numerators are built from vector and tensor generalizations of the
scalar building block (\ref{eq5.84}),
\begin{align}
t_4^\mu(A,B,C) &= \ep_A^\mu t_4(B,C) +\ep_B^\mu  t_4(A,C) + \ep_C^\mu t_4(A,B)
\notag \\
t_4^{\mu \nu}(A,B,C,D) &= (\ep_A^\mu \ep_B^\nu+ \ep_A^\nu\ep_B^\mu) t_4(C,D) + (A,B|A,B,C,D) 
 \label{eq5.85}
\\
t_4^{\mu \nu \lambda}(A,B,C,D,E) &= (\ep_A^\mu \ep_B^\nu \ep_C^\lambda + {\rm symm}(\mu,\nu,\la)) t_4(D,E) + (A,B,C|A,B,C,D,E) \, ,
\notag
\end{align}
which are again symmetric under exchange of multiparticle labels
$t_4^{\ldots}(A,B,\ldots)= t_4^{\ldots}(B,A,\ldots)$ and were firstly considered
in the context of one-loop superstring amplitudes with reduced supersymmetry \cite{Berg:2016wux, Berg:2016fui}.  
With these definitions, the BCJ numerators following from the correlator
(\ref{eq4.7}) include\footnote{Similar to the earlier examples, the
  permutation sums in (\ref{eq5.87}) include all $ t^\mu_4(ij,\ldots)$
  with $1{\leq} i{<}j {\leq} 4$ and
  $t_4(ijk,\ldots) + t_4(kji,\ldots)$ with
  $1{\leq} i{<}j {<}k{\leq} 4$.}  
\begin{align}
-N\oneloop_{\text{hyp}}(+,1,2,3,-) &= \ell_\mu t_4^\mu(1,2,3) - \frac{1}{2} \Big\{ t_4(12,3) +  t_4(13,2) +  t_4(23,1) \Big\}
 \label{eq5.86}
 \\
 -N\oneloop_{\text{hyp}}(+,1,2,3,4,-) &= \frac{1}{2}\ell_\mu \ell_\nu t_4^{\mu \nu}(1,2,3,4) -  \frac{1}{2}\ell_\mu \Big\{ t^\mu_4(12,3,4) +  (1,2|1,2,3,4) \Big\} \notag \\
 &+  \frac{1}{6} \Big\{ t_4(123,4) + t_4(321,4)  + (4\leftrightarrow 1,2,3) \Big\} \label{eq5.87} \\
 &+ \frac{1}{4} \Big\{ t_4(12,34) + t_4(13,24) + t_4(14,23) \Big\}   \cr
 &  + \widehat{t}_8(1,2,3,4) - \frac{1}{12} N\oneloop\maxsusy(+,1,2,3,4,-) \, ,
\notag
\end{align}
where the quantity $\widehat{t}_8(1,2,3,4)$ in the last line generalizes (\ref{eq5.81}) to half-maximal supersymmetry
and does {\it not} coincide with the gauge-invariant $t_8(1,2,3,4)$:
\beq
\widehat{t}_8(1,2,3,4) = \frac{1}{12} \Big\{(k_2^{\mu}{-}k_1^{\mu}) t_4^\mu(12,3,4) + (1,2|1,2,3,4) \Big\}
 - \frac{1}{24} t_4^{\mu \nu}(1,2,3,4) \sum_{j=1}^4 k_j^\mu k_j^\nu \, .
  \label{eq5.88}
\eeq
By comparison with (\ref{eq5.5}) and (\ref{eq5.7}), the triangle and
box numerator with half-maximal supersymmetry share the combinatorics
of maximally supersymmetric pentagon and hexagon numerators. The
examples (\ref{eq5.86}) and (\ref{eq5.87}) have been known from
\cite{Berg:2016fui, He:2017spx}, and the one-loop string-amplitude
prescription implies\footnote{More specifically, this property follows
  from the sum over spin structures in the RNS-prescription for
  one-loop amplitudes of conventional strings and ambitwistor
  strings. Depending on the amount of spacetime supersymmetry, the
  partition functions for given spin structures conspire to eliminate
  $2$ or $4$ singularities from the fermionic two-point functions when
  performing the spin sum \cite{Berg:2016wux, He:2017spx}.} that
$n$-point amplitudes with half-maximal supersymmetry generally have
the same complexity as $(n{+}2)$-point amplitudes with maximal
supersymmetry \cite{Berg:2016wux}. Accordingly, the following pentagon
numerator is inspired by the maximally supersymmetric heptagon
numerator (\ref{eq5.82}),
\begin{align}
&-N\oneloop_{\text{hyp}}(+,1,2,3,4,5,-)=  \frac{1}{6} \ell_\mu \ell_\nu \ell_\lambda t_4^{\mu\nu\lambda}(1,2,3,4,5)  - \frac{1}{4} \ell_\mu \ell_\nu \Big\{
t^{\mu \nu}_4(12,3,4,5)  + (1,2|1,2,3,4,5) \Big\} \notag \\
& \quad + \frac{1}{4} \ell_\mu \Big\{ t^\mu_4({12},{34},5) +t^\mu_4({13},{24},5)+t^\mu_4({14},{23},5)+(5\leftrightarrow 1,2,3,4)\Big\} \notag \\
& \quad + \frac{1}{6} \ell_\mu \Big\{ t^\mu_4({123},{4},5) +t^\mu_4({321},{4},5)+(4,5|1,2,3,4,5) \Big\}  
+ \ell_\mu \widehat t^\mu_{8}(1,2,3,4,5)  \notag \\
& \quad - \frac{1}{2} \Big\{ \widehat t_{8}(12,3,4,5) +(1,2|1,2,3,4,5) \Big\} 
 -  \frac{1}{12}  \Big\{  t_4({123},45)  + t_4({321},45)   +(4,5|1,2,3,4,5) \Big\}  \notag \\
&\quad + \frac{1}{12} \Big\{ {-} t_4(1234,5) + 
t_4(4321,5) + t_4(1423,5) + t_4(2314,5)  +(5\leftrightarrow 1,2,3,4)\Big\} \notag \\
  &\quad - \frac{1}{96} \Big\{ \Delta_4(12|3,4,5)+ (1,2|1,2,3,4,5)\Big\} - \frac{1}{12} N\oneloop\maxsusy(+,1,2,3,4,5,-) 
    \, , \label{eq5.89}
\end{align}
where by analogy with (\ref{eq5.83}) and (\ref{newdel})
\begin{align}
\widehat t^\mu_{8}(1,2,3,4,5) &=  \frac{1}{12} \Big\{ (k_2^\lambda {-} k_1^\lambda) t_4^{\mu \lambda}(12,3,4,5) + (1,2|1,2,3,4,5)
\Big\} 
-\frac{1}{24} t_4^{\mu\nu \lambda}(1,2,3,4,5) \sum_{j=1}^5 k_j^\nu k_j^\lambda \notag
\\
\widehat t_{8}(12,3,4,5) &= 
 \frac{1}{12} \Big\{ (k_3^\mu {-} k_{12}^\mu) t_4^\mu(123,4,5) + (k_4^\mu {-} k_3^\mu) t_4^\mu(12,34,5) + {\rm cyc}(3,4,5) \Big\}   \label{eq5.90} \\
                              &-\frac{1}{24} t_4^{\mu\nu}(12,3,4,5) \Big\{ k_{12}^\mu k_{12}^\nu + \sum_{j=3}^5 k_j^\mu k_j^\nu  \Big\} \notag\\
  \Delta_4(12|3,4,5) &= 2 (k_1 \cdot k_2) \Big( (\ep_1 \cdot \ep_3)(\ep_2 \cdot(k_3 {-} k_1)) - (\ep_2 \cdot \ep_3)(\ep_1 \cdot(k_3 {-} k_2)) \Big)t_4(4,5)+ (3 \leftrightarrow 4,5) \notag \,.
\end{align}
The appearance of the maximally supersymmetric pentagon numerator (\ref{eq5.5})
in the last line of (\ref{eq5.89}) generalizes the $t_8(1,2,3,4)$
in the last line of (\ref{eq5.87}) to five points.
The derivation of this new five-point result has been greatly
facilitated by the representation (\ref{eq4.6}) of the correlator
induced by forward limits.

%%%%%%%%%%%%%%%%%%%%%%%%%%%%%%%%%%%%%%%%%%%%%%%%%%% 
%%%%%%%%%%%%%%%%%%%%%%%%%%%%%%%%%%%%%%%%%%%%%%%%%%% 
%%%%%%%%%%%%%%%%%%%%%%%%%%%%%%%%%%%%%%%%%%%%%%%%%%% 
%%%%%%%%%%%%%%%%%%%%%%%%%%%%%%%%%%%%%%%%%%%%%%%%%%% 

\subsection{Parity-odd examples}
\label{sec:MPF.4}

The forward-limit prescription (\ref{eq2.8new}) for parity-odd
correlators can also be lined up with compact BCJ numerators in terms
of multiparticle fields. On top of the simplest non-vanishing numerator (\ref{odd.6}) at multiplicity
$n=\frac{D}{2}$, the correlator (\ref{odd.8}) at $\frac{D}{2}{+}1$ points leads to
the BCJ numerators
\begin{align}
-iN\oneloop_{\rm odd}(+,1,2,\ldots,\tfrac{D}{2}{+}1,-)  &= \Big\{(\ell \cdot \ep_2)  \varep_D(\ell,\ep_{1},f_3,\ldots,f_{D/2+1}) {+} (2\leftrightarrow3,4,\ldots,\tfrac{D}{2}{+}1)\Big\} \notag\\
&\quad - \frac{1}{2} \Big\{  \varep_D(\ell,\ep_{12},f_3,\ldots,f_{D/2+1}) {+} (2\leftrightarrow3,4,\ldots,\tfrac{D}{2}{+}1)\Big\}
\label{eq5.91} \\
&\quad - \frac{1}{2} \Big\{  \varep_D(\ell,\ep_{1},f_{23},\ldots,f_{D/2+1}) {+} (2,3|2,3,\ldots,\tfrac{D}{2}{+}1)\Big\}  \, .\notag
\end{align}
Similar expressions are expected at higher multiplicity, where (\ref{eq2.8new}) manifests that no more
than $n{+}1{-}\frac{D}{2}$ powers of $\ell$ can occur in $n$-point numerators. In $D=4$, this leads to a power 
counting of $\ell^{n-3}$ in $N\oneloop_{\rm odd}$ that exceeds the $\ell^{n-2}$ in the parity-even part
of numerators of chiral ${\cal N}=1$ SYM inferred from (\ref{eq4.fourd}).

%%%%%%%%%%%%%%%%%%%%%%%%%%%%%%%%%%%%%%%%%%%%%%%% 
%%%%%%%%%%%%%%%%%%%%%%%%%%%%%%%%%%%%%%%%%%%%%%%% 
\section{Summary and outlook}
\label{summary}

In this work, we have constructed streamlined representations of
one-loop correlators in various gauge theories by taking forward
limits of tree-level correlators. Our results are driven by new
representations of two-fermion correlators at tree level which closely
resemble their bosonic counterparts. The combination of their forward
limits therefore manifests all supersymmetry cancellations, and the
power counting of loop momenta follows from representation-theoretic
identities between Lorentz traces over vector and spinor indices.

Our results apply to gauge-theory correlators in arbitrary even
dimensions and with any combination of adjoint scalars, fermions and
gauge bosons running in the loop. Also in the non-supersymmetric case,
we expand the correlators in terms of Parke--Taylor factors in a
subset of the external legs accompanied by Pfaffians. It is then
straightforward to extract BCJ numerators \wrt linearized propagators
by rearranging the Parke--Taylor factors according to well-established
tree-level techniques in the ${\rm YM}{+}\phi^3$ theory.

A variety of interesting follow-up questions is left for the future,
for instance:
\begin{itemize}
\item The strategy of this work calls for an application to higher-loop 
correlators, starting from the two-loop case on a bi-nodal Riemann sphere 
  \cite{Geyer:2016wjx, Geyer:2018xwu}. It remains to identify suitable
  representations of tree-level correlators to perform multiple
  forward limits, and the four-fermion correlator in appendix
  \ref{app:4ferm} could be a convenient starting point. The gluing
  operators of \cite{Roehrig:2017gbt} and the discussion of
  double-forward limits in \cite{Geyer:2019hnn} will give crucial
  guidance in this endeavor.
\item The Parke--Taylor decompositions of the one-loop correlators in
  this work lead to BCJ numerators \wrt linearized propagator in the
  loop momenta. Their algorithmic recombination to quadratic
  propagators is still an open problem (see \cite{Gomez:2016cqb, Gomez:2017lhy, Gomez:2017cpe, Ahmadiniaz:2018nvr, Agerskov:2019ryp}
  for recent progress in this direction) and has not yet been understood at the
  level of the $(n{+}2)$-point tree-level building blocks.  We hope
  that our representations of BCJ numerators in general gauge theories
  provide helpful case studies to (i) pinpoint the key mechanisms in
  the conversion to quadratic propagators (ii) offer a way to preserve
  the BCJ duality in this process.
\item The description of our one-loop BCJ numerators in terms of
  multiparticle fields has not yet been generalized to arbitrary
  multiplicity. Even though the Berends--Giele currents for tree-level
  subdiagrams in BCJ gauge are available to all multiplicity
  \cite{Lee:2015upy, Bridges:2019siz}, their composition rules in
  one-loop numerators involve additional structures. An
  all-multiplicity construction of one-loop BCJ numerators from
  multiparticle fields is likely to shed new light on the
  long-standing questions concerning a kinematic
  algebra.
\end{itemize}

%%%%%%%%%%%%%%%%%%%%%%%%%%%%%%%%%%%%%% 
%%%%%%%%%%%%%%%%%%%%%%%%%%%%%%%%%%%%%% 
%%%%%%%%%%%%%%%%%%%%%%%%%%%%%%%%%%%%%% 
%%%%%%%%%%%%%%%%%%%%%%%%%%%%%%%%%%%%%%

\acknowledgments

We are grateful to Henrik Johansson, Carlos Mafra, Lionel Mason, Gustav Mogull, Ricardo Monteiro and Yong Zhang for combinations of inspiring discussions and collaboration on related topics. AE is supported by the Knut and Alice Wallenberg Foundation under 
KAW 2018.0116, {\it From Scattering Amplitudes to Gravitational Waves.}  SH is supported 
in part by NSF of China under Grant No.\ 11947302 and 11935013. 
OS is supported by the European Research Council under ERC-STG-804286 UNISCAMP.
FT is supported in part by the Knut and
Alice Wallenberg Foundation under grant KAW 2013.0235, and the Ragnar S{\"o}derberg
Foundation (Swedish Foundations’ Starting Grant). 

%%%%%%%%%%%%%%%%%%%%%%%%%%%%%%%%%%%%%% 
%%%%%%%%%%%%%%%%%%%%%%%%%%%%%%%%%%%%%% 
%%%%%%%%%%%%%%%%%%%%%%%%%%%%%%%%%%%%%% 
%%%%%%%%%%%%%%%%%%%%%%%%%%%%%%%%%%%%%% 

\appendix

%%%%%%%%%%%%%%%%%%%%%%%%%%%%%%%%%%%%%% 
%%%%%%%%%%%%%%%%%%%%%%%%%%%%%%%%%%%%%% 
%%%%%%%%%%%%%%%%%%%%%%%%%%%%%%%%%%%%%% 
%%%%%%%%%%%%%%%%%%%%%%%%%%%%%%%%%%%%%% 

\section{One-loop integrands with linear propagators and CHY formulas}
\label{app:linprop}

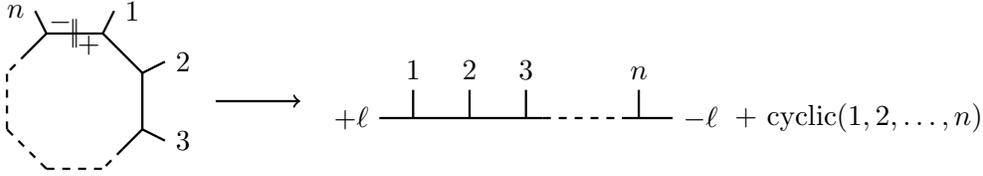
\begin{figure}
\begin{tikzpicture} [scale=0.75, line width=0.30mm]
\begin{scope}[xshift=-0.8cm]
\draw (0.5,0)--(-0.5,0);
\draw (-0.5,0)--(-0.85,-0.35);
\draw [dashed](-0.85,-0.35)--(-1.2,-0.7);
\draw (0.5,0)--(1.2,-0.7);
\draw[dashed] (-1.2,-1.7)--(-1.2,-0.7);
\draw (1.2,-1.7)--(1.2,-0.7);
\draw (1.2,-1.7)--(0.85,-2.05);
\draw[dashed] (0.85,-2.05)--(0.5,-2.4);
\draw[dashed] (-0.5,-2.4)--(0.5,-2.4);
\draw[dashed] (-0.5,-2.4)--(-1.2,-1.7);
\draw (-0.5,0)--(-0.7,0.4)node[left]{$n$};
\draw (0.5,0)--(0.7,0.4)node[right]{$1$};
\draw (1.2,-0.7)--(1.6,-0.5)node[right]{$2$};
\draw (1.2,-1.7)--(1.6,-1.9)node[right]{$3$};
\draw (0,0) node{$| \! |$};
\draw (-0.25,0.2)node{$-$};
\draw (0.25,-0.2)node{$+$};
\end{scope}
%%%%%
%\draw[->] (2,0.5) .. controls (5.5,0.5) .. (5.5,-0.1);
%%%%%
\draw[-> ](1.7,-1.2)  -- (3.2,-1.2);
%%%%%
\begin{scope}[xshift=1.7cm, yshift=0.5cm]
\draw(11.4,-2)node{$+ \ {\rm cyclic}(1,2,\ldots,n)$};
\draw (2.9,-2)node[left]{$+\ell$} -- (5.8,-2);
\draw (7.2,-2) -- (8.1,-2)node[right]{$-\ell$};
\draw (3.5,-2) -- (3.5,-1.5)node[above]{$1$};
\draw (4.5,-2) -- (4.5,-1.5)node[above]{$2$};
\draw (5.5,-2) -- (5.5,-1.5)node[above]{$3$};
\draw[dashed] (5.8,-2) -- (7.2,-2);
\draw (7.5,-2) -- (7.5,-1.5)node[above]{$n$};
%
%\draw(3.1,-2)node{$>$}node[above]{$\ell$};
%\draw(7.9,-2)node{$<$}node[above]{$-\ell$};
\end{scope}
\end{tikzpicture}
\caption{Interpretation of the partial-fraction representation of loop integrals as $(n{+}2)$-point tree-level diagrams.\label{figcutting}}
\end{figure}

Throughout the paper, we adopt a non-standard representation for Feynman integrals of one-loop amplitudes, 
which naturally arises in one-loop CHY formulas or forward limits of tree amplitudes. Repeated partial-fraction
manipulations of the standard Feynman propagators $( \ell+K)^2$ in 
one-loop integrals (with some linear combination $K$ of external momenta)
allow to eliminate the reference to $\ell^2$ from all propagators except for
one~\cite{Geyer:2015bja, Geyer:2015jch}. It suffices to show the result for the 
massless $n$-gon ($s_{12\ldots p,\pm \ell} = \sum_{1\leq i<j}^p s_{ij} 
\pm\ell \cdot k_{12\ldots p}$ with $k_{12\ldots p}=\sum_{j=1}^p k_j$),
\begin{align}
&\int {2^{n-1} \ {\rm d}^D \ell \over \ell^2 (\ell{+}k_1)^2 (\ell{+}k_{12})^2 \ldots (\ell {+} k_{12\ldots n-1})^2}
= \sum_{i=0}^{n-1} \int { 2^{n-1} \ {\rm d}^D \ell \over (\ell {+} k_{12\ldots i})^2 } \prod_{j \neq i} {1\over (\ell {+} k_{12\ldots j})^2 - (\ell {+} k_{12\ldots i})^2 }\notag \\
& \ \ \ \ \ = \sum_{i=0}^{n-1} \int { {\rm d}^D \ell \over \ell^2}   \prod_{j=0}^{i-1} {1\over s_{j+1,j+2,\ldots,i,-\ell}} \prod_{j=i+1}^{n-1} {1\over s_{i+1,i+2,\ldots,j,\ell} } \label{ngonPF}\, ,
\end{align}
where we have performed an $i$-dependent shift of the loop momentum
$\ell \rightarrow \ell{-} k_{12\ldots i}$ to uniformly obtain $\ell^2$ as the
only quadratic propagator in the second line. 
As visualized in figure~\ref{figcutting}, each term in the sum over $i$
can be interpreted as one way of opening up the $n$-gon and
is associated with an $(n{+}2)$-point tree diagram involving off-shell 
momenta $ \pm  \ell$  \cite{He:2015yua, Cachazo:2015aol}.
Each of the cubic diagrams can have different kinematic numerators, leaving a total of $n!$ inequivalent $n$-gon numerators.

The manipulations in \eqref{ngonPF} also apply to one-loop integrals with massive
momenta $k_A,k_B,k_C$ and $k_D$ such as $k_A = k_{a_1}+k_{a_2}+\ldots+k_{a_p}$ for $A=a_1a_2\ldots a_p$.
E.g.\ a massive box admits the following four-term representation:
\begin{align}
\int {8 \ \dd^D \ell  \over \ell^2 (\ell{+}k_A)^2 (\ell {+} k_{AB})^2 (\ell {+} k_{ABC})^2}
= \int {\dd^D \ell \over \ell^2} \Big( {1\over s_{A,\ell} s_{AB,\ell} s_{D,-\ell} } + {\rm cyc}(A,B,C,D) \Big)  
\end{align}
These rearrangements uniquely decompose the one-loop integrand for color-ordered single-trace
amplitudes into $n$ terms dubbed {\it partial integrands} \cite{He:2016mzd}, similar to the
decomposition \eqref{ngonPF} of the $n$-gon. Each partial integrand can be interpreted as the
forward limit of a color-ordered $(n{+}2)$-point tree amplitude with off-shell momenta, where
for instance the momenta of the two legs between $n$ and $1$ in figure \ref{figcutting} are identified as
$\ell$ and $- \ell$ \cite{Cachazo:2015aol}. Although it is an open problem to perform loop
integrals over linearized propagators, the above rearrangements of loop integrals have to yield 
the same result as integrating the quadratic propagators.

Such integrals naturally arise from one-loop CHY formulas, which can be obtained by performing forward limits on tree-level CHY formulas, or by localizing the $ \tau$ integral of ambitwistor-string formulas 
at genus one~\cite{Adamo:2013tsa} 
at the cusp $ \tau  \rightarrow i  \infty$, where the torus degenerates to
a nodal sphere~\cite{Geyer:2015bja, Geyer:2015jch}. A general formula for {\it e.g.} 
one-loop amplitudes of gravity and gauge theories in $D$ spacetime dimensions reads
(with the normalization factor $\mathscr{N}$ from the $(n{+}2)$-point tree amplitude (\ref{eq:CHY}))
\begin{align} 
{\cal M}^{(1)}_{L \otimes R}&= 
\mathscr{N} \int {\dd^D \ell \over \ell^2} \int  \frac{\dd \sigma_+ \dd \sigma_- \dd^n\sigma_i}{\text{Vol}[\text{SL}(2,\mathbb{C})]}\, \prod_{i=+,-,1}^n{} \! \! \! \! ' \ \ \delta ( E_i)%{ (\ell \cdot k_i) \over \sigma_i} + \sum_{j=1 \atop{j \neq i}}^n {k_i \cdot k_j \over \sigma_{ij}} 
 \, {\cal I}^{(1)}_L(\ell) \, {\cal I}^{(1)}_R(\ell) \ , \label{loopSUGRA}
\end{align}
where $E_i$ is the $i$-th tree-level scattering equation of $(n{+}2)$ points and we take forward limit by $k_\pm \to \pm \ell$. As indicated by the prime, three of the equations are redundant due to the ${\rm SL}(2,\mathbb C)$ symmetry. 
For gauge theories, one of the two half integrands ${ \cal I}^{(1)}_L( \ell)
 \rightarrow  { \cal I}^{(1)}_{U(N)}$
 \begin{align}\label{fixgaugeint}
{\cal I}^{(1)}_{U(N)}= \sum_{\rho \in S_{n-1}} {\rm Tr} (t^{a_1} t^{a_{\rho(2)}} t^{a_{\rho(3)}}  \ldots t^{a_{\rho(n)}}) \, \PT^{(1)}(1,\rho(2,3,\ldots,n)) \ ,
\end{align}
is a sum of color traces accompanied by one-loop analogues $\PT^{(1)}$ of Parke--Taylor factors,
\begin{align}\label{loopPT}
\PT^{(1)}(1,2,\ldots,n) \equiv  {1\over \sigma_{-,1} \sigma_{1, 2} \sigma_{2, 3} \ldots \sigma_{n-1,n} \sigma_{n,+}\sigma_{+,-}} + {\rm cyc}(1,2,\ldots,n) \ .
\end{align}
Throughout the paper we are interested in the other, polarization-dependent half-integrand ${\cal I}^{(1)}_R \to {\cal I}^{(1)}_{\rm SYM} (\ell)$ with our main results in (\ref{eq3.22}) and (\ref{eq4.6}). By expanding it as a linear 
combination of $(n{+}2$)-point Parke-Taylor factors, the coefficients become BCJ master numerators for one of the $n$ terms in the decomposition of one-loop $n$-gon in \eqref{ngonPF}. In terms of the two
half-integrands, one-loop amplitudes \eqref{loopSUGRA}  in gauge theory and (super-)gravity are obtained as ${ \cal
M}^{(1)}_{U(N)  \otimes { \rm SYM}}$ and ${ \cal M}^{(1)}_{{ \rm SYM}  \otimes { \rm SYM}}$, respectively.

Note that these integrals with linearized propagators not only naturally appear from CHY formulas, 
but also enter the Q-cut representation of loop amplitudes~\cite{Baadsgaard:2015twa}.
Such representations provide a well-defined notion of ``loop integrands" for non-planar
diagrams and generic theories\footnote{Also see \cite{Tourkine:2019ukp} for the emergence of global 
loop integrands from the field-theory limit of string amplitudes.} and offer valuable perspectives
on the structure of loop amplitudes. 
It also allows one to generalize KLT and BCJ relations to one loop~\cite{He:2016mzd,He:2017spx}.
 
\section{Conformal field theory and tree-level correlators}
\label{bigapp:CFT}

%%%%%%%%%%%%%%%%%%%%%%%%%%%%%%%%%%%%%% 
%%%%%%%%%%%%%%%%%%%%%%%%%%%%%%%%%%%%%% 
%%%%%%%%%%%%%%%%%%%%%%%%%%%%%%%%%%%%%% 
%%%%%%%%%%%%%%%%%%%%%%%%%%%%%%%%%%%%%% 

\subsection{CFT basics}
\label{app:CFT}

In the worldsheet conformal field theory (CFT) of the RNS formalism in
$D=10$, the free-field OPEs relevant for the correlators of gluon
vertex operators (\ref{eq2.1}) read
\begin{align}
\psi^\mu(\sigma) \psi^\nu(0) &\sim \frac{\eta^{\mu \nu} }{\sigma} \, ,
&e^{p \phi}(\sigma) e^{q \phi}(0) &\sim \sigma^{-pq} e^{(p+q)\phi} \label{OPE1}
\\
%%%
P^\mu(\sigma) X^\nu(0) &\sim -i \frac{ \eta^{\mu \nu} }{\sigma} \, ,
&P^\mu(\sigma) e^{ik\cdot X}(0) &\sim 
 \frac{k^\mu e^{ik\cdot X}}{\sigma} \ . \label{OPE2}
\end{align}
The spin field in the fermion vertex (\ref{eq2.2}) interacts with
worldsheet spinor $\psi^\mu$ via
\begin{equation}
S_\alpha(\sigma) S^\beta(0) \sim \frac{ \delta_\alpha^\beta }{\sigma^{5/4}} \ , \ \ \ \ \ \
S_\alpha(\sigma) S_\beta(0) \sim \frac{ \gamma^\mu_{\alpha \beta} \psi_\mu(0)}{\sqrt{2}\sigma^{3/4}} \ .
\label{OPEferm}
\end{equation}
As a result of the OPEs, we have two- and three-point correlation
functions ($\sigma_{ij} = \sigma_i{-}\sigma_j$) 
\begin{align}
\langle S_\alpha e^{-\phi/2}(\sigma_1) S^\beta e^{-3\phi/2}(\sigma_2) \rangle &=
\frac{ \delta_\alpha^\beta  }{\sigma_{12}^2}
\label{CFTx} 
\\
\langle S_\alpha e^{-\phi/2}(\sigma_1) S_\beta e^{-\phi/2}(\sigma_2) \psi^\mu e^{-\phi}(\sigma_3) \rangle &=
\frac{ \gamma^\mu_{\alpha \beta}}{ \sqrt{2}\sigma_{12} \sigma_{13} \sigma_{23} } \, ,
\label{CFTa} 
\end{align}
see \cite{Hartl:2010ks} for higher-point spin-field correlators in
various dimensions.

The conformal fields $\psi^\mu$ and $S_\alpha$ are primary fields of a
Kac-Moody current algebra at level $k=1$ with generators
$\psi^\mu \psi^\nu$. By Kac-Moody Ward identities, current insertions
in a correlator can be removed by summing over all OPE singularities
such as \cite{Friedan:1985ge, Cohn:1986bn, Kostelecky:1986xg}
\begin{align}
\psi^\mu \psi^\nu(\sigma) \psi^\lambda(0) &\sim \frac{2 \psi^{[\mu} \eta^{\nu]\lambda}(0)}{\sigma} \ , \ \ \ \ \ \ \psi^\mu \psi^\nu(\sigma) S_\alpha(0) \sim -\frac{\gamma^{\mu \nu}{}_\alpha{}^\beta S_\beta(0)}{2\sigma} 
\label{CFTc}
\\
\psi^\mu \psi^\nu (\sigma) \psi^\lambda \psi^\rho(0) &\sim \frac{ 2 \eta^{\lambda[\nu} \eta^{\mu]\rho} }{\sigma^2}+
 \frac{ 2 (\eta^{\lambda[\nu} \psi^{\mu]} \psi^\rho - \eta^{\rho[\nu} \psi^{\mu]} \psi^\lambda )(0)}{\sigma} \, ,
 \label{CFTd}
\end{align} 
with the normalization conventions
$2 \eta^{\lambda[\nu} \eta^{\mu]\rho}= \eta^{\lambda \nu} \eta^{\mu
  \rho} - \eta^{\lambda \mu} \eta^{\nu \rho}$ for antisymmetrization
brackets.  Hence, current-algebra techniques can be used to
straightforwardly compute spin-field correlators with any number of
$ \psi_{\mu}\psi_\nu$ insertions. In this way, the contributions
$\sim f^{\mu \nu} \psi_{\mu}\psi_\nu(\sigma)$ of bosonic vertex
operator (\ref{eq2.1}) in the zero picture can be addressed in
presence of spin fields.

%%%%%%%%%%%%%%%%%%%%%%%%%%%%%%
%%%%%%%%%%%%%%%%%%%%%%%%%%%%%%
%%%%%%%%%%%%%%%%%%%%%%%%%%%%%%

\subsection{Bosonic correlators and the Pfaffian}
\label{revpsi}

When the bosonic correlator (\ref{eq2.5}) is evaluated as the reduced
Pfaffian in (\ref{eq2.6}), the antisymmetric $2n\times 2n$ matrix
$\Psi_{\{12\ldots n\}}$ is organized into $n\times n$ blocks $\mathsf{A},\mathsf{B}$ and
$\mathsf{C}$ \cite{Cachazo:2013iea}
\beq\label{psimatrix}
\Psi = \left(
         \begin{array}{cc}
           \mathsf{A} &  -\mathsf{C}^T \\
           \mathsf{C} & \mathsf{B} \\
         \end{array}
       \right)_{2n\times 2n} 
\eeq
with $\mathsf{C}^T$ denoting the transpose of $\mathsf{C}$.
The entries of the $n \times n$ matrices $\mathsf{A}$, $\mathsf{B}$, $\mathsf{C}$ are given by\footnote{The $\mathsf{C}$ matrix defined here differs from that in the original CHY paper~\cite{Cachazo:2013hca} and appendix A of~\cite{AlexFei}. In particular, $\mathsf{C}_{\text{here}}=-\mathsf{C}_{\text{there}}$ such that the expansions~\eqref{eq2.7}, \eqref{eq2.8} and~\eqref{eq2.11} do not involve alternating signs, {\it cf.} the corresponding equations in~\cite{AlexFei}.}
\beq\label{blocks}
\mathsf{A}_{ij} = \begin{cases} \displaystyle \frac{k_i\cdot k_j}{\sigma_{ij}} & i\neq j\, ,\\
\displaystyle \quad ~~ 0 & i=j \, ,\end{cases} \qquad \mathsf{B}_{ij} = \begin{cases} \displaystyle \frac{\ep_i\cdot \ep_j}{\sigma_{i j}} & i\neq j\,,\\
\displaystyle \quad ~~ 0 & i=j \, ,\end{cases} \qquad \mathsf{C}_{ij} = \begin{cases} \displaystyle -\frac{\ep_i \cdot k_j}{\sigma_{i j}} \Bigg. &\quad i\neq j \, ,\\
\displaystyle \sum_{p=1 \atop{p\neq i}}^n \frac{\ep_i \cdot k_p}{\sigma_{ip}} & \quad i=j\, .\end{cases}
\eeq
We define the Pfaffian of a $2n\times 2n$ anti-symmetric matrix as
\begin{align}
\Pf \Psi=(-1)^{\frac{n(n+1)}{2}}\frac{1}{2^n n!}\sum_{\rho\in S_{2n}}\text{sign}(\rho)\prod_{i=1}^{n}\Psi_{\rho(2i-1)\rho(2i)}\,.  
\end{align}
As a consequence of momentum conservation and scattering equations,
the matrix $\Psi$ has two null vectors such that $\Pf \Psi=0$. The
reduced Pfaffian in (\ref{eq2.6}), by contrast, yields a non-vanishing
bosonic correlator on the support of momentum conservation and
scattering equations.

The diagonal terms of the $\mathsf{C}$-matrix in (\ref{blocks}) arise when the
first term $\sim \ep^\mu P_\mu(\sigma)$ in $V^{(0)}$ contracts the
plane waves of the remaining vertex operators, 
\beq
\Big \langle \ep_j^\mu P_\mu(\sigma_j)  \prod_{i=1}^n e^{i k_i \cdot X(\sigma_i) } \Big \rangle
= \mathsf{C}_{jj} \Big \langle \prod_{i=1}^n e^{i k_i \cdot X(\sigma_i) } \Big \rangle
\, , \ \ \ \ \ \
\mathsf{C}_{jj} = \sum^n_{i=1 \atop {i\neq j}} \frac{ \epsilon_j \cdot k_i }{\sigma_{ji} }\, ,
\label{Pcorr}
\eeq
see (\ref{OPE2}) for the underlying OPEs. Accordingly, when multiple
$V^{(0)}$ contribute through the conformal field $P_\mu$, the
plane-wave correlators relevant to any number of bosons and fermions
evaluate to
\beq
\Big \langle \Big(  \prod_{j=1}^m \ep_j^\mu P_\mu(\sigma_j)  \Big) \prod_{i=1}^n e^{i k_i \cdot X(\sigma_i) } \Big \rangle
= \Big(  \prod_{j=1}^m \mathsf{C}_{jj} \Big) \Big \langle \prod_{i=1}^n e^{i k_i \cdot X(\sigma_i) } \Big \rangle  \, .
\label{Pcorr2}
\eeq
This is the CFT origin of those term in the correlators (\ref{eq2.7}),
(\ref{eq2.8}) and (\ref{eq2.11}), where the Pfaffian $\Pf \Psi_A$
contributes via products of the $\mathsf{C}_{jj}$ for all the labels in the set
$A$ \cite{Mason:2013sva}. The admixtures of the $\mathsf{A}$- and $\mathsf{B}$-blocks in (\ref{blocks}) as
well as the non-diagonal $\mathsf{C}_{ij}$ at $i\neq j$ will be discussed in
the next subsections.

%%%%%%%%%%%%%%%%%%%%%%%%%%%%%%
%%%%%%%%%%%%%%%%%%%%%%%%%%%%%%
%%%%%%%%%%%%%%%%%%%%%%%%%%%%%%

\subsection{Two-fermion correlators}
\label{app:2ferm}

For the first representation (\ref{eq2.8}) of the fermionic vertex
operator, the three-point example spelt out in (\ref{eq2.ex}) is an
immediate consequence of the spin-field correlator (\ref{CFTa}). We
shall now derive the contributions from the additional insertions of
$V^{(0)}_j(\sigma_j)$ at $n \geq 4$ points from the recursive
techniques outlined above.

At four points, the first term
$V^{(0)}_2(\sigma_2) \rightarrow \ep_2 \cdot P(\sigma_2)$ can be
straightforwardly addressed via (\ref{Pcorr}) and yields
$\mathsf{C}_{22}=\Psi_{\{2\} } $. Together with the accompanying spin-field
correlator (\ref{CFTa}), we obtain
$\Psi_{\{2\} } {\cal I}^\tree_{\rm 2f}(1_f,3_f, \ferm{4})$ and reproduce
the first term in (\ref{eq2.ex1}).  The second term
$V^{(0)}_2(\sigma_2) \rightarrow \frac{1}{2} f^{\mu \nu}_2 \psi_\mu
\psi_\nu(\sigma_2)$ requires the summation of the OPEs (\ref{CFTc})
according to the Kac--Moody Ward identity
\begin{align}
\langle &S_\alpha(z_1) \psi^\mu \psi^\nu(z_2) S_\beta(z_3) \psi^\la(z_4) \rangle = - \frac{\gamma^{\mu \nu}{}_\alpha{}^\delta}{2 \sigma_{21}} \langle S_\delta(z_1)S_\beta(z_3) \psi^\la(z_4) \rangle 
\notag\\
&\ \ \ \ - \frac{\gamma^{\mu \nu}{}_\beta{}^\delta}{2 \sigma_{23}} \langle S_\alpha(z_1)S_\delta(z_3) \psi^\la(z_4) \rangle
+ \frac{ 2 }{\sigma_{24}} \langle S_\alpha(z_1)S_\beta(z_3) \psi^{[\mu}(z_4) \rangle \eta^{\nu] \la } \label{simpcorr} 
\\
&= \frac{1}{2\sqrt{2} \sigma^{3/4}_{13} \sigma^{1/2}_{14} \sigma^{1/2}_{34} } \Big\{
\Big( \frac{1}{\sigma_{24}} - \frac{1}{\sigma_{21}} \Big) (\gamma^{\mu \nu} \gamma^\la)_{\alpha \beta}
+\Big( \frac{1}{\sigma_{24}} - \frac{1}{\sigma_{23}} \Big) (\gamma^{\mu \nu} \gamma^\la)_{\beta \alpha}
\Big\} \, . \notag
\end{align}
In passing to the last line, we have inserted the three-point
correlator (\ref{CFTa}) and used the gamma-matrix identity 
$\gamma^\mu_{\alpha\beta} \eta^{\nu \la} - \gamma^{\nu}_{\alpha \beta} \eta^{\mu \la} = \frac{1}{2}
(\gamma^{\mu \nu} \gamma^\la)_{\alpha \beta} + \frac{1}{2}
(\gamma^{\mu \nu} \gamma^\la)_{ \beta\alpha} $. Upon contraction with
$\frac{1}{2} \chi_1^\alpha f_2^{\mu \nu} \chi_3^\beta \ep_4^\la$ and
dressing with the superghost correlator, this reproduces the last two
terms in (\ref{eq2.ex1}) and completes the derivation of the
four-point correlator
\begin{align}
{\cal I}^\tree_{\rm 2f}(1_f, 2,3_f, \ferm{4})
&= \Pf \Psi_{\{2\}} \frac{(\chi_1 \slash \! \! \! \ep_4 \chi_3)}{\sigma_{14} \sigma_{43} \sigma_{31}}+\frac{ (\chi_1 \slash \! \! \! f_2 \slash \! \! \! \ep_4 \chi_3)  }{ \sigma_{12} \sigma_{24} \sigma_{43} \sigma_{31}}
+\frac{ (\chi_1  \slash \! \! \! \ep_4 \slash \! \! \! f_2 \chi_3)  }{ \sigma_{14} \sigma_{42} \sigma_{23} \sigma_{31}}  \, .
\end{align}
Starting from five points, we encounter double-insertions of the
current $ \psi_\mu \psi_\nu$, and Kac--Moody Ward identities involve
the double-pole in their OPE (\ref{CFTd}). These double-poles are
attributed to the Pfaffians in (\ref{eq2.7}), (\ref{eq2.8}) \&
(\ref{eq2.11}) and yield the entries of the $\mathsf{A}$- and $\mathsf{B}$-blocks in
$\Pf \Psi_{ \{12\ldots m\} } = \prod_{j=1}^m \mathsf{C}_{jj} + {\cal
  O}(\mathsf{A}_{ij},\mathsf{B}_{ij})$. More specifically, the expression (\ref{eq2.8})
for the five-point correlator involves the two-particle Pfaffian
\begin{align}
\Pf \Psi_{\{2,3\}} = \mathsf{C}_{22} \mathsf{C}_{33} - \mathsf{C}_{23} \mathsf{C}_{32} - \mathsf{A}_{23} \mathsf{B}_{23} = \mathsf{C}_{22} \mathsf{C}_{33} + \frac{ (\ep_2 {\cdot} k_3)(\ep_3{\cdot} k_2) - (k_2{\cdot} k_3)(\ep_2 {\cdot} \ep_3) }{\sigma_{23}^2} \, ,
\label{pf23a}
\end{align}
where the last two terms arise from the double-pole terms in the Ward identity
\begin{align}
\langle S_\alpha(z_1) \psi^\mu \psi^\nu(z_2) \psi^\la \psi^\rho(z_3) S_\beta(z_4) \psi^\tau(z_5) \rangle^{\te{double}}_{\te{poles}}
= \frac{2 \eta^{\la [ \nu } \eta^{\mu] \rho} }{\sigma_{23}^2} \langle S_\alpha(z_1)  
S_\beta(z_4) \psi^\tau(z_5) \rangle\, .
\label{pf23b}
\end{align}
The simple-pole terms in turn are given by
\begin{align}
\langle &S_\alpha(z_1) \psi^\mu \psi^\nu(z_2) \psi^\la \psi^\rho(z_3) S_\beta(z_4) \psi^\tau(z_5) \rangle^{\te{simple}}_{\te{poles}} =
 - \frac{\gamma^{\mu \nu}{}_\alpha{}^\delta}{2 \sigma_{21}}\langle S_\delta(z_1) \psi^\la \psi^\rho(z_3) S_\beta(z_4) \psi^\tau(z_5) \rangle \notag
 \\
& \!
 - \frac{\gamma^{\mu \nu}{}_\beta{}^\delta}{2 \sigma_{24}}\langle S_\alpha(z_1) \psi^\la \psi^\rho(z_3) S_\delta(z_4) \psi^\tau(z_5) \rangle
+\frac{2}{\sigma_{23}} \Big( \eta^{\la [\nu} \langle S_\alpha(z_1)  \psi^{\mu]}\psi^\rho(z_3) S_\beta(z_4) \psi^\tau(z_5) \rangle - (\lambda {\leftrightarrow} \rho) \Big)\notag \\
& \! + \frac{2}{\sigma_{25}} \langle S_\alpha(z_1)  \psi^\la \psi^\rho(z_3) S_\beta(z_4) \psi^{[ \mu}(z_5 ) \rangle \eta^{\nu] \tau}
\end{align}
and recurse to the simpler correlators we have already evaluated in
(\ref{simpcorr}). Once the permutations of
$\eta^{\la \nu} \gamma^{\mu \rho}$ are rewritten as
a commutator $[\gamma^{\mu \nu},\gamma^{\la \rho}]$, we arrive~at
\begin{align}
\langle S_\alpha(z_1)& \psi^\mu \psi^\nu(z_2) \psi^\la \psi^\rho(z_3) S_\beta(z_4) \psi^\tau(z_5) \rangle^{\te{simple}}_{\te{poles}} = \frac{1}{4 \sqrt{2} z_{14}^{3/4} z_{15}^{1/2} z_{45}^{1/2} } \\
\times \Big\{ & \frac{ z_{15} }{z_{12} z_{23} z_{35} } ( \gamma^{\mu \nu } \gamma^{\la \rho} \ga^\tau)_{\alpha \beta} 
+ \frac{ z_{15} z_{45} }{z_{12} z_{25} z_{34} z_{35} } (  \gamma^{\mu \nu }\ga^\tau  \gamma^{\la \rho} )_{\alpha \beta} 
+ \frac{ z_{45} }{z_{42} z_{23} z_{35} } ( \ga^\tau \gamma^{\la \rho}  \gamma^{\mu \nu })_{\alpha \beta}  \notag \\
%%%
%%%
+\,&\frac{ z_{15} }{z_{13} z_{32} z_{25} } (  \gamma^{\la \rho}  \gamma^{\mu \nu }\ga^\tau)_{\alpha \beta}  
+ \frac{ z_{15} z_{45} }{z_{13} z_{35} z_{24} z_{25} } (  \gamma^{\la \rho} \ga^\tau   \gamma^{\mu \nu })_{\alpha \beta} 
+ \frac{ z_{45} }{z_{43} z_{32} z_{25} } ( \ga^\tau  \gamma^{\mu \nu }  \gamma^{\la \rho} )_{\alpha \beta}\Big\} \notag
\end{align}
after partial-fraction manipulations of the form
$(\sigma_{12} \sigma_{23})^{-1} + {\rm cyc}(1,2,3) = 0$.  Upon
contraction with
$\frac{1}{4} \chi_1^\alpha f_2^{\mu \nu} f_3^{\la \rho} \chi_4^\beta
\ep_5^\tau$ and dressing with the superghost correlator, this
reproduces the $A= \emptyset$ contribution to (\ref{eq2.8}). Terms
with $A= \{2\}$ and $A= \{3\}$ are easily checked by combining
(\ref{Pcorr}) with (\ref{simpcorr}), and the Pfaffian associated with
$A=\{2,3\}$ has been determined in (\ref{pf23a}) and (\ref{pf23b}).
This completes the derivation of the five-point correlator.

The detailed five-point calculation exemplifies the CFT origin of the
gamma-matrix products in the $n$-point correlator (\ref{eq2.8}): They
arise from the simple poles in the OPEs (\ref{CFTc}) and (\ref{CFTd})
that govern the recursive evaluation of spin-field correlators with an
arbitrary number of Lorentz-current insertions.  The latter capture
the contributions
$V^{(0)}_j(\sigma_j) \rightarrow \frac{1}{2} f^{\mu \nu}_j \psi_\mu
\psi_\nu(\sigma_j)$ which are converted to
$ \frac{1}{4} f^{\mu \nu}_j (\gamma_{\mu \nu})_\alpha{}^\beta
S_\beta(\sigma_1)/\sigma_{1j}$ when performing the OPE (\ref{CFTc})
with a spin field. The double-poles in the OPE (\ref{CFTd}) among
Lorentz currents in turn promote the contributions
$\prod_{j \in A} \mathsf{C}_{jj}$ from
$V^{(0)}_j(\sigma_j) \rightarrow\epsilon_j^\mu P_\mu(\sigma_j)$ to
$\Pf \Psi_A$ along the lines of (\ref{pf23a}).

Iterating these OPEs leads to products of gamma matrices, where the
multiplication order is correlated with the labels of the accompanying
$\sigma_{ij}^{-1}$. Partial-fraction manipulations and the commutators
of $\gamma^{\mu \nu}$ can be used to arrive at the same number of
gamma matrices and at a chain-structure
$(\ldots \sigma_{ij}\sigma_{jk}\sigma_{kl}\ldots)^{-1}$ in each
term. By analyzing the combinatorics of this algorithm and keeping in
mind that the correlator does not depend on the order in which the
$\psi^\mu \psi^\nu$ are eliminated via Ward identities, one arrives at
the $n$-point expression in (\ref{eq2.8}). The same logic has been used in deriving the
$n$-point tree-level correlator in the pure-spinor formalism
\cite{Mafra:2011nv}, where the double-pole contributions have been
absorbed to redefine the kinematic factors of the simple poles and to
eventually obtain multiparticle superfields.

The same way of applying Kac--Moody Ward identities gives rise to the
alternative form (\ref{eq2.11}) of the two-fermion correlator.  For
instance, the three-point correlator in (\ref{eq2.ex2}) follows from
the same use of Ward identities that eliminated a single Lorentz
current in (\ref{simpcorr}). On the one hand, the three-point
correlator involving fermionic ghost pictures $V^{(-1/2)} V^{(-3/2)}$
shares certain intermediate steps with the four-point correlator from
$V^{(-1/2)} V^{(-1/2)}$.  On the other hand, we can give the same kind
of all-multiplicity results (\ref{eq2.8}) and (\ref{eq2.11}) for both
ghost-picture assignments. The discussion in sec.~\ref{sec:loop.1}
illustrates that (\ref{eq2.11}) due to $n{-}2$ insertions of $V^{(0)}$
instead of $n{-}3$ is more suitable to manifest the interplay with the
bosonic correlator (\ref{eq2.7}) upon forward limits.

These techniques to successively remove insertions of
$\psi^\mu \psi^\nu$ from the correlator are universal to the $SO(D)$
Kac-Moody symmetry of the RNS model \cite{Friedan:1985ge, Cohn:1986bn,
  Kostelecky:1986xg} in any number of spacetime dimensions $D$. Since
the Clifford algebra (\ref{eq2.9}) also takes the same form in any
number of dimensions, the structure of the gamma-matrix product in the
two-fermion correlators (\ref{eq2.8}) and (\ref{eq2.11}) is universal
to any even value of $D$. The only $D$-dependent aspect of these
correlators is the relative chirality of the fermion wavefunctions
$\chi_j$ which can be understood from the three-point correlator (\ref{CFTa}) for
lower-dimensional spin fields that initiates the recursion based on
Ward identities. The $D$-dimensional three-point correlator
is nonzero in case of alike chiralities in
$D=2 \ \textrm{mod} \ 4$ and opposite chiralities in
$D=4 \ \textrm{mod} \ 4$, see \eg section 3 of
\cite{Hartl:2010ks}. Hence, the two-fermion correlators in
(\ref{eq2.8}) and (\ref{eq2.11}) can be used in any even $D\leq 10$
provided that one of the chiralities is flipped in
$D=4 \ \textrm{mod} \ 4$.

%%%%%%%%%%%%%%%%%%%%%%%%%%%%%%
%%%%%%%%%%%%%%%%%%%%%%%%%%%%%%
%%%%%%%%%%%%%%%%%%%%%%%%%%%%%%

\subsection{Four-fermion correlators}
\label{app:4ferm}

The recursive computation of two-fermion correlators can be straightforwardly extended to
the four-fermion case. In this case, Ward identities reduce correlators with Lorentz-current insertions
to the basic spin-field correlator 
\begin{align}
\langle S_\alpha e^{-\frac{\phi}{2}}(\sigma_1) S_\beta e^{-\frac{\phi}{2}}(\sigma_2) 
S_\gamma e^{-\frac{\phi}{2}}(\sigma_3) S_\delta e^{-\frac{\phi}{2}}(\sigma_4) \rangle &= \frac{1}{2\sigma_{41} \sigma_{23} }\Big( \frac{ \gamma_{\mu \alpha \beta} \gamma^\mu_{\gamma \delta}}{\sigma_{12} \sigma_{34}} + \frac{ \gamma_{\mu  \alpha \gamma } \gamma^\mu_{\beta \delta} }{\sigma_{13} \sigma_{24}}
\Big) \, .
\label{CFTb}
\end{align}
Note that this result is specific to $D=10$ dimensions, see \cite{Kostelecky:1986xg, Hartl:2010ks} 
for the tensor structure of lower dimensional four-spin-field correlators.
Permutation invariance under exchange of
$(\alpha,\sigma_1) \leftrightarrow (\beta,\sigma_2)$ is obscured on the right-hand side
of (\ref{CFTb}) but can be checked using the gamma matrix
identity $\gamma_{\mu(\alpha \beta} \gamma^\mu_{\gamma)\delta}=0$ in
ten dimensions. It can be manifested by rewriting the correlator as a
reduced determinant with entries
$\gamma^\mu_{\alpha \beta}/\sigma_{12}$.

As an immediate consequence of (\ref{CFTb}), the four-fermion correlator is given by
\begin{align}
 \langle V^{(-\frac 1 2)}_1(\sigma_1) V^{(-\frac 1 2)}_2(\sigma_2) V^{(-\frac 1 2)}_3(\sigma_3) V^{(-\frac 1 2)}_4(\sigma_4)\rangle  
&=\frac{(\chi_1 \gamma^\mu \chi_2) (\chi_3 \gamma_\mu \chi_4)}{4 \sigma_{12} \sigma_{23} \sigma_{34} \sigma_{41} }+\frac{(\chi_1 \gamma^\mu \chi_3) (\chi_2 \gamma_\mu \chi_4)}{4\sigma_{13} \sigma_{32} \sigma_{24} \sigma_{41}}\,.
\label{E40}
\end{align}
Additional bosonic vertex operators yield the same contributions of $\mathsf{C}_{jj}$ from 
$V^{(0)}_j(\sigma_j) \rightarrow\epsilon_j^\mu P_\mu(\sigma_j)$ and the same $\oslashed{f}_j$ contractions 
from $V^{(0)}_j(\sigma_j) \rightarrow  \frac{1}{2} f^{\mu \nu}_j \psi_\mu \psi_\nu(\sigma_j)$
as detailed in the two-fermion case. For instance, the five-point correlator is obtained in the following form
\begin{align}
&\langle V^{(-\frac 1 2)}_1(\sigma_1) V^{(-\frac 1 2)}_2(\sigma_2) V^{(-\frac 1 2)}_3(\sigma_3) V^{(-\frac 1 2)}_4(\sigma_4) V^{(0)}_5(\sigma_5) \rangle 
\label{E41}  \\
 %%%
&  = \frac{1}{  4 \sigma_{12} \sigma_{23} \sigma_{34} \sigma_{41}  }  \Big[  \Pf \Psi_{\{ 5\} } (\chi_1 \gamma^\mu \chi_2) (\chi_3 \gamma_\mu \chi_4)
+  \frac{(\chi_1 \slash \! \! \! f_{5} \gamma^\mu \chi_2)(\chi_3 \gamma_\mu \chi_4)}{\sigma_{15}  }+\frac{ (\chi_1  \gamma^\mu \slash \! \! \! f_{5} \chi_2)(\chi_3 \gamma_\mu \chi_4) }{\sigma_{52}   }  \notag \\
&\ \ \ \ \ \ \ \ \ \ \ \ \ \ \ \ \ \ \ \ \ \ \ \ +\frac{ (\chi_1  \gamma^\mu \chi_2)(\chi_3 \slash \! \! \! f_{5} \gamma_\mu \chi_4) }{\sigma_{35}  }{+}\frac{ (\chi_1  \gamma^\mu  \chi_2)(\chi_3 \gamma_\mu \slash \! \! \! f_{5} \chi_4) }{\sigma_{54} } \Big] + (2\leftrightarrow 3) \,.\notag
\end{align}
Note that the exchange of $2$ and $3$ acts on both the $\chi_j$ and on the punctures in the four-point
Parke--Taylor factor as well as the $\sigma_{ij}^{-1}$ inside the square brackets. One may eliminate
one of the field-strength contractions via
\begin{align}
(\chi_1 \slash \! \! \! f_{5} \gamma^\mu \chi_2)(\chi_3 \gamma_\mu \chi_4)
{-} (\chi_1  \gamma^\mu \slash \! \! \! f_{5} \chi_2)(\chi_3 \gamma_\mu \chi_4) 
{+}(\chi_1  \gamma^\mu \chi_2)(\chi_3 \slash \! \! \! f_{5} \gamma_\mu \chi_4) 
{-}(\chi_1  \gamma^\mu  \chi_2)(\chi_3 \gamma_\mu \slash \! \! \! f_{5} \chi_4) = 0
\label{falloff}
\end{align}
to manifest the quadratic falloff as $\sigma_5 \rightarrow \infty$, but we chose to display (\ref{E41})
in the more symmetric form, where the generalization to higher multiplicity is more apparent.
Similar to the two-fermion case, the general formula is then given by a sum over all subsets $A$ of the
bosons $\{5,6,\ldots, n\}$ along with $\Pf \Psi_{A}$. For a fixed choice of $A$, it remains to sum over all 
possibilities to insert gamma-matrix contracted field strengths $\oslashed{f_j}$ of the bosons in the complement of
$A$ adjacent to the four fermion wavefunctions.

To simplify the notation, let us define a ``field-strength-inserted" fermion wave function, ${\cal X}_{i,B_i}$ for a fermion $i$ and a set of bosons $B_i=\{b_1, b_2, \ldots, b_p\}$,
\beq
{\cal X}^\alpha_{i, B_i}\equiv\sum_{\omega\in S_{p}} \frac{(\slash\!\!\! {f}_{\omega(b_1)} \slash\!\!\! {f}_{\omega(b_2)} \cdots \slash\!\!\! {f}_{\omega(b_p)})^\alpha{}_\beta \chi^\beta_i}{\sigma_{\omega(b_1), \omega(b_2)} 
\sigma_{\omega(b_2), \omega(b_3)} \ldots \sigma_{\omega(b_{p{-}1}), \omega(b_p)} \sigma_{\omega(b_p), i}}\,,
\eeq
where we sum over permutations of $B_i$. In \eqref{E41}, we have one of the simplest examples
\beq \label{notexample}
{\cal X}_{i,b}= \frac{ \slash\!\!\! {f}_b \chi_i}{\sigma_{b,i}} =  \frac{ \chi_i\slash\!\!\! {f}_b }{\sigma_{i,b}} \ , \ \ \ \ \ \ {\cal X}_{i, \{b_1, b_2\}}= \frac{\slash\!\!\! {f}_{b_1} \slash\!\!\! {f}_{b_2} \chi_i }{ \sigma_{b_1,b_2} \sigma_{b_2, i}} + (b_1\leftrightarrow b_2)
= \frac{ \chi_i \slash\!\!\! {f}_{b_2} \slash\!\!\! {f}_{b_1}}{ \sigma_{i,b_2} \sigma_{b_2, b_1}} + (b_1\leftrightarrow b_2) \ .
\eeq
With this definition, the numerator $(\chi_1 \gamma^\mu \chi_2)( \chi_3 \gamma_\mu \chi_4)$ in (\ref{E40}) is generalized to $({\cal X}_{1, B_1} \, \gamma^\mu \,{\cal X}_{2, B_2} ) \ $ $\times({\cal X}_{3, B_3}\, \gamma_\mu \, {\cal X}_{4, B_4})$, which has four sets of field-strength insertions $B_1, B_2, B_3, B_4$, associated with fermions $1,2,3,4$, respectively. This is symmetric for bosons in each set $B_i$ ($i=1,2,3,4$), and by the gamma-matrix identity
\beq
(\gamma^{\la \rho} \gamma_\mu)_{\alpha \beta} \gamma^\mu_{\gamma\delta}
 - ( \gamma_\mu \gamma^{\la \rho})_{\alpha \beta} \gamma^\mu_{\gamma\delta} 
 + \gamma^\mu_{\alpha \beta} (\gamma^{\la \rho} \gamma_\mu)_{\gamma \delta}
 -  \gamma^\mu_{\alpha \beta} ( \gamma_\mu \gamma^{\la \rho})_{\gamma \delta} = 0
\eeq
underlying (\ref{falloff}) has the correct SL$_2$ weights for the $\sigma_j$ of all the bosons involved. 
Now it becomes clear how to write down the general form of the $n$-point correlator with four fermions:
\begin{align}
 \Big \langle &\Big( \prod_{i=1}^4 V^{(-\frac 1 2)}_i(\sigma_i)  \Big) \, \Big( \prod_{j=5}^n V^{(0)}_j(\sigma_j) \Big) \Big \rangle= \frac{1}{4} \sum_{ \{5,6,\ldots,n\} = A \atop{\cup B_1\cup\ldots \cup B_4}} \! \! \! \Pf \Psi_A \label{E4} \\
 &\times
 \left(\frac{ ({\cal X}_{1, B_1} \gamma^\mu {\cal X}_{2, B_2} )( {\cal X}_{3, B_3} \gamma_\mu {\cal X}_{4, B_4} )}{ \sigma_{12} \sigma_{23} \sigma_{34} \sigma_{41}}+\frac{ ({\cal X}_{1, B_1} \gamma^\mu {\cal X}_{3, B_3} )
 ({\cal X}_{2, B_2} \gamma_\mu {\cal X}_{4, B_4})}{  \sigma_{13} \sigma_{32} \sigma_{24} \sigma_{41} }\right)\, .\notag
\end{align}
It would be interesting to apply double-forward limits of this result to supersymmetric
two-loop amplitudes \cite{Geyer:2019hnn}.

%%%%%%%%%%%%%%%%%%%%%%%%%%%%%%
%%%%%%%%%%%%%%%%%%%%%%%%%%%%%%
%%%%%%%%%%%%%%%%%%%%%%%%%%%%%%

\section{Details of gamma-matrix traces}
\subsection{Decomposition of $\trs \to \trv$}
\label{app:traces}
Here we present a derivation of
the decomposition of $\trs$ in terms of $\trv$ given in
(\ref{eq3.11}).  First, we remind the reader of a well-known recursive
formula for calculating $\gamma$ traces.  Using that formula, we will
find relative signs and the overall factor for the length-$n$ $\trv$
within the length-$n$ $\trs$.  Then we show how the multitrace terms
arise from the recursive calculation of the $\gamma$ traces.  For
notational simplicity, we will focus on the sequential ordering of
labels $\trs(1,2,\ldots ,n)$ in (\ref{eq3.4}), with the understanding that that
other orderings can be reached by application of suitable
permutations.

The parity-even piece of a generic-length $\gamma$ trace, in arbitrary
dimension, can be computed using
\begin{equation}
  \treven{ \gamma_{\mu_1} \gamma_{\mu_2} \dots \gamma_{\mu_n}} = \sum_{j=2}^n (-1)^j \eta_{\mu_1 \mu_j} \treven{\gamma_{\mu_2} \dots \miss{\gamma}_{\mu_j} \dots \gamma_{\mu_n}} \, ,
  \label{TRrec}
\end{equation}
where $\tr(\gamma_{\mu_2} \dots \miss{\gamma}_{\mu_j} \dots
  \gamma_{\mu_n}  )$ is the trace of $n-2$ $\gamma$s with
$\gamma_{\mu_j}$ removed.  The recursion ends with
$\tr(id_{\text{CA}})$ which depends on the representation
of the Clifford algebra, and therefore carries the $D$ dependence of
the traces.  We can use this formula to evaluate the $\trs$, using
(\ref{eq2.10}) to rewrite
\begin{equation}
  \trs(1,2,\ldots,n) = 2^{-n} f_1^{\mu_1 \nu_1} \dots f_n^{\mu_n \nu_n} \treven{ \gamma_{\mu_1} \gamma_{\nu_1} \dots \gamma_{\mu_n} \gamma_{\nu_n}} \label{TRfexpl}
\end{equation}
and noting that $\eta_{\mu \nu} f^{\mu \nu}=0$ means that all terms that generate
$\eta_{\mu_i \nu_i}$ will not contribute to $\trs$.  To see the
patterns relevant to the $\trv$ decomposition, we will need to make
some clever use of the cyclic properties of $\tr$.  The maximal-length
$\trv$ terms can be written in terms of $\eta$ contractions as
\begin{equation}
  \frac{1}{2} \left( \trv(1,\sigma(2), \dots,\sigma(n)) + (-1)^{n} \trv(1 ,\sigma(n), \dots ,\sigma(2))\right) =  f_1^{\mu_1 \nu_1} \dots f_n^{\mu_n \nu_n} \eta_{\nu_1 \mu_{\sigma(2)}}\dots \eta_{\nu_{\sigma(n)} \mu_1}
  \label{TRcontr}
\end{equation}
with $\sigma \in S_{n-1}$, and the explicit reversal contribution is
included to demonstrate the factor of $2^{-j}$ in (\ref{eq2.10}).  The
$\eta_{\nu_1 \mu_{\sigma(2)}}$ term can be directly sourced out of
(\ref{TRrec}) by rotating the $\tr$ in (\ref{TRfexpl}) using cyclicity
so that $\gamma_{\nu_1}$ is the first in the string.  Then, since
$\gamma_{\mu_{\sigma(2)}}$ will always occupy the \emph{even} slots in
the trace, this term always carries a $+$.  The final
$\eta_{\nu_{\sigma(n)} \mu_1}$ can always be chosen as the last step
of the recursion (\ref{TRrec}), and thus also always carries a $+$.  However,
$\gamma_{\mu_1}$ remaining in the $\tr$ until the end is vitally
important, as it is what breaks the symmetry between the two
intermediate cases: $\sigma(i)$ coming before $\sigma(j)$ in
$1,2, \dots ,n$, or coming after.  If $\sigma(i)$ comes first, then the
$\tr$ can be cycled such that $\gamma_{\nu_{\sigma(i)}}$ is at the
front of the trace, and this cycling will never put $\gamma_{\mu_1}$
between $\gamma_{\nu_{\sigma(i)}}$ and $\gamma_{\mu_{\sigma(j)}}$.
Since each pair of $\gamma_\mu \gamma_\nu$ can be removed in adjacent
steps of this recursion, there will always be an even number of
$\gamma$ between $\gamma_{\nu_{\sigma(i)}}$ and
$\gamma_{\mu_{\sigma(j)}}$, and thus (\ref{TRrec}) will provide a $+$
contribution.  On the other hand, when $\sigma(i)$ comes after
$\sigma(j)$, the process of cycling $\gamma_{\nu_{\sigma(i)}}$ to the
front will always leave $\gamma_{\mu_1}$ between
$\gamma_{\nu_{\sigma(i)}}$ and $\gamma_{\mu_{\sigma(j)}}$.  As in the
previous case, there will always be an even number of $\gamma$s
removed between $\sigma(i)$ and $\sigma(j)$, but now $\gamma_{\mu_1}$
shifts the counting by $1$, so (\ref{TRrec}) will introduce a $-$
sign.  We collect all of the resulting signs into the $\ord$ function
introduced in (\ref{eq3.12}) to get
\begin{equation}
  \trs(1,2,\ldots,n) \Big|_{\text{single}} = 
  2^{D/2-n-2} \sum_{\rho \in S_{n-1}}\ord^{id}_{\rho}\trv(1,\rho)
\end{equation}
which provides the leading trace term from (\ref{eq3.11}). Notably,
the reversed $\trv$ from (\ref{TRcontr}) is included as one of the
elements of $\rho$.

The recursive realization of the $\tr(\gamma \dots)$ in (\ref{TRrec})
also naturally generates the multi-Lorentz-trace terms.  Each of the
subtraces can be resolved, one at a time, in the same method as above.
The $\gamma$ not participating in the targeted subtrace always cycle
together, and thus only shift the counting between targeted $\gamma$
by an even number, never changing the sign.  Each $\trv$ picks up a
factor of $\frac{1}{2}$ as in (\ref{TRcontr}) to account for the
reversal overcount, leading to the factor of $2^{-j}$ in
(\ref{eq3.11}).

\subsection{Higher $t_{2n}$ tensors from $D=10$ SYM}
\label{not12}

This appendix gives more details on the permutation symmetric tensors
$t_{2n}$ defined in (\ref{3.t2n}). More specifically, we will
determine the coefficients of $\trv(1,2,\dots ,n)$ once the spinor
traces are rewritten in terms of vectorial ones via
(\ref{eq3.11}). This will allow to verify the cancellation of the
six-trace from the exceptionally simple expression (\ref{simpt12}) for
$t_{12}$.

Using (\ref{eq3.11}), we can count the $+$ and $-$ contributions of
the longest $\trv(1,2,\dots ,n)$ to the permutation sum (\ref{3.t2n})
defining $t_{2n}$.  Since $t_{2n}$ is fully permutation symmetric, it
suffices to count the number of permutations in $S_{n-1}$ that
generate a positive coefficient for $\trv(1,2,\dots ,n)$ vs those that
generate a negative one.  These counts can be expressed directly in
terms of the Eulerian numbers
\begin{equation}
  \eulern{i}{k} = \sum_{j=0}^{k+1} (-1)^j \binom{i+1}{j} (k-j+1)^i \,,
\end{equation}
which count the number of permutations of length $i$ that have $k$
\emph{permutation ascents}; adjacent labels in the permutation $\rho$
that have $\rho_j < \rho_{j{+}1}$ are a \emph{permutation ascent}.
This is exactly the information needed by the $\ord^\rho_\sigma$ sign
(\ref{eq3.12}), and as such those terms with $k$ even will carry a $+$
sign, while $k$ odd will carry a $-$.\footnote{The sign of $\ord$ is
  actually set by $i-k$, but since $i$ is even, $i-k$ and $k$ have the
  same parity.}  The symmetric tensor $t_{2n}(f_1, f_2 , \dots ,f_n)$
will contain the term $\trv(1,2,\ldots,n)$ with a coefficient given by
\begin{align}
  \coef(t_{2n}(f_1, f_2, f_3, \dots, f_n) , \trv(1,2,\ldots,n)) &= \frac{2}{(n{-}1)!}\left[ 2 -   2^{3-n}\sum_{k=0}^{n-1} (-1)^k \eulern{n-1}{k}\right] \,.
\end{align}
The additional overall factor of $2$ is due to the parity properties
(\ref{eq3.p}). As a necessary condition for the simplification
(\ref{simpt12}) of $t_{12}$, the case with $n=6$ gives
\begin{align}
  \coef(t_{12}(f_1, f_2 ,\dots, f_6), \trv(1,2,\ldots,6)) &=\frac{1}{60} \left[  2 -   2^{-3}\sum_{k=0}^{5} (-1)^k \eulern{5}{k}\right] \nonumber \\
  &=\frac{1}{60}\left[2 - \frac{1}{8} \left(68 - 52\right)\right] = 0\, ,
\end{align}
so there is \emph{no} contribution of $\trv(1,2,\ldots,6)$ to
correlators (\ref{eq3.22}) of $D=10$ SYM up to and including seven
points.  However, \emph{all} other even $n$ admit length-$n$ Lorentz
traces.

\subsection{Parity-odd traces}
\label{sec:po-trace}
In this appendix, we derive the parity-odd $\gamma$ trace expansion
used in (\ref{odd.7}).  Namely, our goal is to work out an evaluation
of $\trodd{\gamma^{\mu_1} \dots}$ defined by (\ref{deftrodd}) in terms of $\varepsilon^{\mu_i \dots}$.
We start by making the standard identification
of tensor structures\footnote{This identification is based on the representation of the $2^{D/2}\times 2^{D/2}$
chirality matrix $\Gamma_{D+1}$ in terms of antisymmetrized products $\varepsilon^{\mu_1 \mu_2 \dots \mu_D}\Gamma_{\mu_1} \Gamma_{\mu_2} \ldots \Gamma_{\mu_D}$ of Dirac gamma matrices $\Gamma_{\mu_j}$.}
\begin{align}
  \trodd{\gamma^{\mu_1} \gamma^{\mu_2} \ldots \gamma^{\mu_D} } = i 2^{D/2-1} \varepsilon^{\mu_1 \dots \mu_D}
  = \frac{i}{D!} \varepsilon^{\nu_1 \dots \nu_D} \left(\treven{\gamma_{\nu_1}\dots \gamma_{\nu_D} \gamma^{\mu_1} \dots \gamma^{\mu_D}} \right)
\end{align}
which gives us the natural extension to more $\gamma$ in $\trodd{\ldots}$
\begin{equation}
  \trodd{\gamma^{\mu_1} \dots \gamma^{\mu_{D{+}1}} \gamma^{\mu_{D{+}2}}} = 
  \frac{i}{D!} \varepsilon^{\nu_1 \dots \nu_D} \left(\treven{\gamma_{\nu_1}\dots \gamma_{\nu_D} \gamma^{\mu_1} \dots \gamma^{\mu_{D{+}1}} \gamma^{\mu_{D{+}2}}} \right) \,.
\end{equation}
From here, we could directly run the recursive evaluation from
(\ref{TRrec}) on the right-hand side.  However, it is worth pointing out an
interesting feature of the calculation: the evaluation order will
fully contract the $\varepsilon^{\nu \dots}$ first, and then leave
behind $\tr(\gamma^{\mu_i} \gamma^{\mu_j})$ that are not contracted into the
$\varepsilon$.  Thus, running the recursive evaluation until the
$\varepsilon$ is completely contracted, we find
\begin{equation}
  \trodd{\gamma^{\mu_1} \dots \gamma^{\mu_{D{+}1}} \gamma^{\mu_{D{+}2}}} =
 \! \! \! \! \! \! \! \!  \sum_{A^{(D)} \cup B^{(2)} = \{1,2, \dots, D{+}2\}}  \! \! \! \! \! \! \! \!  i(-1)^{\sk^A_B}\varepsilon^{\mu_{A_1} \dots \mu_{A_D}} \left(\treven{\gamma^{\mu_{B_1}} \gamma^{\mu_{B_2}}}\right)\, ,
  \label{eq:po-d2}
\end{equation}
where the summation range $A^{(D)} \cup B^{(2)} = \{1,2,\dots, D{+}2\}$
follows our convention of $A$ and $B$ being disjoint ordered subsets
of $\{1,2,\dots, D{+}2\}$, with the additional constraint that $A$ has
length $D$, and $B$ length $2$.  The sign $(-1)^{\sk^A_B}$ compensates
for skipping over the $\gamma^{\mu_{B_i}}$ as the $\gamma^{\mu_{A_j}}$
are paired with $\gamma_{\nu_k}$, ensuring that all of the terms in
the remaining $B$ trace eventually have the correct relative signs.
For the simple case in (\ref{eq:po-d2}), $\sk^A_B$ is the number of
$A_i$ between the two elements of $B$, which can in turn be reduced to
the representation given in~(\ref{odd.7}).

In order to generalize this computation to larger numbers $(D+2j)$ of $\gamma$s with
$j\geq 2$, we need to more carefully account for $\sk$.  As mentioned, it needs to restore the
signs required in (\ref{TRrec}) that were dropped when separating the
indices into the $A$ and $B$ set.  A convenient definition for
$\sk^A_B$ that accomplishes this is
\begin{equation}
  \sk^A_B = \sum_{i=1}^{D} \text{number of elements of }B\text{ before } A_{i} \,.
\end{equation}
Note that this definition exactly captures the behavior described by
(\ref{odd.7}): an even separation between the $B_i$ will have
\begin{equation}
  \sk^A_B = 0 + \dots + \underbrace{1+\dots+1}_{\text{even}} + 2+\dots \to (-1)^{\sk^A_B} = 1 \,,
\end{equation}
whereas an odd separation will give
\begin{equation}
  \sk^A_B = 0 + \dots + \underbrace{1+\dots+1}_{\text{odd}} + 2 +\dots \to (-1)^{\sk^A_B} = -1 \,.
\end{equation}
All of these considerations allow us to
generalize the computation fully
\begin{equation}
  \trodd{\gamma^{\mu_1} \dots  \gamma^{\mu_{D{+}2j}}} =\! \! \! \! \! \! \! \!
  \sum_{A^{(D)} \cup B^{(2j)}  \atop{= \{1, 2,\dots, D{+}2j\}}}\! \! \! \! \! \! \! \! i(-1)^{\sk^A_B}\varepsilon^{\mu_{A_1} \dots \mu_{A_D}} \left(\treven{\gamma^{\mu_{B_1}}\dots \gamma^{\mu_{B_{2j}}}}\right) \,.
\end{equation}
Notably, this construction specifically includes (\ref{odd.2}) as
the $j=0$ case using $\tr(id_{\text{CA}}) = 2^{D/2-1}$.

%%%%%%%%%%%%%%%%%%%
%%%%%%%%%%%%%%%%%%%
%%%%%%%%%%%%%%%%%%%

%\bibliographystyle{JHEP}
%\bibliography{cites}{}

\begin{thebibliography}{100}

\bibitem{BCJ}
Z.~Bern, J.~J.~M. Carrasco and H.~Johansson, \emph{{New Relations for
  Gauge-Theory Amplitudes}},
  \href{http://dx.doi.org/10.1103/PhysRevD.78.085011}{\emph{Phys. Rev.} {\bf
  D78} (2008) 085011}, [\href{http://arxiv.org/abs/0805.3993}{{\tt
  0805.3993}}].

\bibitem{loopBCJ}
Z.~Bern, J.~J.~M. Carrasco and H.~Johansson, \emph{{Perturbative Quantum
  Gravity as a Double Copy of Gauge Theory}},
  \href{http://dx.doi.org/10.1103/PhysRevLett.105.061602}{\emph{Phys. Rev.
  Lett.} {\bf 105} (2010) 061602}, [\href{http://arxiv.org/abs/1004.0476}{{\tt
  1004.0476}}].

\bibitem{Bern:2017yxu}
Z.~Bern, J.~J. Carrasco, W.-M. Chen, H.~Johansson and R.~Roiban, \emph{{Gravity
  Amplitudes as Generalized Double Copies of Gauge-Theory Amplitudes}},
  \href{http://dx.doi.org/10.1103/PhysRevLett.118.181602}{\emph{Phys. Rev.
  Lett.} {\bf 118} (2017) 181602}, [\href{http://arxiv.org/abs/1701.02519}{{\tt
  1701.02519}}].

\bibitem{Bern:2019prr}
Z.~Bern, J.~J. Carrasco, M.~Chiodaroli, H.~Johansson and R.~Roiban, \emph{{The
  Duality Between Color and Kinematics and its Applications}},
  \href{http://arxiv.org/abs/1909.01358}{{\tt 1909.01358}}.

\bibitem{Kawai:1985xq}
H.~Kawai, D.~C. Lewellen and S.~H.~H. Tye, \emph{{A Relation Between Tree
  Amplitudes of Closed and Open Strings}},
  \href{http://dx.doi.org/10.1016/0550-3213(86)90362-7}{\emph{Nucl. Phys.} {\bf
  B269} (1986) 1--23}.

\bibitem{Bern:2010yg}
Z.~Bern, T.~Dennen, Y.-t. Huang and M.~Kiermaier, \emph{{Gravity as the Square
  of Gauge Theory}},
  \href{http://dx.doi.org/10.1103/PhysRevD.82.065003}{\emph{Phys. Rev.} {\bf
  D82} (2010) 065003}, [\href{http://arxiv.org/abs/1004.0693}{{\tt
  1004.0693}}].

\bibitem{Bern:2012uf}
Z.~Bern, J.~J.~M. Carrasco, L.~J. Dixon, H.~Johansson and R.~Roiban,
  \emph{{Simplifying Multiloop Integrands and Ultraviolet Divergences of Gauge
  Theory and Gravity Amplitudes}},
  \href{http://dx.doi.org/10.1103/PhysRevD.85.105014}{\emph{Phys. Rev.} {\bf
  D85} (2012) 105014}, [\href{http://arxiv.org/abs/1201.5366}{{\tt
  1201.5366}}].

\bibitem{Bern:2012cd}
Z.~Bern, S.~Davies, T.~Dennen and Y.-t. Huang, \emph{{Absence of Three-Loop
  Four-Point Divergences in N=4 Supergravity}},
  \href{http://dx.doi.org/10.1103/PhysRevLett.108.201301}{\emph{Phys. Rev.
  Lett.} {\bf 108} (2012) 201301}, [\href{http://arxiv.org/abs/1202.3423}{{\tt
  1202.3423}}].

\bibitem{Bern:2013uka}
Z.~Bern, S.~Davies, T.~Dennen, A.~V. Smirnov and V.~A. Smirnov,
  \emph{{Ultraviolet Properties of N=4 Supergravity at Four Loops}},
  \href{http://dx.doi.org/10.1103/PhysRevLett.111.231302}{\emph{Phys. Rev.
  Lett.} {\bf 111} (2013) 231302}, [\href{http://arxiv.org/abs/1309.2498}{{\tt
  1309.2498}}].

\bibitem{Bern:2014sna}
Z.~Bern, S.~Davies and T.~Dennen, \emph{{Enhanced ultraviolet cancellations in
  $\mathcal N=5$ supergravity at four loops}},
  \href{http://dx.doi.org/10.1103/PhysRevD.90.105011}{\emph{Phys. Rev.} {\bf
  D90} (2014) 105011}, [\href{http://arxiv.org/abs/1409.3089}{{\tt
  1409.3089}}].

\bibitem{Bern:2017ucb}
Z.~Bern, J.~J.~M. Carrasco, W.-M. Chen, H.~Johansson, R.~Roiban and M.~Zeng,
  \emph{{Five-loop four-point integrand of $N=8$ supergravity as a generalized
  double copy}},
  \href{http://dx.doi.org/10.1103/PhysRevD.96.126012}{\emph{Phys. Rev.} {\bf
  D96} (2017) 126012}, [\href{http://arxiv.org/abs/1708.06807}{{\tt
  1708.06807}}].

\bibitem{Bern:2018jmv}
Z.~Bern, J.~J. Carrasco, W.-M. Chen, A.~Edison, H.~Johansson, J.~Parra-Martinez
  et~al., \emph{{Ultraviolet Properties of $\mathcal N = 8$ Supergravity at
  Five Loops}}, \href{http://dx.doi.org/10.1103/PhysRevD.98.086021}{\emph{Phys.
  Rev.} {\bf D98} (2018) 086021}, [\href{http://arxiv.org/abs/1804.09311}{{\tt
  1804.09311}}].

\bibitem{Mafra:2011kj}
C.~R. Mafra, O.~Schlotterer and S.~Stieberger, \emph{{Explicit BCJ Numerators
  from Pure Spinors}},
  \href{http://dx.doi.org/10.1007/JHEP07(2011)092}{\emph{JHEP} {\bf 07} (2011)
  092}, [\href{http://arxiv.org/abs/1104.5224}{{\tt 1104.5224}}].

\bibitem{Mafra:2011nv}
C.~R. Mafra, O.~Schlotterer and S.~Stieberger, \emph{{Complete N-Point
  Superstring Disk Amplitude I. Pure Spinor Computation}},
  \href{http://dx.doi.org/10.1016/j.nuclphysb.2013.04.023}{\emph{Nucl. Phys.}
  {\bf B873} (2013) 419--460}, [\href{http://arxiv.org/abs/1106.2645}{{\tt
  1106.2645}}].

\bibitem{Mafra:2014gja}
C.~R. Mafra and O.~Schlotterer, \emph{{Towards one-loop SYM amplitudes from the
  pure spinor BRST cohomology}},
  \href{http://dx.doi.org/10.1002/prop.201400076}{\emph{Fortsch. Phys.} {\bf
  63} (2015) 105--131}, [\href{http://arxiv.org/abs/1410.0668}{{\tt
  1410.0668}}].

\bibitem{He:2015wgf}
S.~He, R.~Monteiro and O.~Schlotterer, \emph{{String-inspired BCJ numerators
  for one-loop MHV amplitudes}},
  \href{http://dx.doi.org/10.1007/JHEP01(2016)171}{\emph{JHEP} {\bf 01} (2016)
  171}, [\href{http://arxiv.org/abs/1507.06288}{{\tt 1507.06288}}].

\bibitem{Mafra:2015mja}
C.~R. Mafra and O.~Schlotterer, \emph{{Two-loop five-point amplitudes of super
  Yang-Mills and supergravity in pure spinor superspace}},
  \href{http://dx.doi.org/10.1007/JHEP10(2015)124}{\emph{JHEP} {\bf 10} (2015)
  124}, [\href{http://arxiv.org/abs/1505.02746}{{\tt 1505.02746}}].

\bibitem{BjerrumBohr:2009rd}
N.~E.~J. Bjerrum-Bohr, P.~H. Damgaard and P.~Vanhove, \emph{{Minimal Basis for
  Gauge Theory Amplitudes}},
  \href{http://dx.doi.org/10.1103/PhysRevLett.103.161602}{\emph{Phys. Rev.
  Lett.} {\bf 103} (2009) 161602}, [\href{http://arxiv.org/abs/0907.1425}{{\tt
  0907.1425}}].

\bibitem{Stieberger:2009hq}
S.~Stieberger, \emph{{Open \& Closed vs. Pure Open String Disk Amplitudes}},
  \href{http://arxiv.org/abs/0907.2211}{{\tt 0907.2211}}.

\bibitem{Tourkine:2016bak}
P.~Tourkine and P.~Vanhove, \emph{{Higher-loop amplitude monodromy relations in
  string and gauge theory}},
  \href{http://dx.doi.org/10.1103/PhysRevLett.117.211601}{\emph{Phys. Rev.
  Lett.} {\bf 117} (2016) 211601}, [\href{http://arxiv.org/abs/1608.01665}{{\tt
  1608.01665}}].

\bibitem{Hohenegger:2017kqy}
S.~Hohenegger and S.~Stieberger, \emph{{Monodromy Relations in Higher-Loop
  String Amplitudes}},
  \href{http://dx.doi.org/10.1016/j.nuclphysb.2017.09.020}{\emph{Nucl. Phys.}
  {\bf B925} (2017) 63--134}, [\href{http://arxiv.org/abs/1702.04963}{{\tt
  1702.04963}}].

\bibitem{Ochirov:2017jby}
A.~Ochirov, P.~Tourkine and P.~Vanhove, \emph{{One-loop monodromy relations on
  single cuts}}, \href{http://dx.doi.org/10.1007/JHEP10(2017)105}{\emph{JHEP}
  {\bf 10} (2017) 105}, [\href{http://arxiv.org/abs/1707.05775}{{\tt
  1707.05775}}].

\bibitem{Tourkine:2019ukp}
P.~Tourkine, \emph{{On integrands and loop momentum in string and field
  theory}},  \href{http://arxiv.org/abs/1901.02432}{{\tt 1901.02432}}.

\bibitem{Casali:2019ihm}
E.~Casali, S.~Mizera and P.~Tourkine, \emph{{Monodromy relations from twisted
  homology}}, \href{http://dx.doi.org/10.1007/JHEP12(2019)087}{\emph{JHEP} {\bf
  12} (2019) 087}, [\href{http://arxiv.org/abs/1910.08514}{{\tt 1910.08514}}].

\bibitem{Cachazo:2013hca}
F.~Cachazo, S.~He and E.~Y. Yuan, \emph{{Scattering of Massless Particles in
  Arbitrary Dimensions}},
  \href{http://dx.doi.org/10.1103/PhysRevLett.113.171601}{\emph{Phys. Rev.
  Lett.} {\bf 113} (2014) 171601}, [\href{http://arxiv.org/abs/1307.2199}{{\tt
  1307.2199}}].

\bibitem{Cachazo:2013iea}
F.~Cachazo, S.~He and E.~Y. Yuan, \emph{{Scattering of Massless Particles:
  Scalars, Gluons and Gravitons}},
  \href{http://dx.doi.org/10.1007/JHEP07(2014)033}{\emph{JHEP} {\bf 07} (2014)
  033}, [\href{http://arxiv.org/abs/1309.0885}{{\tt 1309.0885}}].

\bibitem{Cachazo:2013gna}
F.~Cachazo, S.~He and E.~Y. Yuan, \emph{{Scattering equations and
  Kawai-Lewellen-Tye orthogonality}},
  \href{http://dx.doi.org/10.1103/PhysRevD.90.065001}{\emph{Phys. Rev.} {\bf
  D90} (2014) 065001}, [\href{http://arxiv.org/abs/1306.6575}{{\tt
  1306.6575}}].

\bibitem{Mason:2013sva}
L.~Mason and D.~Skinner, \emph{{Ambitwistor strings and the scattering
  equations}}, \href{http://dx.doi.org/10.1007/JHEP07(2014)048}{\emph{JHEP}
  {\bf 07} (2014) 048}, [\href{http://arxiv.org/abs/1311.2564}{{\tt
  1311.2564}}].

\bibitem{Berkovits:2013xba}
N.~Berkovits, \emph{{Infinite Tension Limit of the Pure Spinor Superstring}},
  \href{http://dx.doi.org/10.1007/JHEP03(2014)017}{\emph{JHEP} {\bf 03} (2014)
  017}, [\href{http://arxiv.org/abs/1311.4156}{{\tt 1311.4156}}].

\bibitem{Adamo:2013tsa}
T.~Adamo, E.~Casali and D.~Skinner, \emph{{Ambitwistor strings and the
  scattering equations at one loop}},
  \href{http://dx.doi.org/10.1007/JHEP04(2014)104}{\emph{JHEP} {\bf 04} (2014)
  104}, [\href{http://arxiv.org/abs/1312.3828}{{\tt 1312.3828}}].

\bibitem{Adamo:2015hoa}
T.~Adamo and E.~Casali, \emph{{Scattering equations, supergravity integrands,
  and pure spinors}},
  \href{http://dx.doi.org/10.1007/JHEP05(2015)120}{\emph{JHEP} {\bf 05} (2015)
  120}, [\href{http://arxiv.org/abs/1502.06826}{{\tt 1502.06826}}].

\bibitem{Casali:2015vta}
E.~Casali, Y.~Geyer, L.~Mason, R.~Monteiro and K.~A. Roehrig, \emph{{New
  Ambitwistor String Theories}},
  \href{http://dx.doi.org/10.1007/JHEP11(2015)038}{\emph{JHEP} {\bf 11} (2015)
  038}, [\href{http://arxiv.org/abs/1506.08771}{{\tt 1506.08771}}].

\bibitem{Cachazo:2014xea}
F.~Cachazo, S.~He and E.~Y. Yuan, \emph{{Scattering Equations and Matrices:
  From Einstein To Yang-Mills, DBI and NLSM}},
  \href{http://dx.doi.org/10.1007/JHEP07(2015)149}{\emph{JHEP} {\bf 07} (2015)
  149}, [\href{http://arxiv.org/abs/1412.3479}{{\tt 1412.3479}}].

\bibitem{Chiodaroli:2014xia}
M.~Chiodaroli, M.~Gunaydin, H.~Johansson and R.~Roiban, \emph{{Scattering
  amplitudes in $ \mathcal{N}=2 $ Maxwell-Einstein and Yang-Mills/Einstein
  supergravity}}, \href{http://dx.doi.org/10.1007/JHEP01(2015)081}{\emph{JHEP}
  {\bf 01} (2015) 081}, [\href{http://arxiv.org/abs/1408.0764}{{\tt
  1408.0764}}].

\bibitem{Chiodaroli:2017ngp}
M.~Chiodaroli, M.~Gunaydin, H.~Johansson and R.~Roiban, \emph{{Explicit
  Formulae for Yang-Mills-Einstein Amplitudes from the Double Copy}},
  \href{http://dx.doi.org/10.1007/JHEP07(2017)002}{\emph{JHEP} {\bf 07} (2017)
  002}, [\href{http://arxiv.org/abs/1703.00421}{{\tt 1703.00421}}].

\bibitem{Geyer:2015bja}
Y.~Geyer, L.~Mason, R.~Monteiro and P.~Tourkine, \emph{{Loop Integrands for
  Scattering Amplitudes from the Riemann Sphere}},
  \href{http://dx.doi.org/10.1103/PhysRevLett.115.121603}{\emph{Phys. Rev.
  Lett.} {\bf 115} (2015) 121603}, [\href{http://arxiv.org/abs/1507.00321}{{\tt
  1507.00321}}].

\bibitem{Geyer:2015jch}
Y.~Geyer, L.~Mason, R.~Monteiro and P.~Tourkine, \emph{{One-loop amplitudes on
  the Riemann sphere}},
  \href{http://dx.doi.org/10.1007/JHEP03(2016)114}{\emph{JHEP} {\bf 03} (2016)
  114}, [\href{http://arxiv.org/abs/1511.06315}{{\tt 1511.06315}}].

\bibitem{Cachazo:2015aol}
F.~Cachazo, S.~He and E.~Y. Yuan, \emph{{One-Loop Corrections from Higher
  Dimensional Tree Amplitudes}},
  \href{http://dx.doi.org/10.1007/JHEP08(2016)008}{\emph{JHEP} {\bf 08} (2016)
  008}, [\href{http://arxiv.org/abs/1512.05001}{{\tt 1512.05001}}].

\bibitem{Geyer:2016wjx}
Y.~Geyer, L.~Mason, R.~Monteiro and P.~Tourkine, \emph{{Two-Loop Scattering
  Amplitudes from the Riemann Sphere}},
  \href{http://dx.doi.org/10.1103/PhysRevD.94.125029}{\emph{Phys. Rev.} {\bf
  D94} (2016) 125029}, [\href{http://arxiv.org/abs/1607.08887}{{\tt
  1607.08887}}].

\bibitem{Geyer:2017ela}
Y.~Geyer and R.~Monteiro, \emph{{Gluons and gravitons at one loop from
  ambitwistor strings}},
  \href{http://dx.doi.org/10.1007/JHEP03(2018)068}{\emph{JHEP} {\bf 03} (2018)
  068}, [\href{http://arxiv.org/abs/1711.09923}{{\tt 1711.09923}}].

\bibitem{Geyer:2018xwu}
Y.~Geyer and R.~Monteiro, \emph{{Two-Loop Scattering Amplitudes from
  Ambitwistor Strings: from Genus Two to the Nodal Riemann Sphere}},
  \href{http://dx.doi.org/10.1007/JHEP11(2018)008}{\emph{JHEP} {\bf 11} (2018)
  008}, [\href{http://arxiv.org/abs/1805.05344}{{\tt 1805.05344}}].

\bibitem{Geyer:2019hnn}
Y.~Geyer, R.~Monteiro and R.~Stark-Much{\~a}o, \emph{{Two-Loop Scattering
  Amplitudes: Double-Forward Limit and Colour-Kinematics Duality}},
  \href{http://dx.doi.org/10.1007/JHEP12(2019)049}{\emph{JHEP} {\bf 12} (2019)
  049}, [\href{http://arxiv.org/abs/1908.05221}{{\tt 1908.05221}}].

\bibitem{He:2016mzd}
S.~He and O.~Schlotterer, \emph{{New Relations for Gauge-Theory and Gravity
  Amplitudes at Loop Level}},
  \href{http://dx.doi.org/10.1103/PhysRevLett.118.161601}{\emph{Phys. Rev.
  Lett.} {\bf 118} (2017) 161601}, [\href{http://arxiv.org/abs/1612.00417}{{\tt
  1612.00417}}].

\bibitem{He:2017spx}
S.~He, O.~Schlotterer and Y.~Zhang, \emph{{New BCJ representations for one-loop
  amplitudes in gauge theories and gravity}},
  \href{http://dx.doi.org/10.1016/j.nuclphysb.2018.03.003}{\emph{Nucl. Phys.}
  {\bf B930} (2018) 328--383}, [\href{http://arxiv.org/abs/1706.00640}{{\tt
  1706.00640}}].

\bibitem{He:2015yua}
S.~He and E.~Y. Yuan, \emph{{One-loop Scattering Equations and Amplitudes from
  Forward Limit}},
  \href{http://dx.doi.org/10.1103/PhysRevD.92.105004}{\emph{Phys. Rev.} {\bf
  D92} (2015) 105004}, [\href{http://arxiv.org/abs/1508.06027}{{\tt
  1508.06027}}].

\bibitem{Cardona:2016bpi}
C.~Cardona and H.~Gomez, \emph{{Elliptic scattering equations}},
  \href{http://dx.doi.org/10.1007/JHEP06(2016)094}{\emph{JHEP} {\bf 06} (2016)
  094}, [\href{http://arxiv.org/abs/1605.01446}{{\tt 1605.01446}}].

\bibitem{Cardona:2016wcr}
C.~Cardona and H.~Gomez, \emph{{CHY-Graphs on a Torus}},
  \href{http://dx.doi.org/10.1007/JHEP10(2016)116}{\emph{JHEP} {\bf 10} (2016)
  116}, [\href{http://arxiv.org/abs/1607.01871}{{\tt 1607.01871}}].

\bibitem{Gomez:2016bmv}
H.~Gomez, \emph{{$\Lambda$ scattering equations}},
  \href{http://dx.doi.org/10.1007/JHEP06(2016)101}{\emph{JHEP} {\bf 06} (2016)
  101}, [\href{http://arxiv.org/abs/1604.05373}{{\tt 1604.05373}}].

\bibitem{Gomez:2016cqb}
H.~Gomez, S.~Mizera and G.~Zhang, \emph{{CHY Loop Integrands from Holomorphic
  Forms}}, \href{http://dx.doi.org/10.1007/JHEP03(2017)092}{\emph{JHEP} {\bf
  03} (2017) 092}, [\href{http://arxiv.org/abs/1612.06854}{{\tt 1612.06854}}].

\bibitem{Gomez:2017lhy}
H.~Gomez, \emph{{Quadratic Feynman Loop Integrands From Massless Scattering
  Equations}}, \href{http://dx.doi.org/10.1103/PhysRevD.95.106006}{\emph{Phys.
  Rev.} {\bf D95} (2017) 106006}, [\href{http://arxiv.org/abs/1703.04714}{{\tt
  1703.04714}}].

\bibitem{Gomez:2017cpe}
H.~Gomez, C.~Lopez-Arcos and P.~Talavera, \emph{{One-loop Parke-Taylor factors
  for quadratic propagators from massless scattering equations}},
  \href{http://dx.doi.org/10.1007/JHEP10(2017)175}{\emph{JHEP} {\bf 10} (2017)
  175}, [\href{http://arxiv.org/abs/1707.08584}{{\tt 1707.08584}}].

\bibitem{Ahmadiniaz:2018nvr}
N.~Ahmadiniaz, H.~Gomez and C.~Lopez-Arcos, \emph{{Non-planar one-loop
  Parke-Taylor factors in the CHY approach for quadratic propagators}},
  \href{http://dx.doi.org/10.1007/JHEP05(2018)055}{\emph{JHEP} {\bf 05} (2018)
  055}, [\href{http://arxiv.org/abs/1802.00015}{{\tt 1802.00015}}].

\bibitem{Agerskov:2019ryp}
J.~Agerskov, N.~E.~J. Bjerrum-Bohr, H.~Gomez and C.~Lopez-Arcos,
  \emph{{Yang-Mills Loop Amplitudes from Scattering Equations}},
  \href{http://arxiv.org/abs/1910.03602}{{\tt 1910.03602}}.

\bibitem{Tsuchiya:1988va}
A.~Tsuchiya, \emph{{More on One Loop Massless Amplitudes of Superstring
  Theories}}, \href{http://dx.doi.org/10.1103/PhysRevD.39.1626}{\emph{Phys.
  Rev.} {\bf D39} (1989) 1626}.

\bibitem{Stieberger:2002wk}
S.~Stieberger and T.~R. Taylor, \emph{{NonAbelian Born-Infeld action and type
  1. - heterotic duality 2: Nonrenormalization theorems}},
  \href{http://dx.doi.org/10.1016/S0550-3213(02)00979-3}{\emph{Nucl. Phys.}
  {\bf B648} (2003) 3--34}, [\href{http://arxiv.org/abs/hep-th/0209064}{{\tt
  hep-th/0209064}}].

\bibitem{Bianchi:2006nf}
M.~Bianchi and A.~V. Santini, \emph{{String predictions for near future
  colliders from one-loop scattering amplitudes around D-brane worlds}},
  \href{http://dx.doi.org/10.1088/1126-6708/2006/12/010}{\emph{JHEP} {\bf 12}
  (2006) 010}, [\href{http://arxiv.org/abs/hep-th/0607224}{{\tt
  hep-th/0607224}}].

\bibitem{Broedel:2014vla}
J.~Broedel, C.~R. Mafra, N.~Matthes and O.~Schlotterer, \emph{{Elliptic
  multiple zeta values and one-loop superstring amplitudes}},
  \href{http://dx.doi.org/10.1007/JHEP07(2015)112}{\emph{JHEP} {\bf 07} (2015)
  112}, [\href{http://arxiv.org/abs/1412.5535}{{\tt 1412.5535}}].

\bibitem{Berg:2016wux}
M.~Berg, I.~Buchberger and O.~Schlotterer, \emph{{From maximal to minimal
  supersymmetry in string loop amplitudes}},
  \href{http://dx.doi.org/10.1007/JHEP04(2017)163}{\emph{JHEP} {\bf 04} (2017)
  163}, [\href{http://arxiv.org/abs/1603.05262}{{\tt 1603.05262}}].

\bibitem{Gomez:2013wza}
H.~Gomez and E.~Y. Yuan, \emph{{N-point tree-level scattering amplitude in the
  new Berkovits` string}},
  \href{http://dx.doi.org/10.1007/JHEP04(2014)046}{\emph{JHEP} {\bf 04} (2014)
  046}, [\href{http://arxiv.org/abs/1312.5485}{{\tt 1312.5485}}].

\bibitem{Lee:2015upy}
S.~Lee, C.~R. Mafra and O.~Schlotterer, \emph{{Non-linear gauge transformations
  in $D=10$ SYM theory and the BCJ duality}},
  \href{http://dx.doi.org/10.1007/JHEP03(2016)090}{\emph{JHEP} {\bf 03} (2016)
  090}, [\href{http://arxiv.org/abs/1510.08843}{{\tt 1510.08843}}].

\bibitem{Mafra:2015vca}
C.~R. Mafra and O.~Schlotterer, \emph{{Berends-Giele recursions and the BCJ
  duality in superspace and components}},
  \href{http://dx.doi.org/10.1007/JHEP03(2016)097}{\emph{JHEP} {\bf 03} (2016)
  097}, [\href{http://arxiv.org/abs/1510.08846}{{\tt 1510.08846}}].

\bibitem{Mafra:2016nwr}
C.~R. Mafra and O.~Schlotterer, \emph{{One-loop superstring six-point
  amplitudes and anomalies in pure spinor superspace}},
  \href{http://dx.doi.org/10.1007/JHEP04(2016)148}{\emph{JHEP} {\bf 04} (2016)
  148}, [\href{http://arxiv.org/abs/1603.04790}{{\tt 1603.04790}}].

\bibitem{Mafra:2018nla}
C.~R. Mafra and O.~Schlotterer, \emph{{Towards the n-point one-loop superstring
  amplitude. Part I. Pure spinors and superfield kinematics}},
  \href{http://dx.doi.org/10.1007/JHEP08(2019)090}{\emph{JHEP} {\bf 08} (2019)
  090}, [\href{http://arxiv.org/abs/1812.10969}{{\tt 1812.10969}}].

\bibitem{Mafra:2018qqe}
C.~R. Mafra and O.~Schlotterer, \emph{{Towards the n-point one-loop superstring
  amplitude. Part III. One-loop correlators and their double-copy structure}},
  \href{http://dx.doi.org/10.1007/JHEP08(2019)092}{\emph{JHEP} {\bf 08} (2019)
  092}, [\href{http://arxiv.org/abs/1812.10971}{{\tt 1812.10971}}].

\bibitem{Roehrig:2017gbt}
K.~A. Roehrig and D.~Skinner, \emph{{A Gluing Operator for the Ambitwistor
  String}}, \href{http://dx.doi.org/10.1007/JHEP01(2018)069}{\emph{JHEP} {\bf
  01} (2018) 069}, [\href{http://arxiv.org/abs/1709.03262}{{\tt 1709.03262}}].

\bibitem{Fu:2017uzt}
C.-H. Fu, Y.-J. Du, R.~Huang and B.~Feng, \emph{{Expansion of
  Einstein-Yang-Mills Amplitude}},
  \href{http://dx.doi.org/10.1007/JHEP09(2017)021}{\emph{JHEP} {\bf 09} (2017)
  021}, [\href{http://arxiv.org/abs/1702.08158}{{\tt 1702.08158}}].

\bibitem{Cardona:2016gon}
C.~Cardona, B.~Feng, H.~Gomez and R.~Huang, \emph{{Cross-ratio Identities and
  Higher-order Poles of CHY-integrand}},
  \href{http://dx.doi.org/10.1007/JHEP09(2016)133}{\emph{JHEP} {\bf 09} (2016)
  133}, [\href{http://arxiv.org/abs/1606.00670}{{\tt 1606.00670}}].

\bibitem{Nandan:2016pya}
D.~Nandan, J.~Plefka, O.~Schlotterer and C.~Wen, \emph{{Einstein-Yang-Mills
  from pure Yang-Mills amplitudes}},
  \href{http://dx.doi.org/10.1007/JHEP10(2016)070}{\emph{JHEP} {\bf 10} (2016)
  070}, [\href{http://arxiv.org/abs/1607.05701}{{\tt 1607.05701}}].

\bibitem{Bjerrum-Bohr:2016axv}
N.~E.~J. Bjerrum-Bohr, J.~L. Bourjaily, P.~H. Damgaard and B.~Feng,
  \emph{{Manifesting Color-Kinematics Duality in the Scattering Equation
  Formalism}}, \href{http://dx.doi.org/10.1007/JHEP09(2016)094}{\emph{JHEP}
  {\bf 09} (2016) 094}, [\href{http://arxiv.org/abs/1608.00006}{{\tt
  1608.00006}}].

\bibitem{Schlotterer:2016cxa}
O.~Schlotterer, \emph{{Amplitude relations in heterotic string theory and
  Einstein-Yang-Mills}},
  \href{http://dx.doi.org/10.1007/JHEP11(2016)074}{\emph{JHEP} {\bf 11} (2016)
  074}, [\href{http://arxiv.org/abs/1608.00130}{{\tt 1608.00130}}].

\bibitem{Teng:2017tbo}
F.~Teng and B.~Feng, \emph{{Expanding Einstein-Yang-Mills by Yang-Mills in CHY
  frame}}, \href{http://dx.doi.org/10.1007/JHEP05(2017)075}{\emph{JHEP} {\bf
  05} (2017) 075}, [\href{http://arxiv.org/abs/1703.01269}{{\tt 1703.01269}}].

\bibitem{AlexFei}
A.~Edison and F.~Teng, \emph{{Efficient Calculation of Crossing Symmetric BCJ
  Tree Numerators}},  \href{http://arxiv.org/abs/2005.03638}{{\tt 2005.03638}}.

\bibitem{Mafra:2014oia}
C.~R. Mafra and O.~Schlotterer, \emph{{Multiparticle SYM equations of motion
  and pure spinor BRST blocks}},
  \href{http://dx.doi.org/10.1007/JHEP07(2014)153}{\emph{JHEP} {\bf 07} (2014)
  153}, [\href{http://arxiv.org/abs/1404.4986}{{\tt 1404.4986}}].

\bibitem{Berends:1987me}
F.~A. Berends and W.~T. Giele, \emph{{Recursive Calculations for Processes with
  n Gluons}}, \href{http://dx.doi.org/10.1016/0550-3213(88)90442-7}{\emph{Nucl.
  Phys.} {\bf B306} (1988) 759--808}.

\bibitem{Bridges:2019siz}
E.~Bridges and C.~R. Mafra, \emph{{Algorithmic construction of SYM
  multiparticle superfields in the BCJ gauge}},
  \href{http://dx.doi.org/10.1007/JHEP10(2019)022}{\emph{JHEP} {\bf 10} (2019)
  022}, [\href{http://arxiv.org/abs/1906.12252}{{\tt 1906.12252}}].

\bibitem{Cachazo:2016njl}
F.~Cachazo, P.~Cha and S.~Mizera, \emph{{Extensions of Theories from Soft
  Limits}}, \href{http://dx.doi.org/10.1007/JHEP06(2016)170}{\emph{JHEP} {\bf
  06} (2016) 170}, [\href{http://arxiv.org/abs/1604.03893}{{\tt 1604.03893}}].

\bibitem{Ramond:1971gb}
P.~Ramond, \emph{{Dual Theory for Free Fermions}},
  \href{http://dx.doi.org/10.1103/PhysRevD.3.2415}{\emph{Phys. Rev.} {\bf D3}
  (1971) 2415--2418}.

\bibitem{Neveu:1971rx}
A.~Neveu and J.~H. Schwarz, \emph{{Factorizable dual model of pions}},
  \href{http://dx.doi.org/10.1016/0550-3213(71)90448-2}{\emph{Nucl. Phys.} {\bf
  B31} (1971) 86--112}.
  
\bibitem{DHoker:1988pdl}
E.~D'Hoker and D.~H. Phong, \emph{{The Geometry of String Perturbation
  Theory}}, \href{http://dx.doi.org/10.1103/RevModPhys.60.917}{\emph{Rev. Mod.
  Phys.} {\bf 60} (1988) 917}.

\bibitem{Friedan:1985ey}
D.~Friedan, S.~H. Shenker and E.~J. Martinec, \emph{{Covariant Quantization of
  Superstrings}},
  \href{http://dx.doi.org/10.1016/0370-2693(85)91466-2}{\emph{Phys. Lett.} {\bf
  B160} (1985) 55--61}.

\bibitem{Friedan:1985ge}
D.~Friedan, E.~J. Martinec and S.~H. Shenker, \emph{{Conformal Invariance,
  Supersymmetry and String Theory}},
  \href{http://dx.doi.org/10.1016/0550-3213(86)90356-1,
  10.1016/S0550-3213(86)80006-2}{\emph{Nucl. Phys.} {\bf B271} (1986) 93--165}.

\bibitem{Knizhnik:1985ke}
V.~G. Knizhnik, \emph{{Covariant Fermionic Vertex in Superstrings}},
  \href{http://dx.doi.org/10.1016/0370-2693(85)90009-7}{\emph{Phys. Lett.} {\bf
  160B} (1985) 403--407}.

\bibitem{Cohn:1986bn}
J.~Cohn, D.~Friedan, Z.-a. Qiu and S.~H. Shenker, \emph{{Covariant Quantization
  of Supersymmetric String Theories: The Spinor Field of the
  Ramond-Neveu-Schwarz Model}},
  \href{http://dx.doi.org/10.1016/0550-3213(86)90053-2}{\emph{Nucl. Phys.} {\bf
  B278} (1986) 577--600}.

\bibitem{Lam:2016tlk}
C.~S. Lam and Y.-P. Yao, \emph{{Evaluation of the Cachazo-He-Yuan gauge
  amplitude}}, \href{http://dx.doi.org/10.1103/PhysRevD.93.105008}{\emph{Phys.
  Rev.} {\bf D93} (2016) 105008}, [\href{http://arxiv.org/abs/1602.06419}{{\tt
  1602.06419}}].

\bibitem{Kostelecky:1986xg}
V.~A. Kostelecky, O.~Lechtenfeld, W.~Lerche, S.~Samuel and S.~Watamura,
  \emph{{Conformal Techniques, Bosonization and Tree Level String Amplitudes}},
  \href{http://dx.doi.org/10.1016/0550-3213(87)90213-6}{\emph{Nucl. Phys.} {\bf
  B288} (1987) 173--232}.

\bibitem{Frost:2017}
H.~Frost,
  \emph{\href{https://people.maths.ox.ac.uk/lmason/Theses/HF-Transfer.pdf}{New
  directions for the ambitwistor string}}.
\newblock Transfer Thesis, University of Oxford, 2017.

\bibitem{Atick:1986rs}
J.~J. Atick and A.~Sen, \emph{{Covariant One Loop Fermion Emission Amplitudes
  in Closed String Theories}},
  \href{http://dx.doi.org/10.1016/0550-3213(87)90075-7}{\emph{Nucl. Phys.} {\bf
  B293} (1987) 317--347}.

\bibitem{Kostelecky:1986ab}
V.~A. Kostelecky, O.~Lechtenfeld, S.~Samuel, D.~Verstegen, S.~Watamura and
  D.~Sahdev, \emph{{The Six Fermion Amplitude in the Superstring}},
  \href{http://dx.doi.org/10.1016/0370-2693(87)90968-3}{\emph{Phys. Lett.} {\bf
  B183} (1987) 299--303}.

\bibitem{Lee:2017ujn}
S.~Lee and O.~Schlotterer, \emph{{Fermionic one-loop amplitudes of the RNS
  superstring}}, \href{http://dx.doi.org/10.1007/JHEP03(2018)190}{\emph{JHEP}
  {\bf 03} (2018) 190}, [\href{http://arxiv.org/abs/1710.07353}{{\tt
  1710.07353}}].

\bibitem{Berkovits:2000fe}
N.~Berkovits, \emph{{Super Poincare covariant quantization of the
  superstring}},
  \href{http://dx.doi.org/10.1088/1126-6708/2000/04/018}{\emph{JHEP} {\bf 04}
  (2000) 018}, [\href{http://arxiv.org/abs/hep-th/0001035}{{\tt
  hep-th/0001035}}].

\bibitem{Green:1982sw}
M.~B. Green, J.~H. Schwarz and L.~Brink, \emph{{N=4 Yang-Mills and N=8
  Supergravity as Limits of String Theories}},
  \href{http://dx.doi.org/10.1016/0550-3213(82)90336-4}{\emph{Nucl. Phys.} {\bf
  B198} (1982) 474--492}.

\bibitem{vanRitbergen:1998pn}
T.~van Ritbergen, A.~N. Schellekens and J.~A.~M. Vermaseren, \emph{{Group
  theory factors for Feynman diagrams}},
  \href{http://dx.doi.org/10.1142/S0217751X99000038}{\emph{Int. J. Mod. Phys.}
  {\bf A14} (1999) 41--96}, [\href{http://arxiv.org/abs/hep-ph/9802376}{{\tt
  hep-ph/9802376}}].

\bibitem{Cachazo:2014nsa}
F.~Cachazo, S.~He and E.~Y. Yuan, \emph{{Einstein-Yang-Mills Scattering
  Amplitudes From Scattering Equations}},
  \href{http://dx.doi.org/10.1007/JHEP01(2015)121}{\emph{JHEP} {\bf 01} (2015)
  121}, [\href{http://arxiv.org/abs/1409.8256}{{\tt 1409.8256}}].

\bibitem{He:2018pol}
S.~He, F.~Teng and Y.~Zhang, \emph{{String amplitudes from field-theory
  amplitudes and vice versa}},
  \href{http://dx.doi.org/10.1103/PhysRevLett.122.211603}{\emph{Phys. Rev.
  Lett.} {\bf 122} (2019) 211603}, [\href{http://arxiv.org/abs/1812.03369}{{\tt
  1812.03369}}].

\bibitem{He:2019drm}
S.~He, F.~Teng and Y.~Zhang, \emph{{String Correlators: Recursive Expansion,
  Integration-by-Parts and Scattering Equations}},
  \href{http://dx.doi.org/10.1007/JHEP09(2019)085}{\emph{JHEP} {\bf 09} (2019)
  085}, [\href{http://arxiv.org/abs/1907.06041}{{\tt 1907.06041}}].

\bibitem{Bern:1994zx}
Z.~Bern, L.~J. Dixon, D.~C. Dunbar and D.~A. Kosower, \emph{{One loop n point
  gauge theory amplitudes, unitarity and collinear limits}},
  \href{http://dx.doi.org/10.1016/0550-3213(94)90179-1}{\emph{Nucl. Phys. B}
  {\bf 425} (1994) 217--260}, [\href{http://arxiv.org/abs/hep-ph/9403226}{{\tt
  hep-ph/9403226}}].

\bibitem{Bern:1993tz}
Z.~Bern and A.~Morgan, \emph{{Supersymmetry relations between contributions to
  one loop gauge boson amplitudes}},
  \href{http://dx.doi.org/10.1103/PhysRevD.49.6155}{\emph{Phys. Rev. D} {\bf
  49} (1994) 6155--6163}, [\href{http://arxiv.org/abs/hep-ph/9312218}{{\tt
  hep-ph/9312218}}].

\bibitem{Bern:1992ad}
Z.~Bern, \emph{{String based perturbative methods for gauge theories}},  in
  \emph{{Proceedings, Theoretical Advanced Study Institute (TASI 92): From
  Black Holes and Strings to Particles}: {Boulder, USA, June 1-26, 1992}},
  pp.~0471--536, 6, 1992.
\newblock \href{http://arxiv.org/abs/hep-ph/9304249}{{\tt hep-ph/9304249}}.

\bibitem{Bern:2013yya}
Z.~Bern, S.~Davies, T.~Dennen, Y.-t. Huang and J.~Nohle,
  \emph{{Color-Kinematics Duality for Pure Yang-Mills and Gravity at One and
  Two Loops}}, \href{http://dx.doi.org/10.1103/PhysRevD.92.045041}{\emph{Phys.
  Rev.} {\bf D92} (2015) 045041}, [\href{http://arxiv.org/abs/1303.6605}{{\tt
  1303.6605}}].

\bibitem{Clavelli:1986fj}
L.~Clavelli, P.~H. Cox and B.~Harms, \emph{{Parity Violating One Loop Six Point
  Function in Type I Superstring Theory}},
  \href{http://dx.doi.org/10.1103/PhysRevD.35.1908}{\emph{Phys. Rev.} {\bf D35}
  (1987) 1908}.

\bibitem{Gross:1987pd}
D.~J. Gross and P.~F. Mende, \emph{{Modular Subgroups, Odd Spin Structures and
  Gauge Invariance in the Heterotic String}},
  \href{http://dx.doi.org/10.1016/0550-3213(87)90489-5}{\emph{Nucl. Phys.} {\bf
  B291} (1987) 653--672}.



\bibitem{Frampton:1983ah}
P.~H. Frampton and T.~W. Kephart, \emph{{Explicit Evaluation of Anomalies in
  Higher Dimensions}}, \href{http://dx.doi.org/10.1103/PhysRevLett.51.232,
  10.1103/PhysRevLett.50.1343}{\emph{Phys. Rev. Lett.} {\bf 50} (1983) 1343}.

\bibitem{Frampton:1983nr}
P.~H. Frampton and T.~W. Kephart, \emph{{The Analysis of Anomalies in Higher
  Space-time Dimensions}},
  \href{http://dx.doi.org/10.1103/PhysRevD.28.1010}{\emph{Phys. Rev.} {\bf D28}
  (1983) 1010}.

\bibitem{Zumino:1983rz}
B.~Zumino, Y.-S. Wu and A.~Zee, \emph{{Chiral Anomalies, Higher Dimensions, and
  Differential Geometry}},
  \href{http://dx.doi.org/10.1016/0550-3213(84)90259-1}{\emph{Nucl. Phys.} {\bf
  B239} (1984) 477--507}.

\bibitem{Green:2013bza}
M.~B. Green, C.~R. Mafra and O.~Schlotterer, \emph{{Multiparticle one-loop
  amplitudes and S-duality in closed superstring theory}},
  \href{http://dx.doi.org/10.1007/JHEP10(2013)188}{\emph{JHEP} {\bf 10} (2013)
  188}, [\href{http://arxiv.org/abs/1307.3534}{{\tt 1307.3534}}].

\bibitem{Berg:2016fui}
M.~Berg, I.~Buchberger and O.~Schlotterer, \emph{{String-motivated one-loop
  amplitudes in gauge theories with half-maximal supersymmetry}},
  \href{http://dx.doi.org/10.1007/JHEP07(2017)138}{\emph{JHEP} {\bf 07} (2017)
  138}, [\href{http://arxiv.org/abs/1611.03459}{{\tt 1611.03459}}].

\bibitem{Chen:2014eva}
W.-M. Chen, Y.-t. Huang and D.~A. McGady, \emph{{Anomalies without an action}},
   \href{http://arxiv.org/abs/1402.7062}{{\tt 1402.7062}}.

\bibitem{Gomez:2013sla}
H.~Gomez and C.~R. Mafra, \emph{{The closed-string 3-loop amplitude and
  S-duality}}, \href{http://dx.doi.org/10.1007/JHEP10(2013)217}{\emph{JHEP}
  {\bf 10} (2013) 217}, [\href{http://arxiv.org/abs/1308.6567}{{\tt
  1308.6567}}].

\bibitem{Garozzo:2018uzj}
L.~M. Garozzo, L.~Queimada and O.~Schlotterer, \emph{{Berends-Giele currents in
  Bern-Carrasco-Johansson gauge for $F^3$- and $F^4$-deformed Yang-Mills
  amplitudes}}, \href{http://dx.doi.org/10.1007/JHEP02(2019)078}{\emph{JHEP}
  {\bf 02} (2019) 078}, [\href{http://arxiv.org/abs/1809.08103}{{\tt
  1809.08103}}].

\bibitem{Mafra:2014gsa}
C.~R. Mafra and O.~Schlotterer, \emph{{Cohomology foundations of one-loop
  amplitudes in pure spinor superspace}},
  \href{http://arxiv.org/abs/1408.3605}{{\tt 1408.3605}}.

\bibitem{Minahan:1987ha}
J.~A. Minahan, \emph{{One Loop Amplitudes on Orbifolds and the Renormalization
  of Coupling Constants}},
  \href{http://dx.doi.org/10.1016/0550-3213(88)90303-3}{\emph{Nucl. Phys.} {\bf
  B298} (1988) 36--74}.

\bibitem{Baadsgaard:2015twa}
C.~Baadsgaard, N.~E.~J. Bjerrum-Bohr, J.~L. Bourjaily, S.~Caron-Huot, P.~H.
  Damgaard and B.~Feng, \emph{{New Representations of the Perturbative
  S-Matrix}},
  \href{http://dx.doi.org/10.1103/PhysRevLett.116.061601}{\emph{Phys. Rev.
  Lett.} {\bf 116} (2016) 061601}, [\href{http://arxiv.org/abs/1509.02169}{{\tt
  1509.02169}}].

\bibitem{Hartl:2010ks}
D.~Haertl and O.~Schlotterer, \emph{{Higher Loop Spin Field Correlators in
  Various Dimensions}},
  \href{http://dx.doi.org/10.1016/j.nuclphysb.2011.03.022}{\emph{Nucl. Phys.}
  {\bf B849} (2011) 364--409}, [\href{http://arxiv.org/abs/1011.1249}{{\tt
  1011.1249}}].

\end{thebibliography}

\providecommand{\href}[2]{#2}\begingroup\raggedright\endgroup

%\providecommand{\href}[2]{#2}\begingroup\raggedright\begin{thebibliography}{10}
%
%
%\end{thebibliography}\endgroup

\end{document}